\documentclass[a4paper,fleqn]{cas-sc}
\usepackage[authoryear]{natbib}
\usepackage[utf8]{inputenc}
\usepackage[T1]{fontenc}
\usepackage{bm}
\usepackage{graphicx}
\usepackage{caption}
\usepackage{soul}
\usepackage{subcaption}
\usepackage{hyperref}
\usepackage{enumitem}
\hypersetup{
    colorlinks=true,
    linkcolor=blue,
    filecolor=magenta,      
    urlcolor=blue,
    citecolor=blue,
    }
\usepackage{tabularray}
\usepackage{algorithm2e}
\usepackage{multirow}
\usepackage{nicematrix}
\usepackage{mathtools,amssymb}

\usepackage{array}

\usepackage{float}
\usepackage{makecell}

\usepackage{pifont}

\usepackage{lineno}

\newcommand{\sif}{stress intensity factor}
\newcommand{\sifs}{stress intensity factors}

\newcommand{\indicator}{\scalebox{1.5}{$\mathbb{1}$}}

\newcommand{\llbracket}{[}
\newcommand{\rrbracket}{]}

\let\svthefootnote\thefootnote
\newcommand\freefootnote[1]{%
  \let\thefootnote\relax%
  \footnotetext{#1}%
  \let\thefootnote\svthefootnote%
}

\newlength{\boxwidth}
\def\btheorem{\begin{theorem}}
\def\etheorem{\end{theorem}}
\def\blemma{\begin{lemma}}
\def\elemma{\end{lemma}}
\def\bproposition{\begin{proposition}}
\def\eproposition{\end{proposition}}
\def\bcorollary{\begin{corollary}}
\def\ecorollary{\end{corollary}}
\def\bdefinition{\begin{definition}}
\def\edefinition{\end{definition}}
\def\bexample{\begin{example}}
\def\eexample{\end{example}}
\def\bremark{\begin{remark}}
\def\eremark{\end{remark}}

\DeclareMathOperator{\sign}{sign}

\newcommand{\be}{\begin{equation}}
\newcommand{\ee}{\end{equation}}
\newcommand{\beq}{\begin{eqnarray}}
\newcommand{\eeq}{\end{eqnarray}}
\newcommand{\bem}{\begin{multline}}
\newcommand{\eem}{\end{multline}}
\newcommand{\ba}{\begin{align}}
\newcommand{\ea}{\end{align}}

\graphicspath{{../}}

\newlist{todolist}{itemize}{2}
\setlist[todolist]{label=$\square$}

\def\tsc#1{\csdef{#1}{\textsc{\lowercase{#1}}\xspace}}
\tsc{WGM}
\tsc{QE}

\begin{document}
\let\WriteBookmarks\relax
\def\floatpagepagefraction{1}
\def\textpagefraction{.001}

\shortauthors{R. Brambilla et al.}  

\title[mode=title]{Curvature-induced fracture mode coupling in doubly curved shells}  
\shorttitle{Curvature-induced fracture mode coupling in doubly curved shells}

\affiliation[1]{organization={Department of Aerospace Structures and Materials},
            addressline={Technische Universiteit Delft}, 
            city={Delft},
            postcode={2629HS}, 
            state={},
            country={The Netherlands}}

\affiliation[2]{organization={Department of Materials Science and Engineering},
            addressline={Technische Universiteit Delft}, 
            city={Delft},
            postcode={2628CM}, 
            state={},
            country={The Netherlands}}

\author[1]{Riccardo Brambilla}%

\author[2]{Siddhant Kumar}%

\cormark[1]

\ead{sid.kumar@tudelft.nl}

\author[1]{Bianca Giovanardi}%

\cormark[1]

\ead{b.giovanardi@tudelft.nl}

\cortext[1]{Senior authors contributed equally.}

\begin{abstract}
Classical shell fracture theories predominantly consider cracks aligned with principal curvature directions in cylindrical and spherical shells under tensile mode or pressure loading, leaving the fracture behavior of arbitrarily oriented cracks in doubly curved shells under general loading poorly understood. Here, we introduce a semi-analytical framework based on Reissner shallow-shell theory to investigate isotropic shells with positive, zero, and negative Gaussian curvature. The formulation computes stress intensity factors, energy release rates, and crack-lip displacements for arbitrarily inclined through-thickness cracks in spheres, ellipsoids, cylinders, hyperboloids, and minimal surfaces under general mechanical loading. We show that shell curvature couples opening, sliding, and tearing, allowing a single loading component to activate multiple fracture modes, unlike in flat plates. This coupling depends strongly on crack orientation and the normal, parallel, and torsional components of the shell curvature. Ellipsoids exhibit weak coupling, hyperboloids generally exhibit the strongest coupling, and cylinders display the most complex orientation-dependent responses. Moreover, in-plane shear can induce crack closure in hyperboloidal shells, suggesting a potential geometric mechanism for enhanced damage resistance. These findings elucidate the role of local curvature in fracture-mode coupling and provide a mechanistic foundation for the analysis of shellular (meta-)structures with engineered curvature distributions.
\end{abstract}


\maketitle

\section{Introduction}

The 1960s-1980s saw extensive research into the fracture mechanics of curved shells, motivated by the need to assess structural integrity and predict failure in applications ranging from aircraft fuselage panels and rocket fairings to ship hulls and nuclear reactors \citep{Maneschy:2025}.
Most studies focused on cylindrical shells and spherical domes under pressure loading. The prevalence of pressure loading also led researchers to focus primarily on cracks subjected to tensile loading and aligned with the principal curvature directions of the shells, such as longitudinal cracks in cylinders. At the time, purely numerical methods, such as finite element analysis, were constrained by limited computational resources, motivating the development of semi-analytical solutions. Structural components were therefore reduced to simplified, idealized geometries, with considerable effort devoted to finding exact solutions to approximate governing equations, rather than approximate solutions to exact equations \citep{Sih:1977}.

In contrast to the relatively simple geometries that shaped classical shell fracture theories, advances in modern manufacturing now enable structures with far more complex curvature profiles, renewing interest in their fracture mechanics. In particular, additive manufacturing has enabled the fabrication of doubly curved structures, including those with negative Gaussian curvature that were uncommon and difficult to  manufacture when classical shell fracture theories were developed. Additively manufactured metamaterials and metastructures, built from carefully engineered architectures such as triply periodic minimal surfaces \citep{Bonatti:2019,AlKetan:2018} and spinodoids \citep{Kumar:2020,Portela:2020}, offer unprecedented control over local curvature and structural response. These architectures have enabled wide-ranging applications in energy absorption \citep{GuellIzard:2019,Anandan:2025,Gandhi:2026}, bone implants \citep{Callen:2021,Deng:2024,Guo:2024,Hsieh:2021,Dong:2021},  piezoelectric transducers \citep{Shi:2024}, etc.
Alongside these advances in manufacturing, growing computational power has enabled numerical approaches to the fracture mechanics of curved shells, including phase-field \citep{Amiri:2014,Kiendl:2016,Ma:2024} and extended finite element methods \citep{Bayat:2026}.

Despite these manufacturing and computational advances, there is still no clear consensus on the role of shell curvature in fracture behavior. Numerous studies \citep{Thakolkaran:2025,Ainger:2027,GuellIzard:2019,Anandan:2025} on architected metamaterials and metastructures have investigated macroscopic mechanical properties, such as fracture toughness and impact energy absorption, without explicitly examining the influence of local curvature. Recent studies \citep{Dhulipala:2025,Guo:2024} have established connections between microstructural curvature distributions and effective mechanical properties, including stiffness, strength, and the relative contributions of stretching- and bending-dominated deformation. However, the role of these curvature distributions in governing fracture behavior remains unexplored. Notably, recent studies \citep{Mitchell:2016,Mirsayar:2026} have investigated the fracture behavior of shells and sheets constrained to curved substrates; however, these systems are fundamentally different from freestanding shells and exhibit substantially different underlying mechanics.

Furthermore, owing to the complex geometries and stress distributions of architected metamaterials and metastructures, cracks aligned with the principal directions of curvature represent only a limiting case. In general, cracks may nucleate at arbitrary orientations relative to the principal curvature directions \citep{Yang:2022}. This motivates, as a first step, a more general understanding of the fracture behavior of inclined cracks in doubly curved shells subjected to complex stress fields.

Although the fracture mechanics of canonical shell geometries, such as cylinders and spheres, under conventional loading conditions has been extensively studied, comparatively little attention has been devoted to arbitrarily inclined cracks, doubly curved geometries, and general loading conditions. The following brief review traces the development of shell fracture theories and highlights the theoretical gaps that motivate the present study.

\textbf{\ul{A brief review of shell fracture theories}}

The earliest contributions in this field were made by  \citet{folias:1965a,folias:1965b,folias:1967,Folias:1969}, who laid the foundations of the Singular Integral Equation method. They proposed extending the Westergaard superposition method \citep{Westergaard:1939}, originally developed for plane stress fracture problems, to curved shells, and demonstrated  the  effect of curvature on the energy release rate of cracked shells.  The analyses were subsequently extended to a broad range of geometries: meridional cracks in spheres \citep{folias:1965a}, axial \citep{folias:1965b}, and circumferential cracks in cylinders \citep{folias:1967}, as well as more complex geometries like cones and ring tori \citep{Folias:1969}.  

Nonetheless, these solutions were obtained only asymptotically, restricting their validity to very small curvature magnitudes. Similarly, doubly curved surfaces were treated only as weighted superpositions of singly curved shell results, rather than as complete boundary value problems.  Furthermore, only cracks aligned with the principal curvature directions were considered.  The loading was also restricted to internal pressure, which produces a membrane stress state.  Thus, bending type loading was excluded from the analysis. Finally, their approach was based on the Kirchhoff shell theory, which assumes that the cross-sections of the shells remain parallel to the mid-surface normal throughout the deformation.  As a result, the transverse shear distributions obtained with the Kirchhoff theory are inaccurate in moderately thick shells \citep{knowles:1960} and exhibit a super-singularity \citep{Sih:1977}.  Specifically, they scale as $r^{-3/2}$ with the distance ($r$) from the crack tip, rather than exhibiting the usual square root singularity of Linear Elastic Fracture Mechanics.  This violates the requirement of a bounded strain energy and invalidates the superposition of bending and membrane stresses assumed by \citet{folias:1965a,folias:1965b,folias:1967,Folias:1969}.

The introduction of dedicated numerical techniques \citep{Erdogan:1969} substantially extended the range of curvatures that could be analyzed by asymptotic methods alone. Nevertheless, the singular integral equations were still solved approximately rather than directly. A systematic approach was later developed to obtain direct solutions.  Integral transforms and change of variables were utilized to reduce the strength of the 
singularities of the stress fields and to smooth the expected oscillating behavior of the integral kernels \citep{Muskhelishvili:1972,Erdogan:1972a}.
Dedicated quadrature rules \citep{Erdogan:1969a,Erdogan:1973}, based on Chebyshev polynomials, eventually reduced the solution of singular integrals equations to that of a linear system \citep{Erdogan:1978}.

Nevertheless, resolving the transverse shear super-singularity required higher order theories.  \citet{Hagendorf:1972} introduced a Reissner-type shell  formulation \citep{Reissner:1947,Naghdi:1957,Sih:1977} for cracked spherical shells. 
Unlike Kirchhoff-type theories, this formulation treats the cross-section rotations as independent variables, thereby eliminating the transverse shear super-singularity. Subsequent comparisons with three-dimensional finite element analyses demonstrated that Reissner shell theories provide greater accuracy than Kirchhoff models within a closed neighborhood surrounding the crack whose  size is comparable to the shell thickness \citep{zucchini:2000}. 

\citet{Hagendorf:1972} also observed that bending loading alone can cause crack closure in shells, necessitating the superposition of sufficient membrane loading to maintain crack opening and meaningfully discuss energy release rates and crack propagation. This phenomenon was already known for flat plates \citep{Sih:1962}, but, to the best of our knowledge, had not previously been demonstrated for curved shells.  Several similar studies, both semi-analytical \citep{Heming:1980} and numerical \citep{Dempsey:1998},  developed formulations to account for crack closure in flat plates subjected to bending loading. By contrast, the corresponding literature for shells remains limited, with only a few studies considering principally embedded cracks in spherical and cylindrical shells. Generally, the stress intensity factors depend on both the shell curvature and the orientation of the bending moment relative to the  shell curvature. %
Moreover, contact is confined to a portion of the crack length, resulting in asymmetric crack opening and closure with respect to the shell mid-surface \citep{Liu:1999}.
\citet{Hagendorf:1972} further noted that Kirchhoff-type assumptions systematically overestimate the bending response for longer cracks.  Finally, they isolated plane stress Stress Intensity Factors as multiplicative factors in the Reissner shell crack tip fields, consistent with the concept of Bueckner's weight function \citep{bueckner:1970}. An analogous equivalence for Kirchhoff shells was established around the same time by \citet{Yuceoglu:1973}.

The introduction of composites further motivated numerous studies investigating the effect of orthotropic materials on  the fracture of curved shells.  The first analysis \citep{Erdogan:1974} of an axial crack in an orthotropic cylinder, with the material principal directions aligned with the principal curvature directions, considered internal pressure loading.  \citet{Yuceoglu:1973} introduced remote torsion loading of cylindrical shells, paving the way for the analysis of loading conditions like in-plane shear, twisting moments, and transverse shear, all of which act antisymmetrically with respect to the crack plane. More recent analyses used energetic arguments to show that the twisting-induced sliding stress intensity factor $K_{II}$ and the transverse stress intensity factor $K_{III}$ must coexist in thin, flat, Reissner plates \citep{Hui:1993}. 

The work of \citet{Hagendorf:1972} and \citet{Yuceoglu:1973} laid the groundwork for analyses of additional geometries using Reissner theories, incorporating orthotropic materials and general loading conditions. Examples include axially cracked cylinders under membrane and bending loading \citep{Krenk:1978},  orthotropic spheres under membrane and bending loading \citep{Delale:1979a},  circumferentially cracked shells under membrane and bending loading \citep{Delale:1979}, the same geometries subjected to in-plane shear, torsion, and transverse shear \citep{Delale:1982}, and finally doubly curved shells under membrane and bending loading \citep{Delale:1983}. Most of these studies, however, focused primarily on the effect of orthotropy rather than curvature. Moreover, nearly all restricted their analyses to cracks aligned with the principal curvature directions of the shells. 

To the best of our knowledge, only two studies have considered cracks embedded in shallow shells at arbitrary angles to principal lines of curvature. \citet{Yahsi:1983} used a superposition approach to investigate an orthotropic cylindrical shell containing an arbitrarily oriented crack. They demonstrated that, regardless of the stress state experienced by the crack, all three stress intensity factors can be nonzero, contrary to what is observed for principal cracks and flat plates. The second study was conducted by \citet{sanders:1983}. Rather than the superposition method introduced by \citet{folias:1965a,folias:1965b,folias:1967,Folias:1969}, they used contour integration and a complex-valued shell model to solve for the stress fields throughout the shell \citep{copley:1969,sanders:1982}. The crack-tip stress fields were subsequently derived asymptotically from the shell solution. However, the shell model used in this formulation has since been shown to be inconsistent under in-plane boundary conditions \citep{chen:2000a}.  Moreover, shallow shell approximations have been shown to introduce only minimal errors when the shell may be regarded as ``infinite'', which is precisely the setting of the superposition approach \citep{elias:1967}.

In light of these gaps, we therefore extend the framework of \citet{Yahsi:1983} and \citet{Yahsi:1981} to doubly curved, shallow shells, considering both ellipsoids (shells with positive Gaussian curvatures: $K>0$) and hyperboloids (shells with negative Gaussian curvatures: $K<0$). We then present a systematic study of prototypical shells with arbitrarily-oriented cracks subjected to tension, bending, in-plane shear, twisting and transverse shear loadings. To isolate the effect of geometry on crack-tip stress concentrations from that of material properties, we restrict the present study to isotropic shells. Our analyses reveal the simultaneous presence of multiple fracture modes across all geometries, even under simple tension stress states around the crack, with all three stress intensity factors $K_I,K_{II},K_{III}$ being generally nonzero. Moreover, we show that both shell curvature and crack orientation relative to the principal curvature directions influence the relative magnitudes of the three fracture modes. Finally, we provide a mechanistic interpretation of the cracked-shell response, elucidating the role of curvature in governing the coupling between fracture modes.  Together, these numerical results and their mechanistic interpretation  advance our understanding of fracture mechanics in curved shells.

\section{Methodology}
\label{sec:methodology}

We now derive a semi-analytic solution for the problem of a doubly curved, thin, shallow shell 
subjected to arbitrary far-field loading and containing a through-the-thickness crack of arbitrary orientation.
We restrict the analysis to small displacements and to homogeneous, isotropic, linear elastic 
materials. 
The formulation is developed within the framework of linear elastic fracture mechanics and Reissner's 
shallow shell theory \citep{Naghdi:1957}, and closely follows the approach described 
by  \citet{Yahsi:1981} for a cylindrical shell containing an arbitrarily oriented crack.

We begin by describing the problem setting and the governing equations in 
Section~\ref{ssec:problem-setting}. 
We then use linear superposition to formulate an equivalent perturbation problem in 
Section~\ref{ssec:superposition}. 
This problem is obtained by decomposing the original cracked-shell problem into the response of an 
uncracked shell subjected to the far-field stress resultants and the response of the cracked shell 
subjected to appropriate crack-surface loading in the absence of far-field loading.
Section~\ref{ssec:semianalyticalSolution} illustrates the procedure used to solve the perturbation 
problem in terms of appropriate auxiliary variables.
Finally, Section~\ref{ssec:flatPlateNormalization} introduces the flat-plate normalization used to 
isolate curvature-induced deviations in the stress intensity factors and energy release rates, and 
discusses the resulting mode-coupling effects.

\subsection{Problem setting}
\label{ssec:problem-setting}

We consider a doubly curved shell with thickness $h$ containing a crack of length $2a$ centered at 
the origin $O$ of a Cartesian, left-handed reference frame $OXYZ$, aligned with the shell principal 
directions.
The crack is inclined with respect to the principal $Y$ direction by a clockwise angle $\beta$.
To facilitate the analysis, we introduce a second Cartesian reference frame $Oxyz$, obtained by 
rotating the $X,Y$ axes clockwise about $Z$ by $\beta$ and normalizing all lengths by the crack 
half-length $a$. 
Thus, $x$ is normal to the crack plane and $y$ is parallel to the crack, while the crack tips are 
located at $y=\pm 1$.
The normalized through-the-thickness coordinate $z$ remains parallel to the $Z$-axis.
We refer to $Oxyz$ as the \emph{crack reference} and to $OXYZ$ as the \emph{principal reference}.

The shell midsurface can be locally approximated by a paraboloid \citep{Calladine:1983},
\begin{equation}
    \label{eqn:Paraboloid}
Z(X,Y)
=
-\frac{1}{2}
\left(
\frac{X^2}{R_X}
+
\frac{Y^2}{R_Y}
\right),
\end{equation}
where $R_X$ and $R_Y$ are the principal radii of curvature at the crack center. The corresponding
principal curvatures are $\kappa_X=1/R_X$ and $\kappa_Y=1/R_Y$, and the local Gaussian curvature is
$K=\kappa_X\kappa_Y$. Depending on the sign of $K$, the local shell geometry is classified as
elliptic ($K>0$), cylindrical ($K=0$), or hyperbolic ($K<0$).

\begin{figure}
    \centering
    \includegraphics[width=\textwidth]{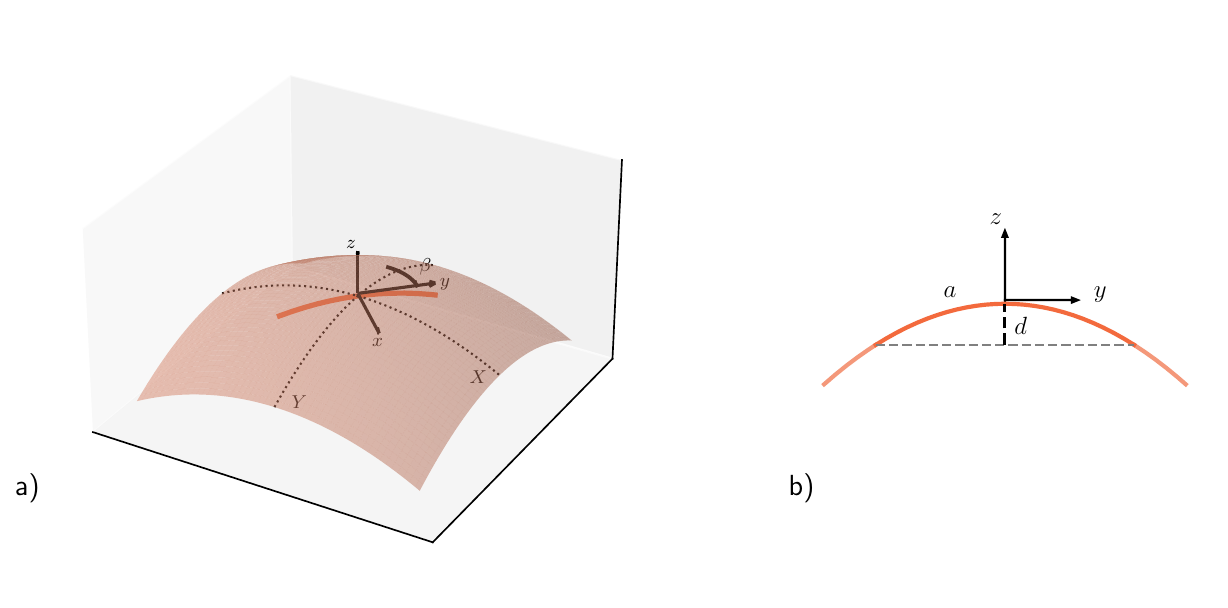}
    \caption{Schematic of the shell geometry. 
    (a) Doubly curved prototypical shell containing a crack with arbitrary orientation $\beta$ with 
    respect to the principal curvature directions. 
    (b) Illustration of the shell shallowness, quantified by the sagitta-to-crack-length ratio 
    $s = d/(2a)$.}
    \label{fig:referenceAndSagitta}
\end{figure}

We restrict the analysis to \emph{shallow} shells. 
Following the fracture-mechanics convention, we quantify shallowness through the ratio $s$ between 
the shell sagitta and the crack length, which provides the characteristic in-plane length scale of 
the problem, see Figure~\ref{fig:referenceAndSagitta}.
Specifically, we consider shells such that $s \leq 0.2$ \citep{Liu:1999}.
Under this assumption, the slopes of the shell midsurface are small, so that
$\partial_X Z(X,Y)\ll 1$ and $\partial_Y Z(X,Y)\ll 1$.
As a result, the Cartesian reference frames introduced above
can be used throughout the analysis, and the through-the-thickness direction can be identified with 
the $Z$ axis.

We model the shell mechanics using shallow shell theory of \citet{Reissner:1947}. This choice is motivated by the fact that Reissner's theory is the simplest shell theory capable of recovering the linear elastic fracture mechanics $r^{-1/2}$ crack-tip stress singularity for modes I, II, and III.
Within this theory, the shell kinematics is described in terms of the midsurface displacements $u_0(X,Y)$, $v_0(X,Y)$, and $w_0(X,Y)$ along the $X$, $Y$, and $Z$ directions, respectively, and two independent cross-section rotations $\beta_X(X,Y)$ and $\beta_Y(X,Y)$. These rotations are defined as the counter-clockwise rotations of the shell cross-section about the $Y$ and $X$ axes, respectively.

Reissner theory further assumes small strains, so that quadratic terms in the displacement gradients 
can be neglected, and moderate rotations. 
Under these assumptions, the membrane component of the strain tensor is given by
\begin{equation}
    \label{eqn:Strains}
    \begin{aligned}
        \varepsilon_{XX}(X,Y)& = \partial_X u_{0}(X,Y) + \partial_X Z(X,Y) \, \partial_X w_0(X,Y), \\
        \varepsilon_{YY}(X,Y)& = \partial_Y v_{0}(X,Y) + \partial_Y Z(X,Y) \, \partial_Y w_0(X,Y), \\
        \varepsilon_{XY}(X,Y)& = \frac{1}{2} \left(\partial_Y u_{0}(X,Y) + \partial_X v_{0}(X,Y) + \partial_X Z(X,Y) \, \partial_Y w_0(X,Y) + \partial_Y Z(X,Y) \, \partial_X w_0(X,Y)\right).
    \end{aligned}
\end{equation}

The bending strain components and transverse shear strains for shallow shells are given by \citep{Naghdi:1957}
\begin{equation}
    \label{eqn:bendingStrains}
    \begin{aligned}
        \kappa_{XX}(X,Y) &= \partial_X \beta_X(X,Y),\\ 
        \kappa_{YY}(X,Y) &= \partial_Y\beta_Y(X,Y), \\ 
        \kappa_{XY}(X,Y) &= \frac{1}{2} \left(\partial_X \beta_Y(X,Y) + \partial_Y\beta_X(X,Y)\right),
    \end{aligned}
\end{equation}
and
\begin{equation}
    \label{eqn:shearStrains}
    \begin{aligned}
        \eta_{XZ}(X,Y) &= \partial_X w_0(X,Y) + \beta_X(X,Y), \\ 
        \eta_{YZ}(X,Y) &= \partial_Y w_0(X,Y) + \beta_Y(X,Y),
    \end{aligned}
\end{equation}
respectively.

The constitutive behavior follows Hooke's law, which is expressed in stress-resultant form by relating
the membrane resultants $N_{ij}$, bending-moment resultants $M_{ij}$, and transverse shear resultants
$V_i$ to the strain measures defined above \citep{Naghdi:1957}:
\begin{subequations}
    \label{eqn:ConstitutiveEquations}
    \begin{align}
        N_{XX}(X,Y)& = \frac{hE}{1 - \nu^2} \left(\varepsilon_{XX}(X,Y) + \nu \, \varepsilon_{YY}(X,Y)\right),\\ 
        N_{YY}(X,Y) &= \frac{hE}{1 - \nu^2} \left(\nu \, \varepsilon_{XX}(X,Y) + \varepsilon_{YY}(X,Y) \right),\\
        N_{XY}(X,Y)&= \frac{hE}{1+ \nu} \, \varepsilon_{XY} (X,Y),
        \label{eqn:InPlaneResShallowShells} 
        \\
        M_{XX}(X,Y)& =D \, \kappa_{XX}(X,Y),\\
        M_{YY}(X,Y) &= D \, \kappa_{YY}(X,Y),\\
        M_{XY}(X,Y) &= D \, (1 - \nu) \, \kappa_{XY}(X,Y),
        \label{eqn:MomResShallowShells} 
        \\
        V_{X}(X,Y)& = h \, B \, \eta_{XZ}(X,Y),\\
        V_{Y}(X,Y)& = h \, B \, \eta_{YZ}(X,Y),
        \label{eqn:TransResShallowShells}
    \end{align}
\end{subequations}
where $E$ and $\nu$ are, respectively, the material's Young's modulus and Poisson's ratio, 
$D = \frac{E h^3}{12 \left(1 - \nu^2\right)}$ is the shell bending stiffness, and 
$B = \gamma \frac{E}{2 \left(1 + \nu\right)}$ is the shear modulus multiplied by the shear 
correction factor $\gamma = \frac{5}{6}$. 
Here, $\gamma$ accounts for the discrepancy between the assumed parabolic through thickness 
distribution of the transverse shear stresses and the actual one, see \cite{Reissner:1947}.

The equilibrium equations of the shell are likewise formulated in resultant form \citep{Naghdi:1956}:
\begin{equation}
    \label{eqn:Equilibrium}
    \begin{aligned}
        \partial_j& N_{ij}(X,Y) = 0,\\
        \partial_i& V_{i}(X,Y) + \partial_j\left(\partial_i Z \, N_{ij}(X,Y)\right) = 0,\\
        \partial_j& M_{ij}(X,Y) - V_i(X,Y) = 0,
    \end{aligned}
\end{equation}
where $i$ and $j$ take values in $\{X,Y\}$ and repeated indices imply summation.

Owing to shallowness, the transverse stress resultants can be omitted in the 
tangential equilibrium equations, while the rotational equilibrium equation about the $Z$-axis 
becomes trivial, see also \cite{Sih:1977,reissner:1984}. 

The stress resultants are related to the stresses with the standard formula:
\begin{equation}
    \label{eqn:StressResultantsDefinition}
    \begin{aligned}
            N_{ij}(X,Y)& = \int_{-h/2}^{h/2} \sigma_{ij}(X,Y,Z) \; \text{d}Z, \\
            M_{ij} (X,Y)& = \int_{-h/2}^{h/2} Z \sigma_{ij}(X,Y,Z) \; \text{d}Z, \\
            V_i(X,Y) &= \int_{-h/2}^{h/2} \sigma_{i Z}(X,Y,Z) \; \text{d}Z, 
    \end{aligned}
\end{equation}
where $i$ and $j$ take values in $\{X,Y\}$.
The integrations in Equation~\eqref{eqn:StressResultantsDefinition} are performed along the $Z$ 
direction, rather than along the shell normal, consistently with the shallow-shell approximation 
\citep{Reissner:1952}.

The reconstruction of the stresses from the stress resultants follows the generalized plane-stress 
assumptions associated with Reissner's theory \citep{Reissner:1947,sih:1971}. 
Specifically, the normal stress $\sigma_{ZZ}$ is considered negligible, the in-plane stresses 
$\sigma_{XX},\sigma_{YY}, \sigma_{XY}$ are assumed to vary linearly through the shell thickness, 
and the transverse shear stresses $\sigma_{XZ},\sigma_{YZ}$ are 
reconstructed using a parabolic through-thickness distribution.
Note that, in Reissner's theory, the transverse shear stresses should be interpreted as weighted 
averages through the shell thickness rather than as pointwise exact quantities 
\citep{Reissner:1947}. 
Performing the integrations in Equation~\eqref{eqn:StressResultantsDefinition} under these generalized 
plane-stress assumptions and inverting the resulting expressions yields explicit relations between 
the stresses and the stress resultants \citep{Reissner:1947}:
\begin{equation}
    \label{eqn:stressFromResultants}
    \begin{aligned}
        \sigma_{ij}(X,Y,Z) &= \frac{N_{ij}(X,Y)}{h} + \frac{12 a}{h}Z \;M_{ij}(X,Y), \\
        \sigma_{iz}(X,Y,Z) &= \frac{3}{2} V_i(X,Y)\left(1 - \left(\frac{2a}{h}Z\right)^2\right),
    \end{aligned}
\end{equation}
where $i$ and $j$ take values in $\{X,Y\}$.

The shell mechanics problem is completed with appropriate boundary conditions representing the 
far-field stress resultants and the traction-free crack surfaces.

\subsection{Superposition and formulation of the perturbation problem}
\label{ssec:superposition}

By linearity of the shallow-shell equations, the cracked-shell problem can be reformulated through
superposition, analogously to the classical approach used to solve the problem of the center-cracked
plate under plane-stress conditions \citep{Westergaard:1939}. 
Through superposition, the cracked-shell response is decomposed into the response of an intact shell 
subjected to the far-field loading and an auxiliary perturbation problem associated with the 
presence of the crack.

\begin{figure}
    \centering
    \includegraphics[width=\textwidth]{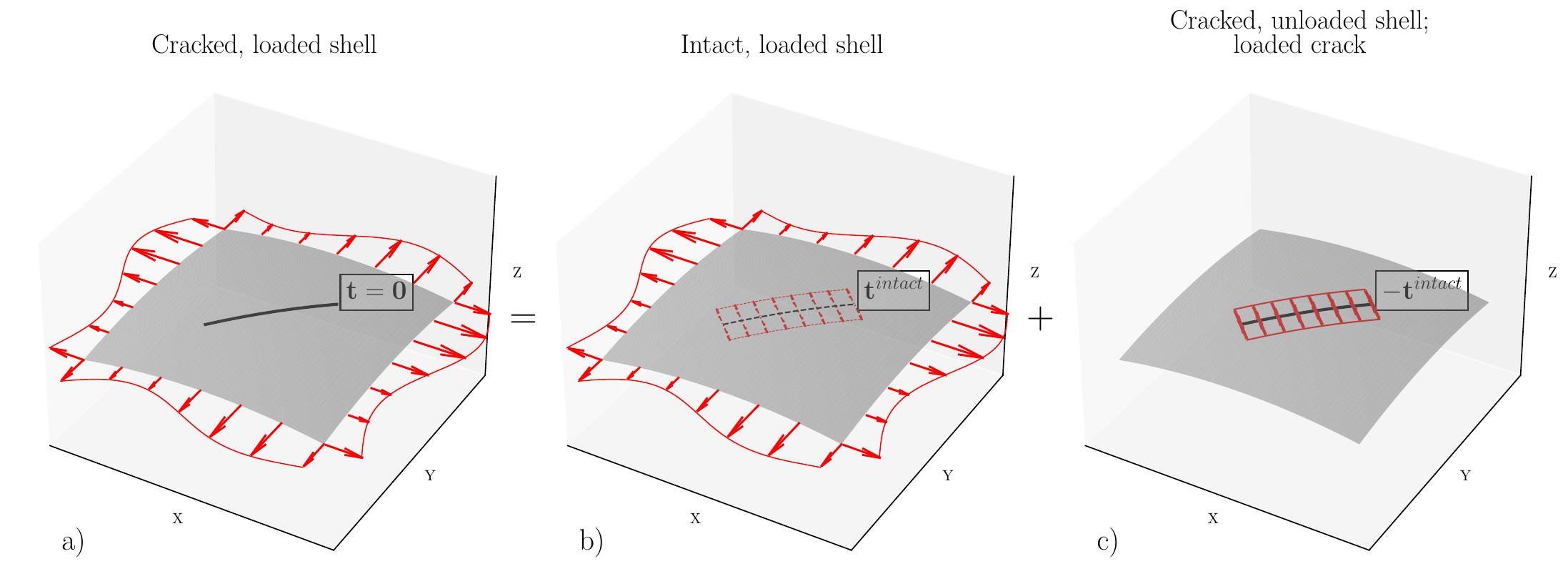}
    \caption{Illustration of the superposition concept. 
    (a) Cracked shell subjected to far-field loading, with traction-free crack surfaces. 
    By linearity of the governing equations, this problem can be decomposed into: 
    (b) the corresponding intact shell subjected to the same far-field loading, with the virtual crack 
    location represented by a dashed line, and 
    (c) an auxiliary cracked shell with unloaded outer boundaries, subjected on the crack surfaces to 
    tractions $\boldsymbol{t}^{\mathrm{residual}}=-\boldsymbol{t}^{\mathrm{intact}}$, equal and 
    opposite to those induced by the intact solution at the virtual crack location.}
    \label{fig:superposition}
\end{figure}

We first consider the original problem: a cracked shell subjected to prescribed loading on its outer
boundaries, with traction-free crack faces, see Figure~\ref{fig:superposition}a. 
The traction vector acting on the crack faces is denoted by
$\boldsymbol{t} = \boldsymbol{\sigma}\boldsymbol{n}$,
where $\boldsymbol{n}$ is the unit normal to the crack surface.
Since the crack faces are traction-free, the original problem satisfies
$\boldsymbol{t}=\boldsymbol{0}$ on the crack faces.

Next, we consider the corresponding intact shell, with the same geometry and boundary loading but
without the crack, see Figure~\ref{fig:superposition}b. 
The intact solution generally produces nonzero tractions $\boldsymbol{t}^{\mathrm{intact}}=\boldsymbol{\sigma}^{\mathrm{intact}}\boldsymbol{n}$ 
on the prospective `virtual' crack faces.

The perturbation induced by the crack is then obtained by considering a cracked shell with unloaded
outer boundaries and by applying, on the crack faces, tractions equal and opposite to those induced
there by the intact solution, see Figure~\ref{fig:superposition}c. 
Thus, the crack-face loading in the perturbation problem is 
$\boldsymbol{t}^{\mathrm{residual}} = -\boldsymbol{t}^{\mathrm{intact}}.$
The solution of this auxiliary problem represents the perturbation introduced by the crack into the
stress state of the intact shell. By construction, superposing this perturbation onto the intact
solution cancels the crack-face tractions and therefore recovers the traction-free condition of the
original cracked-shell problem.

The perturbation problem therefore consists of Equation~\eqref{eqn:Equilibrium} subjected to 
vanishing perturbation fields at infinity and to crack-surface loading prescribed by
\begin{equation}
    \label{eqn:boundaryConditionss}
    \begin{aligned}
        \begin{matrix}
                    \lim_{x\to 0^+} N_{xx}(x,y) & = &  \lim_{x\to 0^-} N_{xx}(x,y) & = & N^0_{xx}(y) \\
                    \lim_{x\to 0^+} N_{xy}(x,y) & = & \lim_{x\to 0^-} N_{xy}(x,y)  & = & N^0_{xy}(y) \\
                    \lim_{x\to 0^+} M_{xx}(x,y) & = & \lim_{x\to 0^-} M_{xx}(x,y)  & = & M^0_{xx}(y) \\
                    \lim_{x\to 0^+} M_{xy}(x,y) & = & \lim_{x\to 0^-} M_{xy}(x,y)  & = & M^0_{xy}(y) \\
                    \lim_{x\to 0^+} V_x(x,y)    & = & \lim_{x\to 0^-} V_x(x,y)     & = & V^0_x(y)
                \end{matrix} \qquad \;\;  \text{ for } \left| y \right| < 1,
    \end{aligned}
\end{equation}
where $N^{0}_{xx}$, $N^{0}_{xy}$, $M^{0}_{xx}$, $M^{0}_{xy}$, and $V^{0}_x$ denote the membrane, 
bending, and transverse shear stress resultants induced by the intact-shell stress field
$\bm{\sigma}^{\mathrm{intact}}$ at the virtual crack location:
\begin{equation}
    \label{eqn:LoadingResultants}
    \begin{aligned}
            N_{xi}^0 (y)& = \int_{-h/2}^{h/2} \sigma_{xi}^{\mathrm{intact}}(0,y,z) \; \text{d}z, \\
            M_{xi}^0 (y)& = \int_{-h/2}^{h/2} z \, \sigma_{xi}^{\mathrm{intact}}(0,y,z) \; \text{d}z, \\
            V_x^0 (y) & = \int_{-h/2}^{h/2} \sigma_{x z}^{\mathrm{intact}}(0,y,z) \; \text{d}z,
    \end{aligned}
\end{equation}
with $i \in \{x,y\}$.

These resultants may depend arbitrarily on $y$, provided that the resulting crack-surface loading is
self-equilibrated. In the present work, however, we restrict the analysis to uniform crack-surface
resultants along the crack length, with the objective of isolating the effect of shell geometry.

\subsection{Solution of the perturbation problem}
\label{ssec:semianalyticalSolution}

We solve the perturbation problem identified in Section~\ref{ssec:superposition} following the
methodology introduced by \citet{Yahsi:1981}. 

We first nondimensionalize the governing equations \eqref{eqn:ConstitutiveEquations} and 
\eqref{eqn:Equilibrium}, and then transform them to the crack reference frame $Oxyz$. 
Details of this derivation are provided in Appendix~\ref{app:EquilibriumToField}. 
This yields
\begin{subequations}
    \label{eqn:FieldEq}
    \begin{align}
        \nabla^4 \phi(x,y) &+ \frac{1}{\lambda^2} \left( \gamma_x \, \partial_{yy}  + \gamma_y \, \partial_{xx}  - 2 \, \gamma_{xy} \, \partial_{xy}  \right)w(x,y) = 0, \label{eqn:sFinalInp} \\
        \nabla^4 w(x,y)& - \lambda^2 \left(1 - \kappa \, \nabla^2 \right)\left( \gamma_x \, \partial_{yy}  + \gamma_y \, \partial_{xx}  - 2 \, \gamma_{xy} \, \partial_{xy}  \right) \phi(x,y) = 0, \label{eqn:sFinalOutp} \\
        \kappa \, &\nabla^2 \psi(x,y) - \psi(x,y) - w(x,y) = 0, \label{eqn:sFinalRot1} \\
        \kappa \, &\frac{1 - \nu}{2} \, \nabla^2 \Omega(x,y) - \Omega(x,y) = 0. \label{eqn:sFinalRot2}
    \end{align}
\end{subequations}

Here, $\phi(x,y)$ is the Airy stress function, $w(x, y)$ is the normalized out-of-plane 
displacement of the shell midsurface. The constants $\lambda^2$, $\kappa$, and $\gamma_{i}$, whose
expressions are reported in the appendix, 
represent the normalized shell thickness, normalized shell transverse stiffness, and 
normalized curvatures, respectively.
In Equation \eqref{eqn:FieldEq}, $\psi(x, y)$ and $\Omega(x, y)$ are two auxiliary functions
introduced to weaken the coupling of the governing equations:
\begin{equation}
    \begin{aligned}
        \Omega(x,y) &= \partial_y\beta_x(x,y) - \partial_x \beta_y(x,y), \\[3pt]
        \psi(x,y)&= \kappa \left(\partial_x \beta_x(x,y) + \partial_y \beta_y(x,y)\right) - w(x,y).
    \end{aligned}
\end{equation}
Equations~\eqref{eqn:sFinalInp} and~\eqref{eqn:sFinalOutp} form a coupled system for $\phi(x,y)$ and
$w(x,y)$. Once $w(x,y)$ is known, Equation~\eqref{eqn:sFinalRot1} can be solved independently for
$\psi(x,y)$, while Equation~\eqref{eqn:sFinalRot2} is fully decoupled from the other three equations.

Applying the biharmonic operator to Equation~\eqref{eqn:sFinalOutp} and substituting the resulting
equation in Equation~\eqref{eqn:sFinalInp} gives
\begin{equation}
    \label{eqn:sEqInw}
        \nabla^8 w 
        + \left(1 - \kappa \, \nabla^2\right) 
        \left[\gamma_x^2 \, \partial_{yyyy} + \gamma_y^2 \, \partial_{xxxx}
        -4 \, \gamma_{xy} \left( \gamma_x \, \partial_{xyyy} + \gamma_y \, \partial_{xxxy}\right) 
        + \left( 4 \, \gamma_{xy}^2 + 2 \, \gamma_x \gamma_y \right) \partial_{xxyy}\right] w = 0.
\end{equation}

We continue the derivation in Fourier space by transforming the fields with respect to the crack
coordinate $y$. 
We define the Fourier transforms $\hat{\phi}(x,\alpha)$, $\hat{w}(x,\alpha)$, $\hat{\psi}(x,\alpha)$, 
and $\hat{\Omega}(x,\alpha)$ of the corresponding fields
$\phi(x,y)$, $w(x,y)$, $\psi(x,y)$, and $\Omega(x,y)$ as
\begin{equation}
    \label{eqn:Transforms}
    \begin{aligned}
        \hat{f}(x, \alpha) = \mathcal{F}(f(x, y)) := \int_{-\infty}^{\infty} f(x,y) \; e^{-i\alpha y} \; \text{d}y.
    \end{aligned}
\end{equation}

In the Fourier domain, the problem is split into the two half-domains $x>0$ and $x<0$, with
appropriate coupling conditions imposed on the crack plane $x=0$. These conditions consist of
Equation~\eqref{eqn:boundaryConditionss} on the crack faces, that is for $|y|<1$, and continuity of
the kinematic fields and stress resultants at and ahead of the crack tips, that is for $|y|\geq 1$.

Substituting $ w(x,y) = \mathcal{F}^{-1}(\hat{w}(x,\alpha))$ in Equation~\eqref{eqn:sEqInw} leads to
an eighth-order ordinary differential equation. 
We seek solutions of the form
\begin{equation}
\hat{w}(x,\alpha) = 
    \left( R^+(\alpha) \, e^{m(\alpha)x}\right) \, \indicator_{x>0} + 
    \left( R^-(\alpha) \, e^{m(\alpha)x}\right) \, \indicator_{x<0}\;,
\end{equation}
where $\indicator_A$ denotes the indicator function of set $A$ and $R^+$, $R^-$, and $m$ are
complex-valued functions of $\alpha$.

The coefficients $m(\alpha)$ are obtained as the roots of the associated characteristic equation
$p(m(\alpha)) = 0$, where 
\begin{equation} 
    \label{eqn:sCharEqYahsi}
    \begin{aligned}
        p(m) &= m^8 - \left(4\,\alpha^2 + \kappa \, \gamma_y^2\right) m^6 - 4\, i\, \kappa\, \gamma_{xy}\, \gamma_y\, \alpha\, m^5 + \nonumber\\
        &+ \left(6\, \alpha^4 + \gamma_y^2 + \kappa\, \alpha^2 \left(\gamma_y^2 + 4\, \gamma_{xy}^2 + 2\, \gamma_x\, \gamma_y \right) \right) m^4 + \nonumber\\
        &- 4\, i\, \alpha\, \gamma_{xy}\, \left( \gamma_y + \kappa\, \alpha^2 \left( \gamma_x + \gamma_y \right)\right) m^3 + \\
        &- \alpha^2 \left(4\, \alpha^4 + 2 \left( 2\, \gamma_{xy}^2 + \gamma_x\, \gamma_y\right) + \kappa\, \alpha^2 \left( \gamma_x^2 + 4\, \gamma_{xy}^2 + 2\, \gamma_x\, \gamma_y \right)\right) m^2 + \nonumber\\
        &-  4\, i\, \alpha^3\, \gamma_x\, \gamma_{xy} \left(1 + \kappa\, \alpha^2 \right) m + \alpha^4 \left(\alpha^4 + \gamma_x^2 \left(1 + \kappa\, \alpha^2\right)\right),
    \end{aligned}
\end{equation}
and the dependence of $m$ on $\alpha$ has been omitted for readability.

The eight roots $m_j$, $j=1,\dots,8$, of the characteristic polynomial cannot be expressed in
closed form as functions of $\alpha$, but are computed numerically for each value of $\alpha$, see
Appendix~\ref{app:CharacteristicPolynomials}. We can therefore write $\hat{w}(x,\alpha)$ in terms
of the unknown functions $R_j^+$ and $R_j^-$, $j=1,\dots,8$, as
\begin{equation} \label{eqn:wHat}
\hat{w}(x,\alpha) = 
    \left( \sum_{j=1}^8 R_j^+(\alpha) \, e^{m_j(\alpha)x} \right) \indicator_{x>0} 
    + \left( \sum_{j=1}^8 R_j^-(\alpha) \, e^{m_j(\alpha)x}\right) \indicator_{x<0}\;,
\end{equation}

These sixteen unknown functions can be reduced to eight by enforcing decay of the perturbation
solution as $x\to\pm\infty$. 
In fact, it can be shown, see appendix, that the roots occur in complex-conjugate pairs and can be 
ordered such that
\begin{equation}
    \begin{aligned}
        \Re \left(m_{j}(\alpha)\right) = - \Re \left(m_{j+4}(\alpha)\right) < 0, \qquad j = 1, \dots,4, \qquad \forall \alpha.
    \end{aligned}
\end{equation}
Since the perturbation displacements must decay at infinity, $\hat{w}$ must decay as
$x\to\pm\infty$. This condition implies
\begin{equation}
\label{eqn:regularityConsequence}
R_1^-(\alpha) = R_2^-(\alpha) = R_3^-(\alpha) = R_4^-(\alpha) = R_5^+(\alpha) = R_6^+(\alpha) = R_7^+(\alpha) = R_8^+(\alpha) \equiv 0 \qquad \forall \alpha.
\end{equation}

We can therefore write $\hat{w}(x, \alpha)$ in terms of eight unknown functions $R_j$,
$j = 1, \dots, 8$, as
\begin{equation}
\label{eqn:splitW}
\hat{w}(x,\alpha) = 
    \left( \sum_{j=1}^4 R_j(\alpha) \, e^{m_j(\alpha)x} \right) \indicator_{x>0} 
    + \left( \sum_{j=5}^8 R_j(\alpha) \, e^{m_j(\alpha)x} \right) \indicator_{x<0}\;,
\end{equation}
where $R_j = R_j^+$, $j = 1, \dots, 4$ and $R_j = R_j^-$, $j = 5, \dots, 8$.

A similar procedure applied to Equation~\eqref{eqn:sFinalRot2} leads to solutions of the form
\begin{equation} \label{eqn:omegaHat}
\hat{\Omega} (x,\alpha) = 
    \left( A_1 (\alpha) e^{r_1(\alpha) x} \right) \indicator_{x>0} 
    + \left( A_2 (\alpha) e^{r_2(\alpha) x} \right) \indicator_{x<0}\;,
\end{equation}
where $A_1$ and $A_2$ are complex-valued functions of $\alpha$. 
In this case, the roots $r_1$ and $r_2$ can be written explicitly as
\begin{equation} \label{eqn:rootsOmega}
r_1(\alpha) = - r_2(\alpha) = - \sqrt{\alpha^2 + \frac{2}{\kappa \left(1 - \nu\right)}}.
\end{equation}

The coupling of the differential equations in \eqref{eqn:FieldEq} implies that
$\hat{\phi}(x,\alpha)$ and $\hat{\psi}(x,\alpha)$ can be expressed in terms of the same roots
$m_j$, $j=1,\dots,8$, appearing in Equation~\eqref{eqn:wHat}. 
Specifically, one has
\begin{equation} \label{eqn:phiHat}
\hat{\phi}(x,\alpha) =
    \left( \sum_{j=1}^4 K_j(\alpha) R_j(\alpha) \, e^{m_j(\alpha)x} \right) \indicator_{x>0} 
    + \left( \sum_{j=5}^8 K_j(\alpha) R_j(\alpha) \, e^{m_j(\alpha)x} \right) \indicator_{x<0}\;,
\end{equation}
with
\begin{equation} \label{eqn:KjYahsi}
K_j(\alpha) = \frac{\left(\gamma_y - \gamma_x\right) \alpha^2 + \gamma_y\,(m_j^2(\alpha) - \alpha^2)}{\lambda^2 \,(m_j^2(\alpha) - \alpha^2)^2} + 2\,i \frac{\gamma_{xy}}{\lambda^2} \frac{\alpha\, m_j(\alpha)}{(m_j^2(\alpha) - \alpha^2)^2}.
\end{equation}
Similarly,
\begin{equation} \label{eqn:psiHat}
\hat{\psi}(x,\alpha) =
    \left( \sum_{j=1}^4 L_j(\alpha) R_j(\alpha) \, e^{m_j(\alpha)x} \right) \indicator_{x>0} 
    + \left( \sum_{j=5}^8 L_j(\alpha) R_j(\alpha) \, e^{m_j(\alpha)x} \right) \indicator_{x<0}\;,
\end{equation}
with
\begin{equation} \label{eqn:Lj}
L_j(\alpha) = \frac{1}{\kappa\, (m_j^2(\alpha) - \alpha^2) -1}.
\end{equation}

The solution of Equation~\eqref{eqn:FieldEq} has therefore been reduced to determining five unknown 
functions of $\alpha$ on each side of the crack plane: $R_1(\alpha),\dots,R_4(\alpha),A_1(\alpha)$ 
for $x>0$, and $R_5(\alpha),\dots,R_8(\alpha),A_2(\alpha)$ for $x<0$. However, the number of unknown 
functions can be halved by reformulating the problem in terms of kinematic jumps.

To this end, we introduce the following unknowns, defined as jumps in displacement derivatives along 
the crack:
\begin{subequations}
    \label{eqn:dispJumps}
    \begin{align}
            g_1 \left( y \right) & = \bigg\llbracket \partial_y u_0 (0,y) -  \frac{\gamma_{xy}}{\lambda^2} \int y  \, \partial_{yy} w_0 (0,y) \text{d}y \bigg\rrbracket, \label{eqn:FirstGreenF} \\
            g_2 \left( y \right) & =  \bigg\llbracket \partial_y \beta_x (0,y)  \bigg\rrbracket, \label{eqn:SecGreenF} \\
            g_3 \left( y \right) & =  \bigg\llbracket \partial_y v_0 (0,y) -  \frac{\gamma_y}{\lambda^2} \, y \,  \partial_{y} w_0 (0,y)  \bigg\rrbracket, \label{eqn:ThirdGreenF} \\
            g_4 \left( y \right) & =  \bigg\llbracket \partial_y \beta_y (0,y)  \bigg\rrbracket, \label{eqn:FourthGreenF} \\
            g_5 \left( y \right) & =  \bigg\llbracket \partial_y w_0 (0,y)  \bigg\rrbracket, \label{eqn:FifthGreenF}
    \end{align}
\end{subequations}
where the jump operator is defined as 
$\llbracket f(0, y) \rrbracket = \lim_{x \to 0^+} f(x, y) - \lim_{x \to 0^-} f(x, y)$.

These new unknowns are Fourier transformed and related to the original set of ten unknown functions 
$R_1(\alpha),\dots,R_8(\alpha)$, $A_1(\alpha),A_2(\alpha)$, see 
Appendix~\ref{app:FourierJumps}. This gives
\begin{equation}
    \label{eqn:newToOldUnknownsMain}
    \begin{aligned}
        R_j (\alpha) &= i \sum_{k =1}^5 B_{jk}(\alpha) \, \hat{g}_k(\alpha) \qquad j = 1, \dots,8 \\
        A_1 (\alpha) &= \sum_{k=1}^5 B_k(\alpha) \, \hat{g}_k(\alpha), \\ 
        A_2 (\alpha) &= \sum_{k=1}^5 C_k(\alpha) \, \hat{g}_k(\alpha),
    \end{aligned}
\end{equation}
where the coefficients $B_{jk}(\alpha)$, $B_k(\alpha)$, and $C_k(\alpha)$ can be computed numerically 
for a given $\alpha$.
We note that this change of variables also reduces the strength of the expected crack-tip 
singularities, see \cite{Muskhelishvili:1972,Yahsi:1983}.

Compatibility ahead of the crack tips requires that the kinematic jumps vanish for $|y|\geq 1$. 
Therefore, the Fourier transform $\hat{g}_k$ of $g_k$ can be simplified as  
\begin{equation}
    \label{eqn:gBackTranform}
    \begin{aligned}
        \hat{g}_k (\alpha) = \int_{-\infty}^\infty g_k(y) \, e^{i \alpha y} \, \text{d}y = \int_{-1}^{1} g_k(y) \, e^{i \alpha y} \, \text{d}y. 
    \end{aligned}
\end{equation}

Next, we reformulate the crack-plane coupling conditions in terms of 
$\hat{g}_k(\alpha)$.
To this end, we first derive the membrane, bending, and transverse shear stress resultants in terms
of $R_1(\alpha), \dots, R_8(\alpha), A_1(\alpha), A_2(\alpha)$, see
Appendix~\ref{app:FourierExpressions}. Using Equation~\eqref{eqn:newToOldUnknownsMain}, we then
rewrite these resultants in terms of $\hat{g}_k(\alpha)$, and finally, using
Equation~\eqref{eqn:gBackTranform}, in terms of $g_k(y)$.
Substituting the resulting expressions into the crack-loading boundary conditions
\eqref{eqn:boundaryConditionss} yields a set of five integral equations
\begin{equation}
    \label{eqn:SingIntegralsMixedLoading}
    \begin{aligned}
        \lim_{x \to 0^+}& \sum_{j =1}^5 \int_{-1}^1 g_j (t) \left[\int_{- \infty}^{\infty} H_{kj}^+\left(\alpha,x\right) e^{i \alpha (t -y)} \, \text{d}\alpha \right] \text{d}t + \\
        + \lim_{x \to 0^-}& \sum_{j =1}^5 \int_{-1}^1 g_j (t) \left[\int_{- \infty}^{\infty} H_{kj}^-\left(\alpha,x\right) e^{i \alpha (t -y)} \, \text{d}\alpha \right] \text{d}t = 2 \, f_k (y), \qquad k = 1,\dots, 5,
    \end{aligned}
\end{equation}
where the kernels $H_{kj}^{\pm}(\alpha,x)$ are reported in Appendix~\ref{app:Kernels}, and
$$
\bm{f}(y) = \left[\tilde{N}^0_{xx}(y),\tilde{N}^0_{xy}(y),\tilde{M}^0_{xx}(y),\tilde{M}^0_{xy}(y),\tilde{V}^0_x(y)\right],
$$
with
\begin{equation}\label{eqn:NormalizedStressResultants}
\tilde{N}^{0}_{xx}=\frac{N^0_{xx}}{E\,h}, \quad
\tilde{N}^{0}_{xy}=\frac{N^0_{xy}}{E\,h}, \quad
\tilde{M}^{0}_{xx}=\frac{M^0_{xx}}{E\,h^2}, \quad
\tilde{M}^{0}_{xy}=\frac{M^0_{xy}}{E\,h^2}, \quad
\tilde{V}^{0}_{x}=\frac{V^0_x}{\gamma\mu h}.
\end{equation}
are the normalized crack-surface loading resultants. 

Equations~\eqref{eqn:SingIntegralsMixedLoading} form a system of singular integral equations. 
The solution approach closely follows the work of  \citet{Erdogan:1972a} and 
\citet{Yahsi:1981}, and is detailed in Appendices~\ref{app:SingularTermExtraction} and~\ref{app:NumericalSingularIntegral}. 
Specifically, following \citet{Yahsi:1981}, we decompose $g_k(t)$ into a bounded unknown 
function $h_k(t)$ and the singular density function $(1-t^2)^{-1/2}$:
\begin{equation}
    \label{eqn:shapeFuncDecomposition}
    \begin{aligned}
        g_k(t) = \frac{h_k(t)}{\sqrt{1-t^2}}, \qquad k = 1, \dots, 5.
    \end{aligned}
\end{equation}
We then discretize the integral equations at $N$ points using a dedicated Gauss--Chebyshev
quadrature scheme, following \citet{Erdogan:1972a}. This yields a $5N \times 5N$ linear system for
the values of $h_k$, $k=1,\dots,5$, evaluated at the $N$ quadrature points.
The resulting linear system can then be solved using standard numerical linear algebra methods,
yielding the values of the bounded functions $h_k$ at the quadrature points.

Once the functions $h_k(t)$ have been obtained, all quantities of interest in the fracture
problem can be evaluated: the crack lip displacement field, the asymptotic crack-tip stress field, 
the stress intensity factors, and the energy release rate, for which we provide detailed derivation
in Appendices~\ref{app:DisplacementFields}, \ref{app:StressFields}, and \ref{app:ERR}, respectively.

\subsection{Flat-plate normalization and curvature-induced mode coupling}
\label{ssec:flatPlateNormalization}

Since the objective of this work is to isolate the effect of curvature, the flat-plate solution
provides a natural reference against which curvature-induced deviations can be measured. 
It is therefore natural to normalize the stress intensity factors and the energy release rate by 
the corresponding quantities for a flat plate subjected to equivalent loading conditions.
To make this comparison meaningful, however, the loading condition must be defined consistently
across flat and curved geometries, as well as crack orientations. 
In a flat plate, the remotely applied loading, when expressed in the crack-local frame, coincides 
with the corresponding crack-surface loading. 
This is not generally the case for curved shells, where the same far-field loading applied to shells 
with different curvature profiles may produce different local loading conditions at the crack surfaces.
To ensure a consistent definition of the loading across shell geometries and crack orientations, 
we therefore prescribe the loading directly on the crack surfaces, through the stress resultants 
defined in Equation~\eqref{eqn:LoadingResultants} and expressed in the crack-local coordinates $x$ and $y$.

After solving the perturbation problem with the solution strategy illustrated in Section \ref{ssec:semianalyticalSolution}, the stress intensity factors can be obtained directly from 
the bounded functions $h_k(t)$ evaluated at the crack tips $y = \pm 1$ through
\begin{subequations}
    \label{eqn:SIFSfromhk}
    \begin{align}
        K^{\pm}_{I} (z) & = - \frac{E}{4} \sqrt{\pi a} \left(h_1(\pm 1) + z \, h_2(\pm 1)\right), \\
        K^{\pm}_{II} (z) & = - \frac{E}{4} \sqrt{\pi a} \left(h_3(\pm 1) + z \, h_4(\pm 1)\right), \\
        K^{\pm}_{III} (z) & = - \frac{3}{4} \gamma \, \frac{E}{2\,(1+\nu)} \, \sqrt{\pi a}  \left( 1 -  \left( \frac{2a}{h}\, z \right)^2 \right) h_5(\pm 1),
    \end{align}
\end{subequations}
where $z\in[-{h}/{(2a)}, {h}/{(2a)}]$.

The \sifs\ are generally not constant through the shell thickness. 
Therefore, evaluating them only at the shell midsurface does not provide a complete description 
of the fracture response in curved Reissner shells. 
Through-the-thickness variations of the opening \sif\ $K_I$ and sliding \sif\
$K_{II}$ indicate nonuniform opening and in-plane sliding responses across the shell thickness,
respectively. These variations may be interpreted as bending-induced opening and twisting-induced 
sliding responses.
In the following, we use the terms \emph{membrane mode I}, \emph{in-plane mode II}, and
\emph{tearing} to refer to the \emph{fracture responses} associated with nonzero values of $K_I$, $K_{II}$, and
$K_{III}$ at the shell midsurface, respectively. 
By contrast, we use the terms \emph{bending mode I} and \emph{twisting mode II} to refer to the 
responses associated with through-the-thickness variations of $K_I$ and $K_{II}$. 
We do not assign a separate interpretation to through-the-thickness variations of the tearing \sif\ 
$K_{III}$, because transverse shear stresses in Reissner's theory should be interpreted as weighted 
through-thickness averages rather than pointwise quantities \citep{Reissner:1947}.

It is therefore useful to introduce the quantities $K_m$, $K_s$, $K_v$, $K_b$, and $K_t$ as 
stress intensity factors individually associated with membrane opening, in-plane sliding, tearing, 
bending opening, and twisting sliding, respectively. 
We define them as:
\begin{equation}
    \label{eqn:virtual-SIFs}
        K_m = K_{I}\left(0\right),  \quad 
        K_s = K_{II}\left(0\right), \quad
        K_v = K_{III}\left(0\right),\quad
        K_b = K_{I}\left(\frac{h}{2a}\right) - K_{I}\left(0\right) , \quad
        K_t = K_{II}\left(\frac{h}{2a}\right) - K_{II}\left(0\right),
\end{equation}
which, for a plane stress, center-cracked, infinite flat plate \citep{Sih:1962,sih:1965} evaluate to 
\begin{equation}
    \label{eqn:referenceSIFS}
    \begin{aligned}
            K^{\mathrm{flat}}_m = \frac{N_{xx}^0}{h} \sqrt{\pi a}, \quad
            K^{\mathrm{flat}}_s = \frac{N_{xy}^0}{h} \sqrt{\pi a}, \quad 
            K^{\mathrm{flat}}_v = \frac{3}{2} \frac{V_x^0}{h} \sqrt{\pi a}, \quad
            K^{\mathrm{flat}}_b = \frac{6 M_{xx}^0}{h^2} \sqrt{\pi a}, \quad 
            K^{\mathrm{flat}}_t = \frac{6 M_{xy}^0}{h^2} \sqrt{\pi a}.
    \end{aligned}
\end{equation}

For a flat plate, these \sifs\ are decoupled: each loading component is uniquely associated with a
single fracture response. 
In curved shells, however, this is no longer the case. 
We therefore distinguish between loading components, characterized by the applied stress resultants, 
and fracture modes, characterized by the resulting stress intensity factors. 
A given loading component may activate multiple \sifs, leading to a mixed-mode response. We
refer to this phenomenon as \emph{mode coupling}.
When mode coupling occurs, we refer to the stress intensity factor associated with the fracture 
mode activated by the same loading in the corresponding flat-plate problem as the \emph{primary} 
stress intensity factor. The other stress intensity factors activated by curvature-induced coupling 
are referred to as \emph{coupled} stress intensity factors.

This coupling can be made explicit by rewriting Equation~\eqref{eqn:SIFSfromhk} in a form that
separates the contribution of each loading component. 
To this end, we use linear superposition and solve the problem independently for each normalized 
loading component $\tilde{N}^{0}_{xx}$, $\tilde{N}^{0}_{xy}$, $\tilde{M}^{0}_{xx}$, 
$\tilde{M}^{0}_{xy}$, and $\tilde{V}^{0}_{x}$ of Equation~\eqref{eqn:NormalizedStressResultants}. 
This yields the corresponding functions $h_k^m(t)$, $h_k^s(t)$, $h_k^v(t)$, $h_k^b(t)$, and 
$h_k^t(t)$, for $k=1,\dots,5$.
The solution for an arbitrary loading scenario can then be expressed as 
$h_k(t)=h_k^m(t)+h_k^s(t)+h_k^v(t)+h_k^b(t)+h_k^t(t)$ and the corresponding \sifs\ at the $y=1$ 
crack tip are
\begin{subequations}
    \label{eqn:SIFSTradMain}
    \begin{align}
        K^+_{I} (z) = 
        & \frac{1}{4 \tilde{N}_{xx}^0} \left(h_1^m(1) + \frac{2z}{h} h_2^m(1)\right) K^{\mathrm{flat}}_m
        + \frac{1}{24 \tilde{M}_{xx}^0} \left(h_1^b(1) + \frac{2z}{h} h_2^b(1)\right) K^{\mathrm{flat}}_b
        + \frac{(1 + \nu)}{3 \gamma \tilde{V}_{x}^0} \left(h_1^v(1) + \frac{2z}{h} h_2^v(1)\right) K^{\mathrm{flat}}_v
        \nonumber\\
        +& \frac{1}{4 \tilde{N}_{xy}^0} \left(h_1^s(1) + \frac{2z}{h} h_2^s(1)\right) K^{\mathrm{flat}}_s
        + \frac{1}{24 \tilde{M}_{xy}^0} \left(h_1^t(1) + \frac{2z}{h} h_2^t(1)\right) K^{\mathrm{flat}}_t \, , \\
        K^+_{II} (z) = 
        & \frac{1}{4 \tilde{N}_{xx}^0} \left(h_3^m(1) + \frac{2z}{h} h_4^m(1)\right) K^{\mathrm{flat}}_m
        + \frac{1}{24 \tilde{M}_{xx}^0} \left(h_3^b(1) + \frac{2z}{h} h_4^b(1)\right) K^{\mathrm{flat}}_b 
        + \frac{(1 + \nu)}{3\gamma \tilde{V}_{x}^0} \left(h_3^v(1) + \frac{2z}{h} h_4^v(1)\right) K^{\mathrm{flat}}_v
        \nonumber\\
        +& \frac{1}{4 \tilde{N}_{xy}^0} \left(h_3^s(1) + \frac{2z}{h} h_4^s(1)\right) K^{\mathrm{flat}}_s
        + \frac{1}{24 \tilde{M}_{xy}^0} \left(h_3^t(1) + \frac{2z}{h} h_4^t(1)\right) K^{\mathrm{flat}}_t \, , \\
        K^+_{III} (z) = 
        & \left( 1 - \left(\frac{2z}{h}\right)^2 \right) \left[
        \frac{3 \gamma}{8 (1+\nu) \tilde{N}_{xx}^0} h_5^m(1) \, K^{\mathrm{flat}}_m 
        + \frac{\gamma}{16 (1+\nu) \tilde{M}_{xx}^0} h_5^b(1) \, K^{\mathrm{flat}}_b 
        + \frac{1}{2 \tilde{V}_{x}^0} h_5^b(1) \, K^{\mathrm{flat}}_v \right.
        \nonumber\\
        +& \left. \frac{3 \gamma}{8 ( 1+ \nu) \tilde{N}_{xy}^0} h_5^s(1)  \, K^{\mathrm{flat}}_s
        + \frac{\gamma}{16 (1+\nu) \tilde{M}_{xy}^0} h_5^t(1)  \, K^{\mathrm{flat}}_t \right],
    \end{align}
\end{subequations}
where the terms have been collected so as to expose the corresponding flat-plate reference \sifs.

Equation~\eqref{eqn:SIFSTradMain} shows that, when only one loading component acts at a time, the
center-cracked flat-plate \sifs\ appear as multiplicative factors. 
The remaining factors therefore quantify the curvature-induced deviation from the flat-plate 
response and reveal which fracture modes are activated by each loading component.

An analogous additive decomposition is not possible for the energy release rate $G$, which depends
quadratically on the \sifs\ and, therefore, on the functions $h_k(t)$. 
Superposition cannot be applied directly to $G$, and combined loading components give rise to 
additional coupling terms.
However, when only one loading component acts at a time, the corresponding center-cracked flat-plate
energy release rate again appears as a multiplicative factor, analogously to the \sifs.

Under the assumptions of self-similar crack growth, infinitesimal crack extension, and sufficient
membrane tension to prevent crack closure \citep{Yahsi:1981}, the energy release rates associated
with individual loading components are obtained from Irwin's formulae \citep{irwin:1957} as
\begin{subequations}
    \label{eqn:ERRsingleLoading}
    \begin{align}
         G_m &= \frac{1}{120 (\tilde{N}_{xx}^0)^2}\left(15 h_1^2(1) + 5 h_2^2(1) + 15h_3^2(1) + 5 h_4^2(1) + \frac{9 \gamma^2}{1 + \nu} h_5^2(1)\right) G^{\mathrm{flat}}_m \, , \\
         G_s &= \frac{1}{120 (\tilde{N}_{xy}^0)^2}\left(15 h_1^2(1) + 5 h_2^2(1) + 15h_3^2(1) + 5 h_4^2(1) + \frac{9 \gamma^2}{1 + \nu} h_5^2(1)\right) G^{\mathrm{flat}}_s \, , \\
         G_b &= \frac{1}{4320 (\tilde{M}_{xx}^0)^2}\left(15 h_1^2(1) + 5 h_2^2(1) + 15h_3^2(1) + 5 h_4^2(1) + \frac{9 \gamma^2}{1 + \nu} h_5^2(1)\right)G^{\mathrm{flat}}_b \, , \\
         G_t &= \frac{1}{4320 (\tilde{M}_{xy}^0)^2}\left(15 h_1^2(1) + 5 h_2^2(1) + 15h_3^2(1) + 5 h_4^2(1) + \frac{9 \gamma^2}{1 + \nu} h_5^2(1)\right)G^{\mathrm{flat}}_t \, , \\
         G_v &= \frac{1}{135 (\tilde{N}_{xx}^0)^2}\left(\frac{15 ( 1 + \nu)}{\gamma^2} h_1^2(1) + \frac{5 ( 1 + \nu)}{\gamma^2} h_2^2(1) + \frac{15 ( 1 + \nu)}{\gamma^2} h_3^2(1) + \frac{5 ( 1 + \nu)}{\gamma^2} h_4^2(1) + 9  h_5^2(1)\right)G^{\mathrm{flat}}_v \, .
    \end{align}
\end{subequations}

\section{Results} \label{sec:results}

We now present the results obtained with the methodology described in Section 
\ref{sec:methodology} for prototypical isotropic shells, with particular emphasis on the interplay 
between shell curvature and fracture mechanics under different loading scenarios.
In particular, we examine how the mechanical response of the shell in the vicinity of the crack 
depends not only on the shell curvature landscape, but also on how the crack is embedded within it, 
thereby determining the curvature effectively \emph{perceived} by the crack.

\begin{figure}
    \centering
    \includegraphics[width=\textwidth]{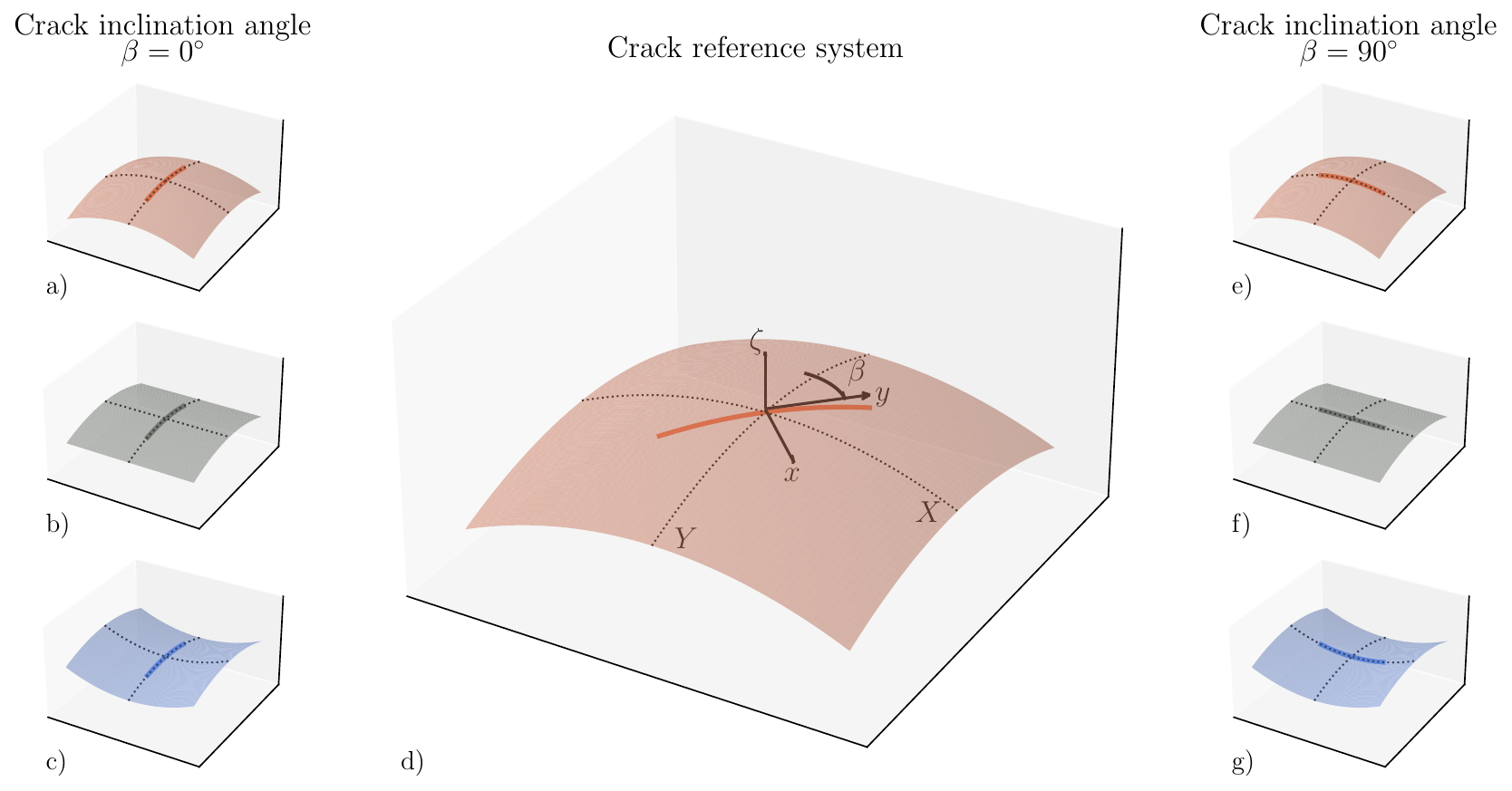}
    \caption{\emph{
    Geometry of the prototypical shells spanning the range of elliptic shells (a,e), cylindrical 
    shells (b,f), and hyperboloids (c,g).
    (d) Embedding of the crack with respect to the shell principal curvature directions $X$ and $Y$ 
    as a function of the crack inclination angle $\beta$. 
    Cracks aligned with $X$ or $Y$, corresponding to $\beta = 90^{\circ}$ and $\beta = 0^{\circ}$, 
    respectively, are referred to as \emph{principal cracks}.
    Note that in the left panels ($\beta = 0^{\circ}$) the curvature parallel to the crack remains 
    constant while the curvature normal to the crack varies. 
    The opposite holds for the right panels ($\beta = 90^{\circ}$).}}
    \label{fig:crack-reference}
\end{figure}

To this end, we consider a set of shell curvature profiles spanning a broad range of Gaussian 
curvatures $K$. See Figure~\ref{fig:crack-reference} for an illustration of the problem setting.
The prototypical shells considered are cylinders ($K = 0$), hyperboloids ($K < 0$) and ellipsoids 
($K > 0$). 
We explore the shell geometry space systematically by fixing one principal curvature, 
$\kappa_Y = 0.2$, while varying the other, $\kappa_X$, 
within a range of values $[-2\,\kappa_Y, 2\,\kappa_Y]$ consistent with 
the shallow-shell assumption of Section \ref{ssec:superposition}.
Within this family of geometries, three notable cases can be highlighted: the sphere 
($\kappa_X = \kappa_Y$), the cylinder ($\kappa_X = 0$), and the minimal surface 
($\kappa_X = -\kappa_Y$).
For each shell curvature profile, we analyze different crack embeddings, by varying the orientation 
of the crack $\beta$ continuously between the two limiting configurations where the crack is aligned 
with the principal directions $Y$ ($\beta = 0^{\circ}$) or $X$ ($\beta = 90^{\circ}$).
In the following, we restrict our attention to thin shells, where curvature effects on the fracture 
response are most prominent. Specifically, we set the shell thickness $h$ normalized by the crack 
half-length $a$ to $h/a = 0.1$. In all analyses, the Poisson's ratio is $\nu = 0.3$.

As discussed at length in Section \ref{ssec:flatPlateNormalization},
a consistent definition of the loading condition 
is required to enable a meaningful comparison of the fracture response across prototypical 
shells with different curvature profiles and crack embeddings. 
In fact, a direct comparison is complicated by the fact that the same far-field loading applied to 
different prototypical shells does not necessarily produce the same local loading condition at the 
crack surfaces. 
To ensure a consistent definition of the loading condition at the crack as the shell curvature profile and 
crack embedding vary, we prescribe the loading directly in terms of stress resultants on the crack 
surfaces.
For example, in the in-plane tensile loading case, a stress resultant normal to the 
crack plane is prescribed irrespective of shell curvature $\kappa_X$ and crack inclination angle 
$\beta$.

In the following, we present results for each normalized crack-surface loading component of Equation \eqref{eqn:NormalizedStressResultants}, 
$\tilde{N}^{0}_{xx}$, $\tilde{N}^{0}_{xy}$, $\tilde{M}^{0}_{xx}$, $\tilde{M}^{0}_{xy}$, and $\tilde{V}^{0}_{x}$, 
considered individually.
Since the governing problem is linear, these elementary solutions can be superposed to represent an
arbitrary loading scenario.

For in-plane shear, bending, twisting, and transverse shear loading components, the individual
response may lead to crack closure and therefore to negative values of the mode I stress 
intensity factor. 
Contact between crack faces is not included in the present formulation. 
Consequently, results for these isolated loading cases should be interpreted as elementary linear 
contributions, which are physically meaningful only when superposed with sufficient membrane tension 
to maintain an open crack.

We first consider the case of membrane tension loading $\tilde{N}^{0}_{xx}$, 
Section~\ref{sec:tensile}.
We then examine in-plane shear loading $\tilde{N}^{0}_{xy}$, Section~\ref{sec:inPlaneShear}, bending 
$\tilde{M}^{0}_{xx}$, Section~\ref{sec:bendingTwisting}, and twisting and transverse shear $\tilde{M}^{0}_{xy},\tilde{V}^{0}_x$, Section~\ref{sec:transverse}. 

\subsection{Membrane tension loading} \label{sec:tensile}

We investigate the response of the prototypical shells under membrane tension loading 
$\tilde{N}^{0}_{xx}>0$.
While this loading produces a pure mode I response for a flat geometry, the response differs 
significantly in the presence of curvature.

\begin{figure}
    \centering
    \includegraphics[width=\textwidth]{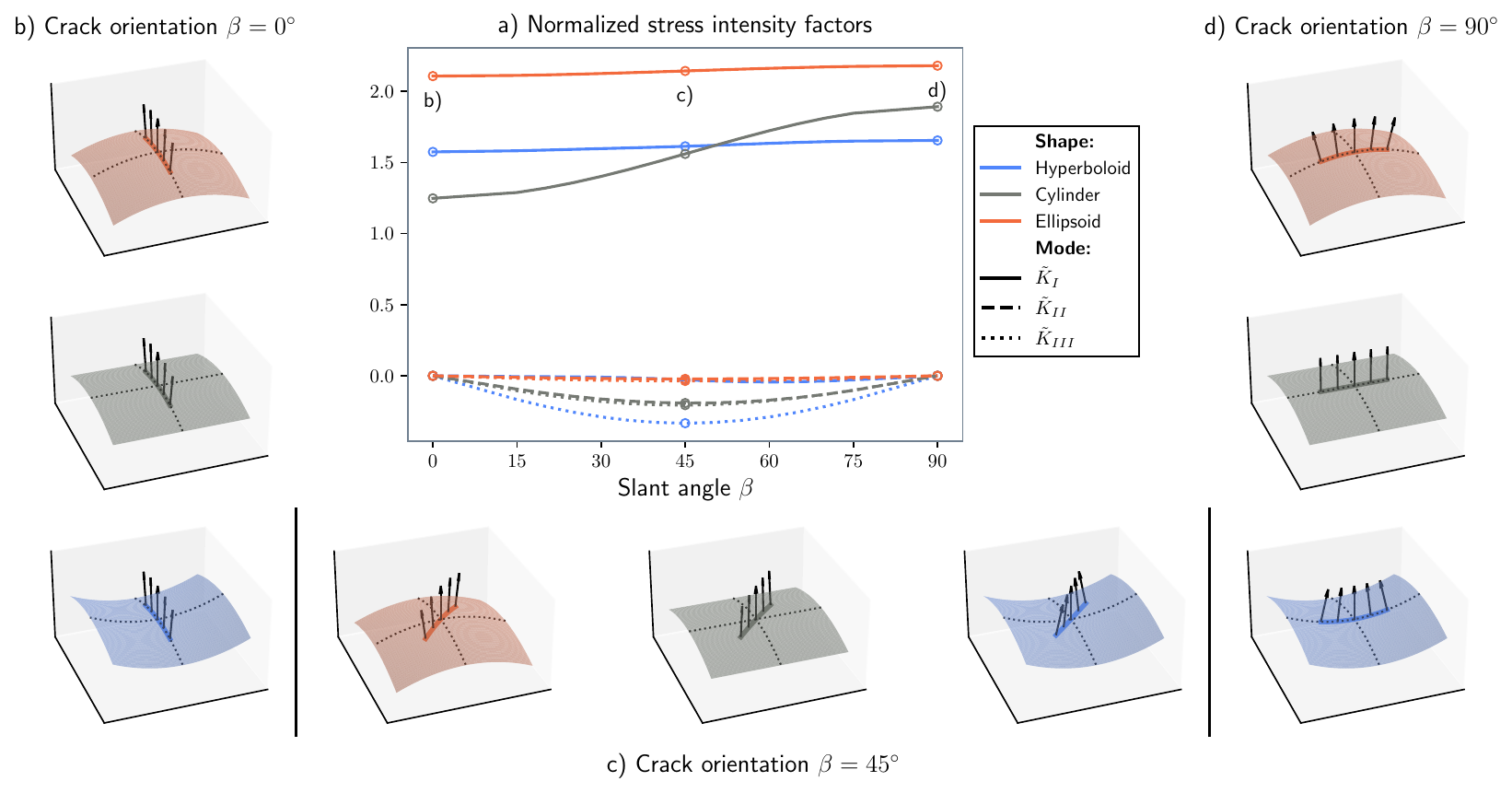}
    \caption{\emph{
    Fracture response of prototypical shells under membrane tension loading as a function of crack 
    orientation $\beta$. 
    (a) Normalized stress intensity factors $\tilde{K}_I$, $\tilde{K}_{II}$ and $\tilde{K}_{III}$ 
    at the shell midsurface ($z = 0$) and at the positive $y$ crack tip.
    (b,c,d) Representative geometries: ellipsoid ($\kappa_X/\kappa_Y = 0.833$), cylinder 
    ($\kappa_X/\kappa_Y = 0$), and hyperboloid ($\kappa_X/\kappa_Y = -0.833$), shown for three crack 
    orientations $\beta = 0^\circ$, $45^\circ$, and $90^\circ$. The arrows indicate the normal 
    vector to the curved shell along the crack.
    For all curvature profiles and crack orientations, the mode I stress intensity factor is 
    consistently larger than its flat-plate counterpart. 
    For the cylindrical and hyperboloidal profiles, whenever the crack is not aligned with a 
    principal curvature direction, all three fracture modes, opening, shearing, 
    and tearing, are activated. 
    Note that, for the cylindrical profile, the curves of the mode II and mode III \sifs\ 
    overlap.}}
    \label{fig:tension-misurface}
\end{figure}

Figure~\ref{fig:tension-misurface} shows the normalized stress intensity factors $\tilde{K}_I$, 
$\tilde{K}_{II}$ and $\tilde{K}_{III}$ at the midsurface of the shell ($z = 0$), as functions of the 
crack orientation $\beta$ for different curvature profiles.
All stress intensity factors are normalized by the corresponding stress intensity factor 
$K^{\mathrm{flat}}_m$ associated with mode I opening response for a center-cracked infinite flat 
plate subjected to the same loading (Equation~\eqref{eqn:referenceSIFS}).
The limiting cases $\beta = 0^\circ$
and $\beta = 90^\circ$ correspond to \emph{principal cracks}, i.e., cracks aligned with the shell
principal curvature directions.

We observe that, for all curvature profiles and crack orientations, the mode I stress intensity
factor $K_I$ is consistently larger than its flat-plate counterpart, with the ellipsoidal profile
exhibiting the largest values. Within the range of curvature profiles considered, $K_I$ exhibits
only a weak dependence on the crack orientation $\beta$, except in the cylindrical case, where the
dependence is more pronounced.
In this latter case, $K_I$ is maximal for $\beta = 90^{\circ}$, corresponding to vanishing curvature 
parallel to the crack, and minimal for $\beta = 0^{\circ}$, corresponding to vanishing curvature 
normal to the crack. 
Notably, for the cylindrical and hyperboloidal profiles, whenever the crack is not aligned with a
principal curvature direction, all three fracture modes, membrane mode I (opening), in-plane mode II (sliding),
and mode III (tearing), activate at the shell midsurface. 
The intensity of modes II and III varies with the crack inclination angle $\beta$ and is maximal 
for $\beta = 45^{\circ}$, corresponding to the configuration most misaligned with the principal 
curvature directions. Note that these \sifs\ are negative, meaning that shear stresses are negative 
on the crack edge with $x>0$ (see coordinate axes in Figure~\ref{fig:crack-reference}).
Mode III is most pronounced in the hyperboloidal profiles, while mode II is most pronounced in 
the cylindrical profiles, where the $K_{II}$ and $K_{III}$ curves overlap.
In contrast, for the ellipsoidal profile, only mode I is activated at the shell 
midsurface in response to membrane tension loading, regardless of the crack orientation.

\begin{figure}
    \centering
    \includegraphics[width=\textwidth]{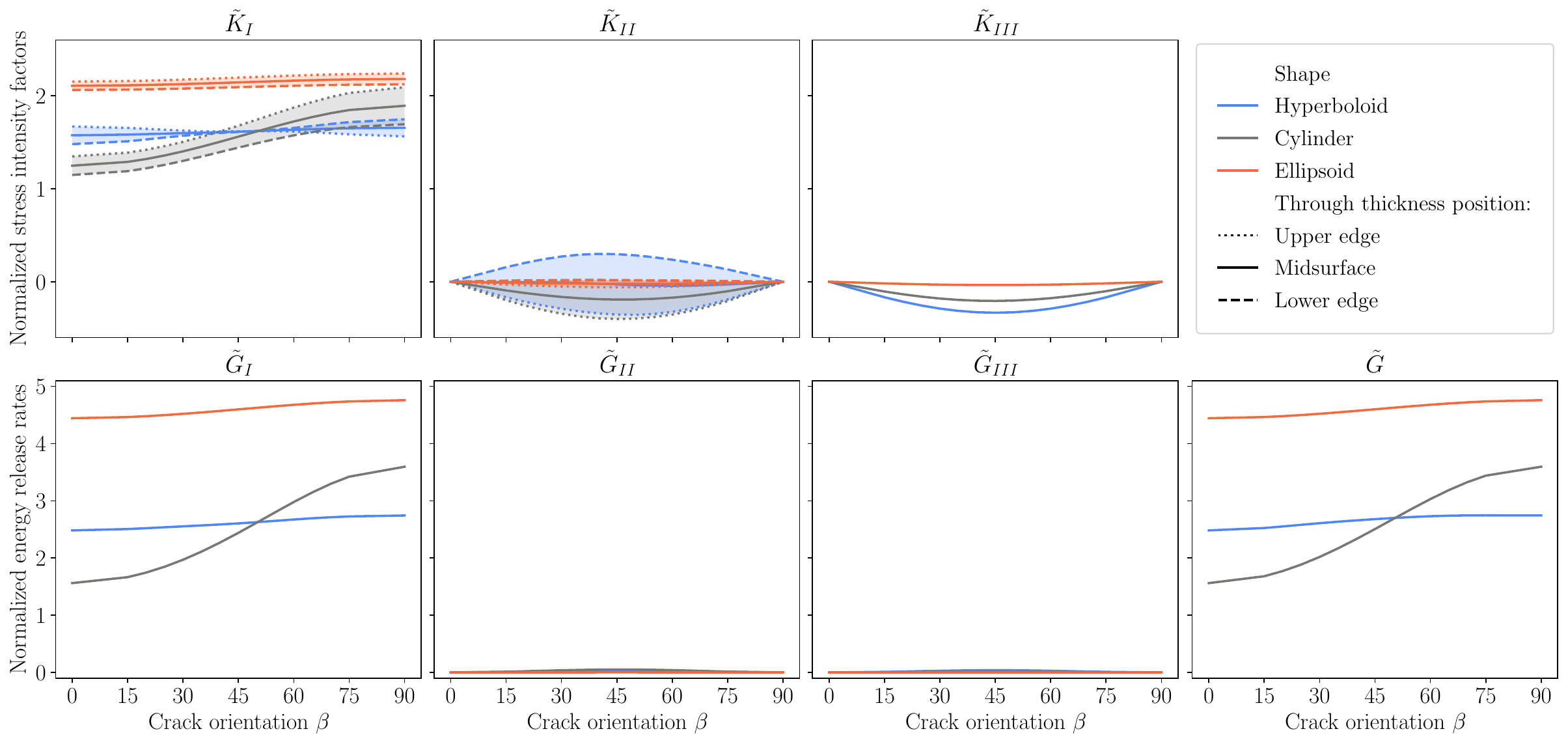}
    \caption{
    \emph{
    Fracture response of prototypical shells under membrane tension loading as a function of crack 
    orientation $\beta$. 
    The shell curvature profiles are: ellipsoid ($\kappa_X/\kappa_Y=0.833$), cylinder ($\kappa_X/\kappa_Y=0$), and hyperboloid ($\kappa_X/\kappa_Y=-0.833$).
    Top row: Normalized stress intensity factors $\tilde{K}_I$, $\tilde{K}_{II}$, and 
    $\tilde{K}_{III}$ at the positive $y$ crack tip and sampled through the thickness 
    (at the midsurface and and at the outer surfaces of the shell).
    The cylindrical and hyperboloidal profiles exhibit the strongest bending opening and twisting 
    sliding responses, respectively.
    The stress intensity factor $\tilde{K}_{III}$ is shown only at the midsurface as it vanishes by 
    construction at the outer shell surfaces due to the Reissner shell assumptions.
    Bottom row: Additive decomposition of the normalized energy release rate $\tilde{G}$ into its 
    opening contribution $\tilde{G}_I$ (including membrane and bending components),  
    sliding contribution $\tilde{G}_{II}$ (including in-plane and twisting components) and tearing 
    contribution $\tilde{G}_{III}$. 
    The contributions of $\tilde{G}_{II}$ and $\tilde{G}_{III}$ to the total energy release rate are 
    negligible, even in the cylindrical and hyperboloidal profiles where substantial mode II and 
    mode III responses are active. 
    Consequently, the total energy release rate is significantly larger for the ellipsoidal profile 
    than for the other ones.}}
    \label{fig:KandERRTension}
\end{figure}

Figure~\ref{fig:KandERRTension} shows the through-the-thickness variations of the normalized 
stress intensity factors as a function of the crack orientation $\beta$ for different curvature 
profiles. We refer to Section \ref{ssec:flatPlateNormalization} for the interpretation of 
through-the-thickness variations of stress intensity factors.
We observe that the ellipsoidal profile exhibits weak mode I opening and negligible twisting 
mode II. 
The hyperboloidal curvature profile exhibits the strongest twisting mode II response together with a moderate 
bending mode I response. 
For this curvature profile, the bending mode I changes sign with the crack orientation angle 
$\beta$ due to the opposite curvature perceived by the crack. 
For $\beta = 45^\circ$, the bending mode I vanishes while the twisting mode II response is maximal.
Finally, the cylindrical curvature profile exhibits all five fracture responses for all non-principal cracks. 
In this case, the bending mode I response is also maximal.

Figure~\ref{fig:KandERRTension} also shows how the five fracture responses, namely membrane 
and bending mode I, in-plane and twisting mode II, and tearing, combine to determine the 
normalized energy release rate $\tilde{G} = G/G_m^{\mathrm{flat}}$ through the additive contributions 
of $\tilde{G}_I = G_I / G_m^{\mathrm{flat}}$ (total opening), 
$\tilde{G}_{II} = G_{II} / G_m^{\mathrm{flat}}$ (total sliding), 
and $\tilde{G}_{III} = G_{III} / G_m^{\mathrm{flat}}$ (tearing).
The energy release rates are normalized by the corresponding energy release rate 
$G_m^{\mathrm{flat}}$ for a center cracked infinite flat plate subjected to the same loading.
We observe that $\tilde{G}_{II}$ and $\tilde{G}_{III}$ contribute negligibly to the total energy 
release rate, even in the cylindrical and hyperboloidal profiles where substantial mode II and mode III 
responses are active.
In addition, the total energy release rate is significantly larger for the ellipsoidal profile than 
for the other curvature profiles. 
This behavior can be understood by noting that, in the cylindrical and hyperboloidal profiles, the 
fracture energy is distributed among multiple coupled fracture modes, whereas in the ellipsoidal 
one it remains concentrated in mode I. 
Since the energy release rate depends quadratically on the \sifs\ 
(Equation~\eqref{eqn:ERRsingleLoading}), the energetic contribution of mode I dominates over those 
associated with modes II and III.
Finally, the cylindrical profile exhibits the strongest dependence of the energy release rate on 
the crack orientation angle $\beta$.

\begin{figure}
    \centering
    \includegraphics[width=\textwidth]{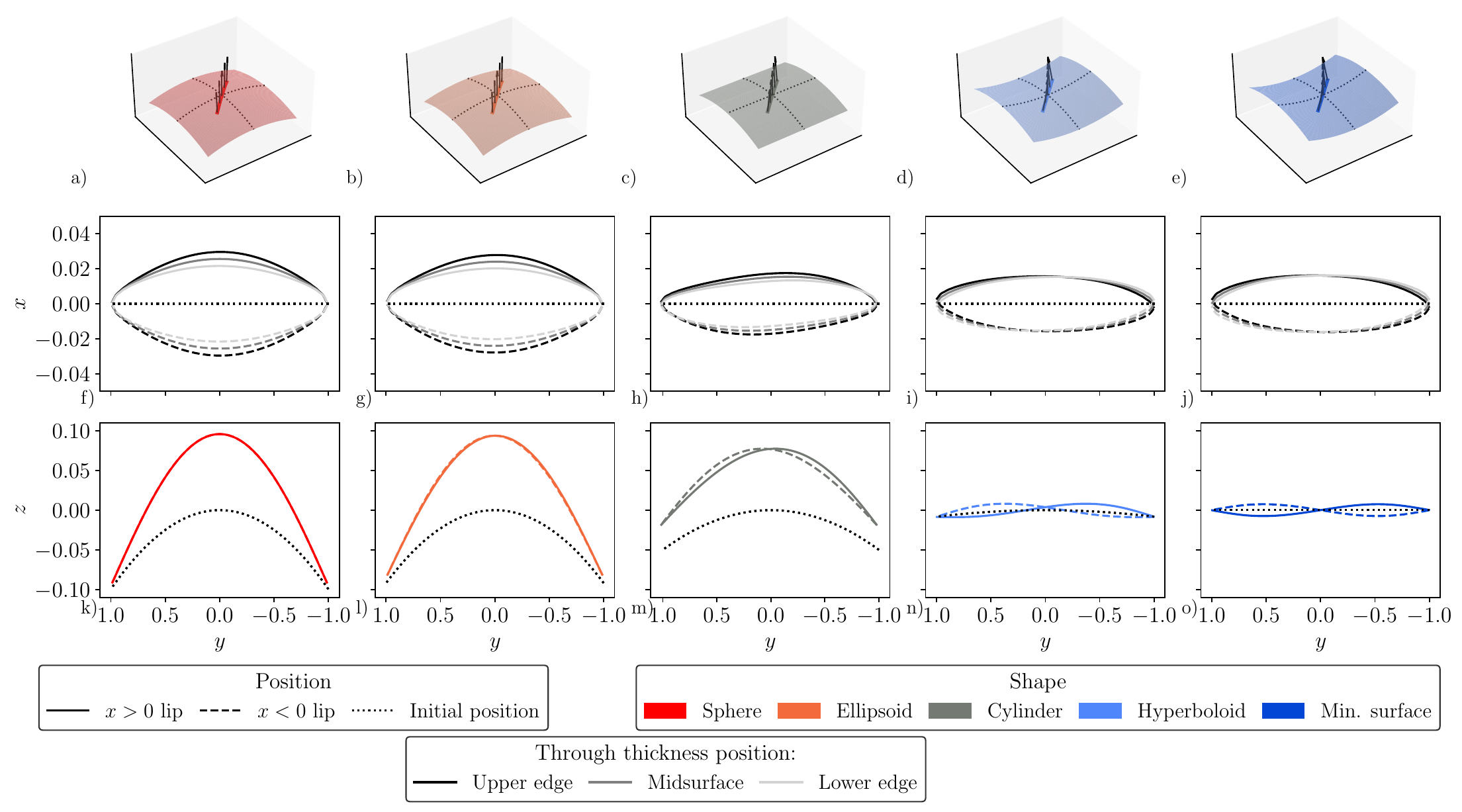}
    \caption{\emph{Crack-lip displacement fields under membrane tension for five representative 
    shells with non-principal cracks ($\beta =45^{\circ}$). 
    (a--e) The reference shell geometries with shell surface normals along the crack: 
    sphere ($\kappa_X/\kappa_Y=1$), ellipsoid ($\kappa_X/\kappa_Y=0.833$), 
    cylinder ($\kappa_X/\kappa_Y=0$), hyperboloid ($\kappa_X/\kappa_Y=-0.833$), 
    and minimal surface ($\kappa_X/\kappa_Y=-1$). 
    (f--j) The projections of the computed crack-lip displacement fields onto the $yx$ plane. 
    (k--o) The corresponding projections onto the $yz$ plane. 
    The sphere and ellipsoid exhibit symmetric in-plane and out-of-plane deformation profiles. 
    The hyperboloid and minimal surface exhibit an in-plane 
    response which combines an overall symmetric shape with antisymmetric components at 
    the crack tips, and a markedly antisymmetric out-of-plane response.
    The cylinder presents a combination of symmetric and antisymmetric deformation 
    both in plane and out of plane.
    Antisymmetric response is always associated with the surface normals twisting out of the crack 
    plane.
    }}
    \label{fig:displacements-tension}
\end{figure}

In Figure~\ref{fig:displacements-tension}, we examine the crack-lip deformation fields for 
representative shell curvature profiles.
We focus on the case $\beta=45^\circ$, for which the crack is maximally misaligned with the 
principal curvature directions, and we set $|\tilde{N}_{xx}^0| = 0.02$ (Equation \eqref{eqn:LoadingResultants}).
The sphere and ellipsoid exhibit symmetric in-plane and out-of-plane deformation profiles. 
The cylinder shows pronounced out-of-plane displacement of the crack lips, together with a
combination of symmetric and antisymmetric deformation components. 
Finally, the hyperboloid and minimal surface exhibit a weakly antisymmetric in-plane response and 
a markedly antisymmetric out-of-plane response.

\subsection{In-plane shear loading}
\label{sec:inPlaneShear}

We now investigate the response of the prototypical shells under in-plane shear loading 
$\tilde{N}^{0}_{xy}>0$ of the crack faces.
In the case of a center-cracked flat plate, this loading leads to a purely antisymmetric displacement parallel to 
the crack, with no opening or out-of-plane displacement \citep{Knott:1973} and friction between the 
crack faces \citep{zehnder:2005}.
In the presence of curvature, however, additional fracture responses are, in general, activated.

\begin{figure}
    \centering
    \includegraphics[width=\textwidth]{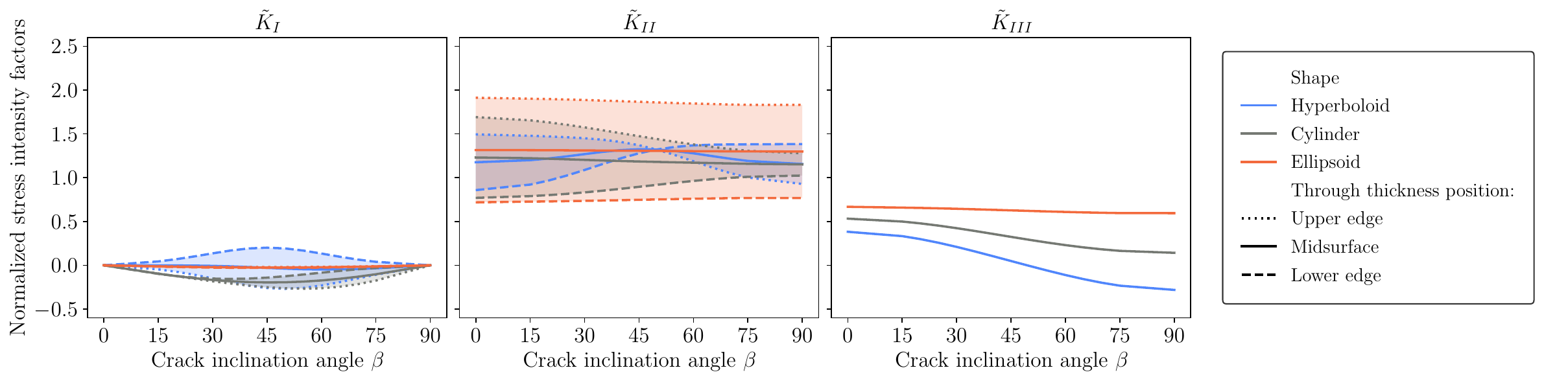}
    \caption{\emph{
    Fracture response of prototypical shells under in-plane shear loading as a function of crack 
    orientation $\beta$. 
    The shell curvature profiles are: ellipsoid ($\kappa_X/\kappa_Y=0.833$), cylinder ($\kappa_X/\kappa_Y=0$), and hyperboloid ($\kappa_X/\kappa_Y=-0.833$).
    The plots show the normalized stress intensity factors $\tilde{K}_I$, $\tilde{K}_{II}$, and 
    $\tilde{K}_{III}$ at the positive $y$ crack tip and sampled through the thickness 
    (at the midsurface and and at the outer surfaces of the shell).
    The in-plane sliding stress intensity factor is 
    consistently larger than the corresponding flat-plate reference value across all three curvature 
    profiles. 
    The ellipsoidal profile exhibits the strongest twisting sliding and tearing responses, with very 
    limited opening. 
    In contrast, cylindrical and hyperboloidal profiles show a stronger dependency on
    the crack orientation, as well as a noticeable opening response. 
    The stress intensity factor $K_{III}$ is shown only at the midsurface as it vanishes by 
    construction at the outer shell surfaces due to the Reissner shell assumptions.}}
    \label{fig:KandERRShear}
\end{figure}

Figure~\ref{fig:KandERRShear} shows the normalized stress intensity factors $\tilde{K}_I$, 
$\tilde{K}_{II}$, and $\tilde{K}_{III}$, sampled at the midsurface and outer edges of the shell, 
as functions of the crack inclination angle $\beta$, for three reference curvature profiles (ellipsoidal,
hyperboloidal and cylindrical). 
All stress intensity factors are normalized by the corresponding 
stress intensity factor $K^{\mathrm{flat}}_s$ associated with in-plane sliding response 
for a center-cracked infinite flat plate subjected to the same loading (Equation~\eqref{eqn:referenceSIFS}).

Like in the membrane tension case, the primary stress intensity factor, this time associated with in-plane 
mode II, is consistently larger than the corresponding center-cracked flat plate reference value across all three 
curvature profiles. 
At the same time, this primary response remains comparable across the different curvature profiles 
and exhibits only a weak dependence on the crack inclination angle.
The activation of the coupled fracture responses, however, varies substantially across curvature 
profiles and crack orientations, with the only consistent features being that principal crack 
orientations do not yield coupled opening responses. 

The ellipsoidal profile exhibits a fracture response that is nearly insensitive to the crack 
inclination angle. 
For this curvature profile, the coupled opening response remains negligible. 
In contrast, this profile exhibits the strongest twisting mode II and tearing responses among all 
the geometries considered.
Similar to ellipsoids, hyperboloids show no membrane opening response across all crack inclinations.
In all other coupled responses, hyperboloids exhibit a strong angular dependence.
Cracks nearly aligned with the principal directions of the shell lead to a limited bending 
response with substantial twisting mode II and tearing.
Strongly inclined cracks, on the other hand, exhibit a 
pronounced bending response with weaker twisting sliding and tearing, eventually reaching zero 
for a critical crack inclination.
In cylinders, for small values of $\beta$, the primary in-plane mode II response is accompanied by negative
membrane mode I, indicating crack closing, with negligible bending, as well as pronounced twisting mode II and tearing. 
As $\beta$ increases, the twisting mode II and tearing responses progressively weaken, while a 
bending mode I component emerges alongside a negative membrane 
mode I at approximately $\beta=30^\circ$ and reaches its maximum around $\beta=60^\circ$. 
As the axial crack configuration ($\beta=90^\circ$) is approached, both mode I
responses vanish, recovering the expected behavior for a principal crack orientation.

We emphasize that the closing response observed for the cylinder represent elementary contributions of in-plane shear loading
within the linear formulation and are physically meaningful only when superposed with sufficient
membrane tension to maintain the crack open. Because the energy release rate depends nonlinearly on
the stress intensity factors, and therefore cannot be obtained by linear superposition, we do not
discuss the energy release rate for this isolated loading case.

\begin{figure}
    \centering
    \includegraphics[width=\textwidth]{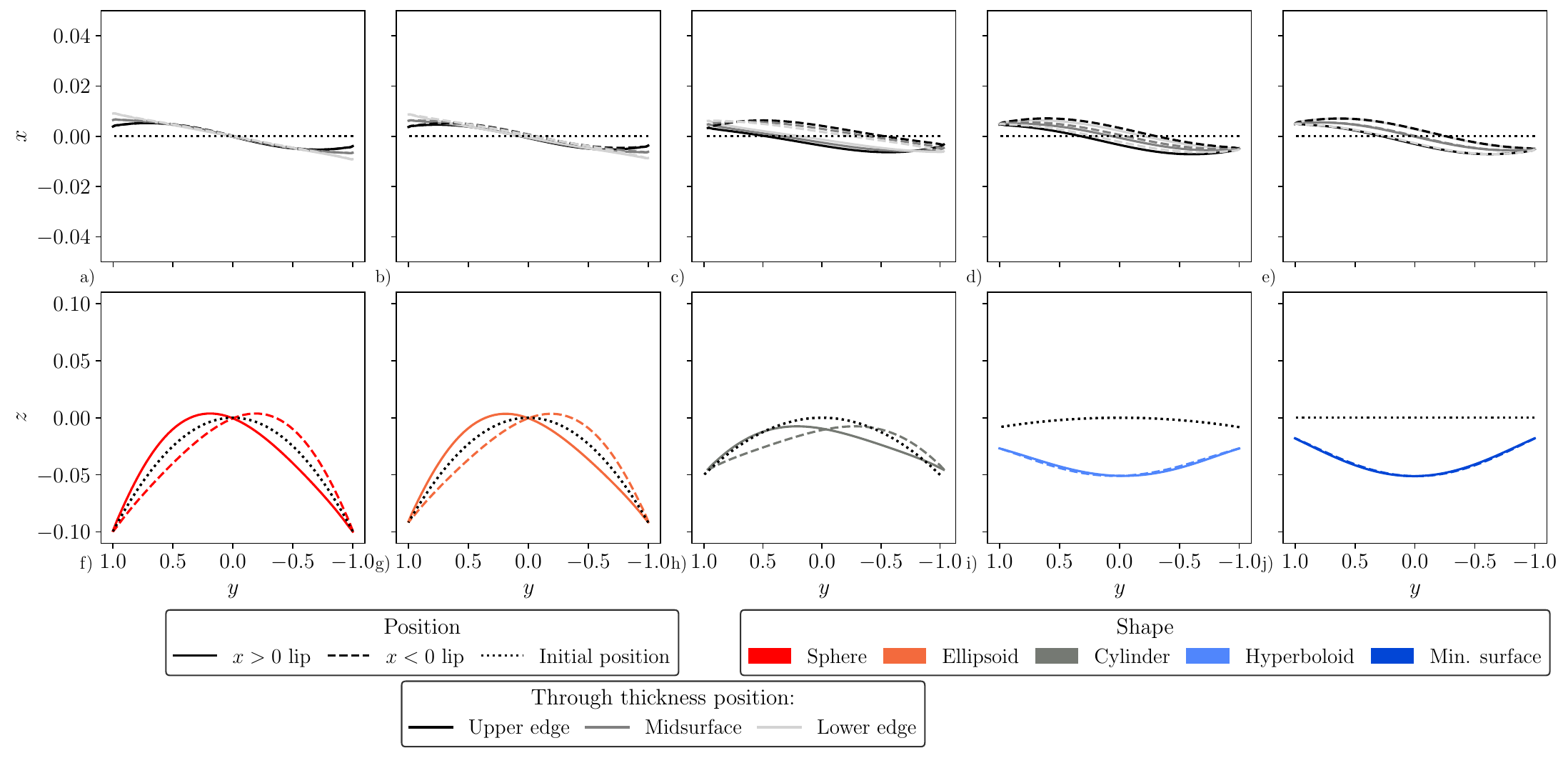}
    \caption{\emph{
    Crack-lip displacement fields under in-plane shear loading for five representative shells with 
    crack orientation $\beta=45^\circ$.
    From left to right, the shell curvature profiles are: sphere ($\kappa_X/\kappa_Y=1$), ellipsoid ($\kappa_X/\kappa_Y=0.833$), cylinder ($\kappa_X/\kappa_Y=0$), hyperboloid 
    ($\kappa_X/\kappa_Y=-0.833$), and minimal surface ($\kappa_X/\kappa_Y=-1$).
    (a--e) The projections of the computed crack-lip displacement fields onto the $yx$ plane. 
    (f--j) The corresponding projections onto the $yz$ plane. 
    The ellipsoid and sphere exhibit negligible opening and a fully antisymmetric out-of-plane 
    deformation. 
    In contrast, the hyperboloid and minimal surface exhibit a symmetric out-of-plane deformation
    and an in-plane deformation profile reminiscent of bending. 
    The cylinder exhibits in-plane closing response along the entire crack 
    length, together with a combination of symmetric and antisymmetric out of plane deformation.
    We recall that, in this setting, deformations involving crack closure are physically
    meaningful only when superposed with sufficient membrane tension to maintain the crack open.
    }}
    \label{fig:DisplacementsShear}
\end{figure}

In Figure~\ref{fig:DisplacementsShear}, we report the crack-lip displacements fields for the same 
five representative shell curvature profiles of Figure~\ref{fig:displacements-tension} and a crack
at an orientation $\beta = 45^\circ$.
Note that, in contrast to the membrane tension case, the loading now acts antisymmetrically with 
respect to the crack plane. 

Analogously to the flat plate response, the ellipsoid and sphere exhibit no in-plane opening. 
The crack tips rotate in opposite directions at the upper 
surface of the shell, producing the S-shape shown in Figure~\ref{fig:DisplacementsShear},
while the crack remains essentially straight at the lower surface.
However, this is accompanied by a fully antisymmetric out-of-plane response. 
The other curvature profiles exhibit a combination of symmetric and antisymmetric deformations, together 
with significant in-plane deformation.
The hyperboloid exhibits an almost perfectly symmetric deformation field.
The in-plane deformation consists in an almost exact symmetric bending response. Therefore, closure is observed at the upper edge, while opening is observed at the lower edge. Similarly, the out-of-plane response is nearly symmetric. In the minimal surface, the deformation is exactly symmetric both in- and out-of-plane.
The cylinder exhibits the largest in-plane displacement and significant out-of-plane displacement.
The in-plane displacement field shows a distinct closure component accompanied by bending and sliding, 
as well as a small antisymmetric twist of the crack tips, similar to that observed in ellipsoids.
The out-of-plane response consists of a superposition of symmetric bending-like
and antisymmetric twisting-like displacements.

\subsection{Bending loading}
\label{sec:bendingTwisting}

In this section, we investigate the response of the prototypical shells under bending loading,
$\tilde{M}^{0}_{xx}>0$.
Bending loading applied to a center-cracked infinite flat plate produces an 
elliptical in-plane deformation field, with equal and opposite interpenetration at the two shell 
edges and zero displacement of the midsurface \citep{Heming:1980}.
The response of the curved shells remains qualitatively close to this flat-plate behavior, with 
substantially weaker curvature-induced mode coupling than observed under membrane tension and 
in-plane shear loading.
As in the flat-plate case, the resulting elementary response involves crack-face closure. 
Therefore, as in the in-plane shear case, we restrict the discussion to the stress intensity factors, 
which can be linearly superposed with an opening loading component sufficient to maintain the crack 
faces separated, and omit the energy release rate.

\begin{figure}
    \centering
    \includegraphics[width=\textwidth]{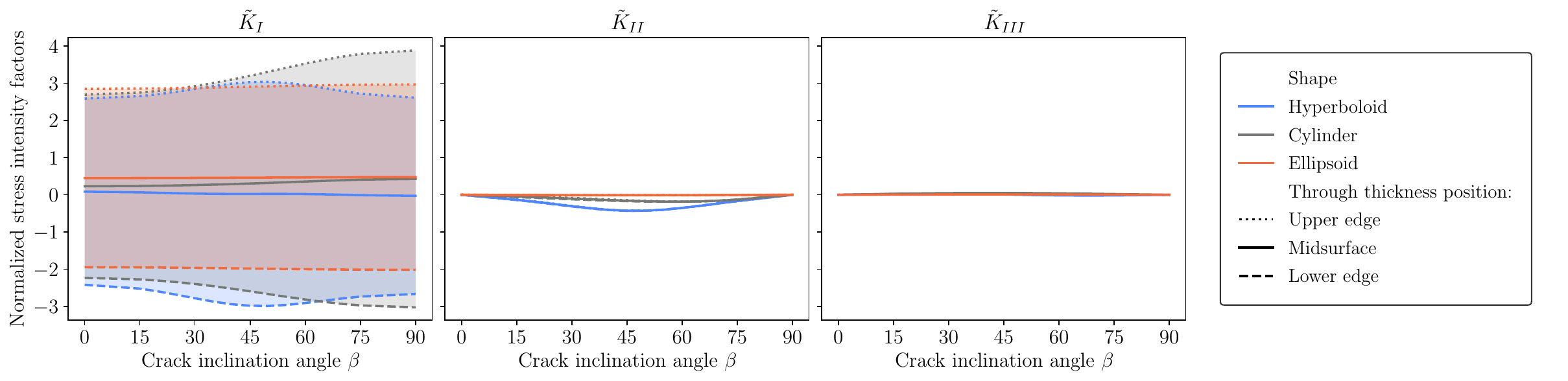}
    \caption{\emph{
    Fracture response of prototypical shells under bending loading as a function of crack 
    orientation $\beta$. 
    The shell curvature profiles are: ellipsoid ($\kappa_X/\kappa_Y=0.833$), cylinder ($\kappa_X/\kappa_Y=0$), and hyperboloid ($\kappa_X/\kappa_Y=-0.833$).
    The plots show the normalized stress intensity factors $\tilde{K}_I$, $\tilde{K}_{II}$, and 
    $\tilde{K}_{III}$ at the positive $y$ crack tip and sampled through the thickness 
    (at the midsurface and and at the outer surfaces of the shell).
    Across all curvature profiles and crack orientations, the response is dominated by the primary 
    bending stress intensity factor, which remains larger than the corresponding center-cracked 
    flat-plate reference value. 
    Twisting mode II and tearing responses are negligible in all cases, while the magnitude of the 
    remaining coupled responses varies with curvature profile and crack orientation.
    The stress intensity factor $K_{III}$ is shown only at the midsurface as it vanishes by 
    construction at the outer shell surfaces due to the Reissner shell assumptions.
    }}
    \label{fig:KandERRbending}
\end{figure}

Figure~\ref{fig:KandERRbending} shows the normalized stress intensity factors $\tilde{K}_I$, 
$\tilde{K}_{II}$, and $\tilde{K}_{III}$, sampled at the shell midsurface and at the upper and lower 
edges of the shell, as functions of the crack inclination angle $\beta$, for three reference 
curvature profiles (ellipsoidal, hyperboloidal and cylindrical). 
All stress intensity factors are normalized by the corresponding stress intensity factor 
$K^{\mathrm{flat}}_b$ associated with bending response for a center-cracked infinite flat plate 
subjected to the same loading (Equation~\eqref{eqn:referenceSIFS}).

As in the previously analyzed loading cases, the response across all curvature profiles and crack 
orientations is dominated by the primary stress intensity factor, which in this case is associated 
with bending. 
For this loading case as well, the primary stress intensity factor is consistently larger than the 
corresponding center-cracked flat-plate reference value.
In addition, for this loading, the tearing and twisting mode II responses remain effectively zero 
for all cases.
The extent of the other coupled responses varies substantially across curvature profiles and crack 
orientations.

The ellipsoidal profile exhibits the largest membrane mode I contribution and, consequently, the 
strongest membrane--bending coupling. 
At the same time, as observed under membrane tension and in-plane shear loading, the ellipsoid shows 
the weakest dependence on the crack inclination angle. 
Finally, for this curvature profile, the in-plane mode II response remains effectively zero irrespective of $\beta$.
The hyperboloid exhibits a slightly stronger bending response than the ellipsoid, with a further 
small increase for inclined cracks. 
Its membrane mode I contribution is negligible, while the in-plane mode II response, although 
still small, is the largest among the curvature profiles considered and reaches its maximum for skew 
cracks.
Finally, the cylinder exhibits a pronounced dependence of the mode I responses on crack orientation. 
The axial crack ($\beta=90^\circ$) produces a stronger response than the circumferential crack 
($\beta=0^\circ$), both in the primary bending component and in the coupled membrane mode I 
contribution.
In contrast, for this curvature profile, the sliding response is negligible regardless of the crack embedding. 

\begin{figure}
    \centering
    \includegraphics[width=\textwidth]{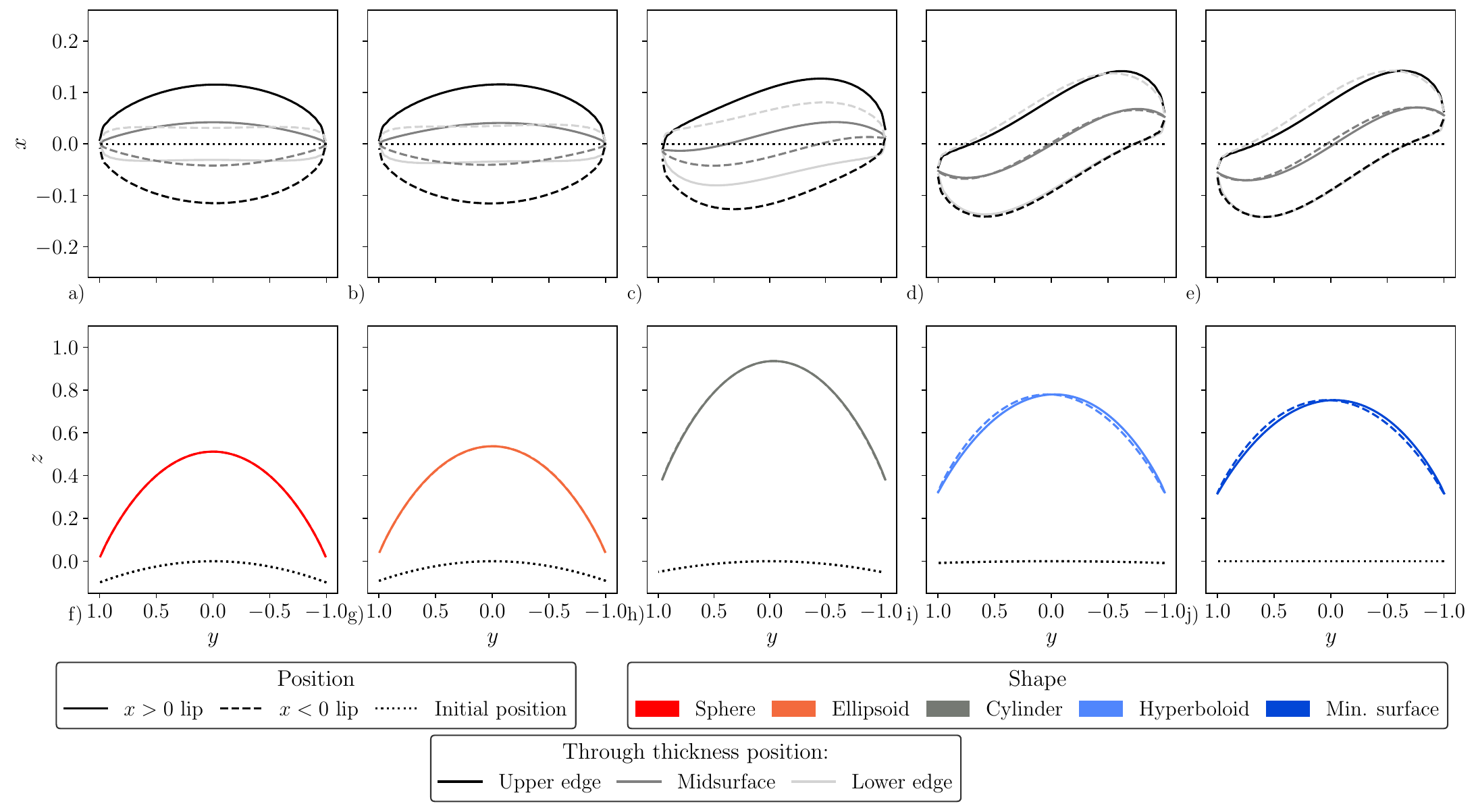}
    \caption{\emph{Crack-lip displacement fields under bending loading for five representative 
    shells with crack orientation $\beta=45^\circ$.
    From left to right, the shell curvature profiles are: sphere ($\kappa_X/\kappa_Y=1$), ellipsoid ($\kappa_X/\kappa_Y=0.833$), cylinder ($\kappa_X/\kappa_Y=0$), hyperboloid 
    ($\kappa_X/\kappa_Y=-0.833$), and minimal surface ($\kappa_X/\kappa_Y=-1$).
    (a--e) The projections of the computed crack-lip displacement fields onto the $yx$ plane. 
    (f--j) The corresponding projections onto the $yz$ plane. 
    The ellipsoid and sphere show a symmetric deformation profile, with the rotation axis located 
    between the mid- and the bottom surface of the shell. 
    Hyperboloid and minimal surface, instead, show a nearly symmetric out-of-plane and a shear-like 
    in-plane deformation.} }
    \label{fig:DisplacementsBending}
\end{figure}

The crack-lip displacement fields for five reference geometries and cracks inclined at an angle 
$\beta = 45^{\circ}$ are presented in Figure~\ref{fig:DisplacementsBending}.

All deformation profiles exhibit the characteristic bending response, with opening at the upper 
crack edge and closure at the lower edge.
The ellipsoid and the sphere exhibit elliptical in-plane deformation profiles with nonzero 
opening at the midsurface. 
For these curvature profiles, the out-of-plane response is symmetric.
The hyperboloid and minimal surface exhibit in-plane deformation profiles that combine the bending 
response of a center-cracked infinite flat plate with the S-shaped profile observed under in-plane 
shear loading in Figure~\ref{fig:DisplacementsShear}.
The crack opening at the midsurface is nearly zero, while opening and closing displacements at the 
outer shell edges are nearly equal in magnitude.
For these curvature profiles, the out-of-plane deformation presents only a weak anti-symmetric 
contribution. 
Finally, the cylinder exhibits an intermediate response. 
Its in-plane deformation combines features of the ellipsoidal response, namely the shifted neutral 
axis and elliptical crack opening, with features of the hyperboloidal response, for which the 
antisymmetric S-shape remains clearly recognizable. 
The midsurface opening is nonzero, while the displacements at the outer shell edges are closer in 
magnitude than in the ellipsoid. 
For this curvature profile, the out-of-plane displacement remains symmetric, as in the ellipsoidal 
geometries.

\subsection{Twisting and transverse loading} \label{sec:transverse}

The two remaining loading components are the twisting moment $\tilde{M}^0_{xy}$ and transverse shear 
loading $\tilde{V}_x^0$. 
As discussed by Hui \emph{et al.} \citep{Hui:1993}, the responses to these loading components are 
inherently coupled even in flat Reissner plates. 
We therefore consider them together in this section.

\begin{figure}
    \centering
    \includegraphics[width=\textwidth]{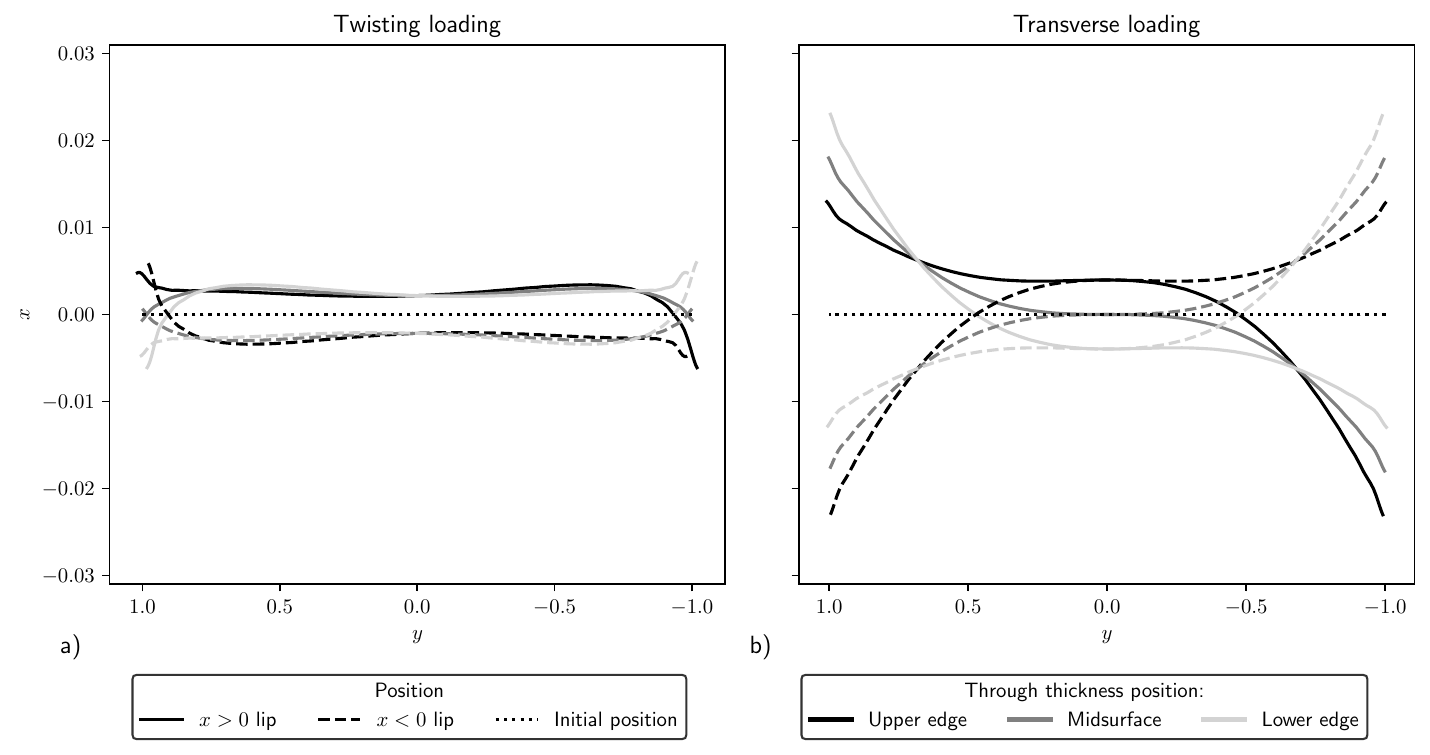}
    \caption{Crack-lip in-plane displacements for a minimal surface ($\kappa_X/\kappa_Y = -1$)
    with a crack embedded at an angle $\beta = 45^\circ$, loaded respectively with a twisting
    moment $\tilde{M}^0_{xy}$ (a) and transverse shear loading $\tilde{V}_x^0$ (b). In both cases, the obtained is not physically compatible, given that the crack lips do not meet at the two crack tips.}
    \label{fig:IncompatibleDisplacements}
\end{figure}

In the presence of curvature, twisting and transverse loadings activate additional coupling mechanisms that 
expose limitations of the present model and solution approach, leading to kinematically incompatible 
displacement fields. 
A detailed discussion of the origin of this incompatibility is provided in 
Appendix~\ref{app:incompatible-displacements}.
Here, we restrict the discussion to the displacement solutions obtained in selected representative 
geometries, with the aim of illustrating the character of the incompatible response.
Figure \ref{fig:IncompatibleDisplacements} shows the deformed in-plane profile for cracks embedded in a minimal surface ($\kappa_X/\kappa_Y = -1$) at an angle $\beta = 45^\circ$. The two crack surfaces are not meeting at the crack tips, creating an incompatible displacement field. The phenomenon is more evident under transverse loading $\tilde{V}_x^0$, shown in \ref{fig:IncompatibleDisplacements}.b.

\section{Discussion} \label{sec:stressCoupling}

In this section we examine in detail the mechanisms by which shell curvature induces the fracture 
mode coupling observed in Section \ref{sec:results}.
We will show in the following that two ingredients are central to this coupling. 
The first is the well-established kinematic interaction between in-plane displacements and 
cross-sectional rotations introduced by curvature, whereby membrane loading generates a bending 
response, and viceversa \citep{Calladine:1983}. 
The second is geometric: when the crack is not aligned with a principal curvature direction, the 
surface normals along the crack no longer remain in the crack plane, breaking the mirror symmetry of 
the crack embedding within the shell.
Together, these effects allow mode I loadings (membrane tension and bending) to activate mode II and 
III responses (tearing, in-plane and twisting sliding), and viceversa.

\begin{figure}
    \centering
    \includegraphics[width=\textwidth]{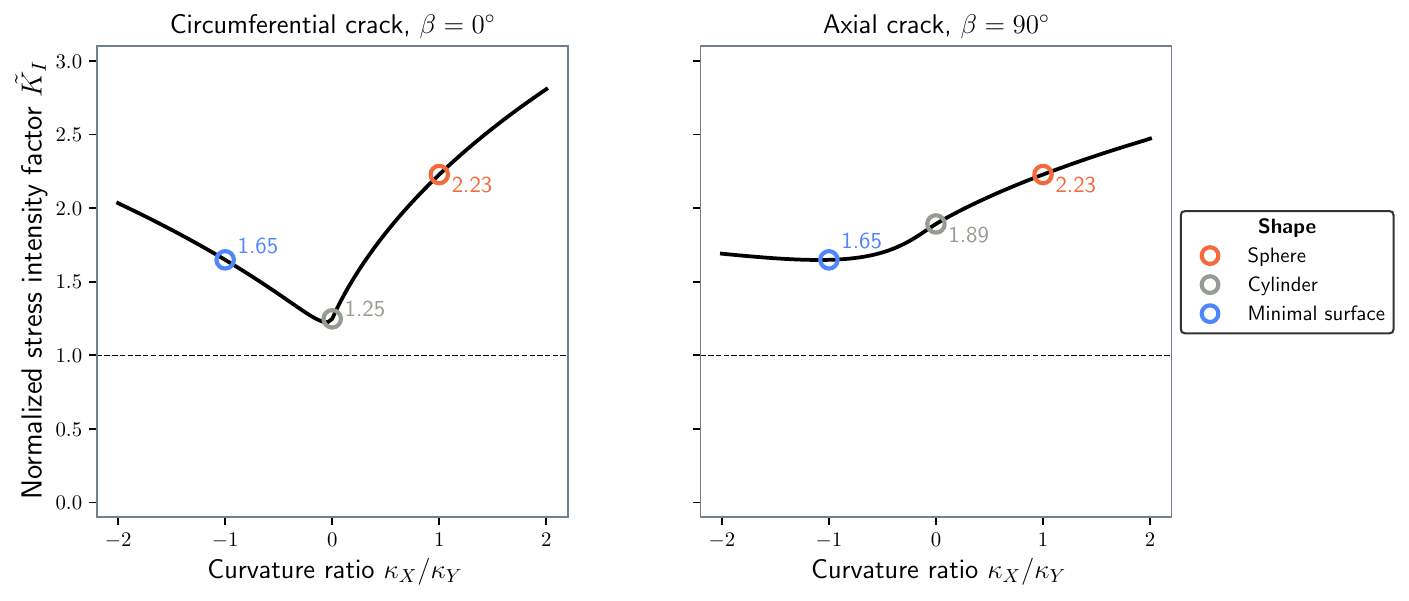}
    \caption{\emph{
    Normalized mode I stress intensity factor $\tilde{K}_I$ at the shell midsurface 
    ($z = 0$) and at the positive $y$ crack tip under membrane tension, as a function of the 
    curvature ratio $\kappa_X/\kappa_Y$ for the two principal crack orientations, 
    $\beta = 0^{\circ}$ and $\beta = 90^{\circ}$.
    The sphere ($\kappa_X/\kappa_Y = 1$), cylinder ($\kappa_X/\kappa_Y = 0$), and minimal surface 
    ($\kappa_X/\kappa_Y = -1$) configurations are highlighted with red, gray, and blue markers, 
    respectively.
    The two crack orientations isolate the effect of varying the curvature normal to the crack 
    ($\beta = 0^{\circ}$) and parallel to the crack ($\beta = 90^{\circ}$), while maintaining the 
    other curvature component constant.
    In both cases, the \sif\ increases monotonically in the ellipsoidal branch ($\kappa_X/\kappa_Y > 0$),
    and has a minimum in the hyperboloidal branch ($\kappa_X/\kappa_Y < 0$) for intermediate values of 
    curvature ratio between the cylinder and the minimal surface.
    The \sif\ is significantly more sensitive to curvature variations normal to the crack than to 
    variations parallel to the crack.}}
    \label{fig:principalSIFS}
\end{figure}

To illustrate the mechanism, it is instructive to begin with membrane-tension loading.
We consider the mode I \sif\ as a function of the 
curvature ratio for principal crack orientations ($\beta = 0^{\circ}$ or $\beta=90^{\circ}$, see 
Figure~\ref{fig:crack-reference}). 
For these crack embeddings, $K_I$ is the only nonzero \sif, as shown previously in 
Figure~\ref{fig:tension-misurface}.
Studying the problem in this setting is particularly insightful, as it allows us to isolate the 
effects of varying the curvature normal to the crack while maintaining the 
curvature parallel to the crack constant ($\beta = 0^{\circ}$), and conversely of varying the 
curvature parallel to the crack while maintaining the curvature normal to the crack constant
($\beta = 90^{\circ}$). 
Figure~\ref{fig:principalSIFS} shows the normalized \sif\ at the shell 
midsurface as a function of the curvature ratio $\kappa_X / \kappa_Y$. 
For both crack orientations, the \sif\ increases monotonically with the curvature ratio in the 
ellipsoidal branch (positive $\kappa_X / \kappa_Y$). 
In the hyperboloidal branch (negative $\kappa_X / \kappa_Y$), the \sif\ first decreases, reaches a 
minimum, and then increases again with increasing magnitude of the curvature ratio, with the 
minimum occurring at different values of $\kappa_X / \kappa_Y$ for the two crack orientations.
However, by comparing the two curves for $\beta = 0^{\circ}$ and $\beta = 90^{\circ}$, we observe 
that the \sif\ is significantly more sensitive to curvature variations normal to the crack 
($\beta = 0^{\circ}$) than to variations parallel to the crack ($\beta = 90^{\circ}$).
In addition, for both crack orientations, the \sif\ is more sensitive to curvature in the 
ellipsoidal branch than in the hyperboloidal branch. 
Finally, we note that for both the minimal surface and the sphere, the \sifs\ corresponding to the 
two principal crack orientations coincide, as required by symmetry.

\begin{figure}
    \centering
    \includegraphics[width=\textwidth]{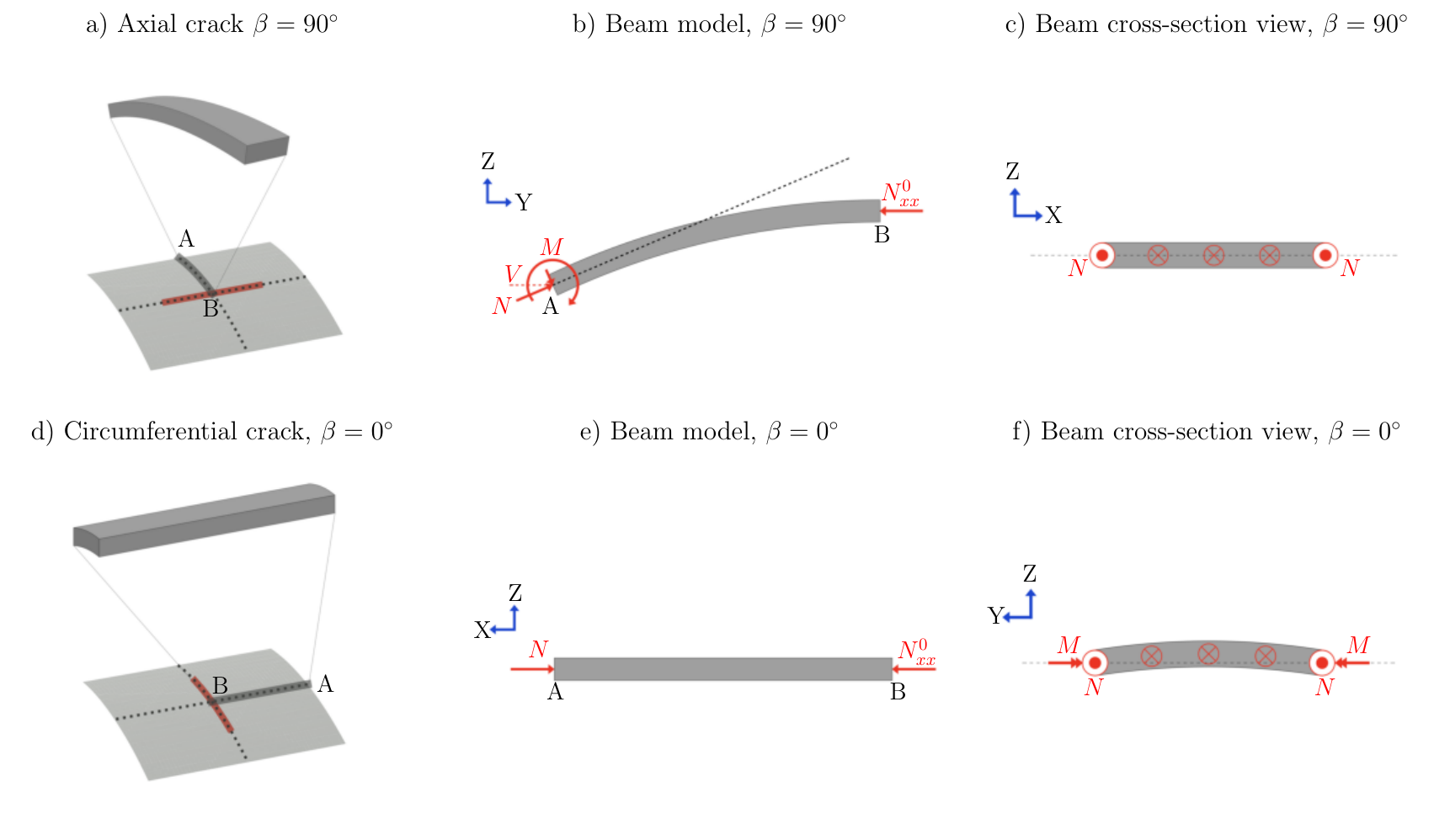}
    \caption{
    \emph{
    Beam analogy used to interpret the mechanics of curvature-induced mode coupling in a cylindrical 
    shell.    
    a), d): Cylindrical shell exhibiting an axial crack ($\beta = 90^\circ$) and a circumferential 
    crack ($\beta = 0^\circ$), respectively. 
    In each case, a section orthogonal to the crack is considered, from which a representative beam 
    segment is identified.
    b), e): Free-body diagrams of the corresponding beam segments.
    c), f): Cross-section geometries of the beam segments.
    For the axial crack, the beam segment is curved in its unstressed configuration and has a 
    symmetric cross-section. 
    In contrast, for the circumferential crack, the beam segment is straight in its unstressed 
    configuration and has an asymmetric cross-section.}}
    \label{fig:beamAnalogy}
\end{figure}

The behavior observed in Figure~\ref{fig:principalSIFS} can be interpreted through the following 
thought experiment.
We consider two \emph{cylindrical} shells, one containing a circumferential crack ($\beta = 0^\circ$) and 
the other an axial crack ($\beta = 90^\circ$). 
We then cut a section of the shell orthogonal to the crack, as illustrated in 
Figure~\ref{fig:beamAnalogy}. 
Within this section, we identify a beam whose one end lies on the crack surface, while the other is 
located sufficiently far from the crack that the local mechanics is unaffected by its presence.
In both cases, the end of the beam lying on the crack surface is subjected to an axial compression 
$N_{xx}^0$ and is free from applied moments, while the opposite end may be regarded as fully clamped, 
consistently with the requirement of the displacement field decaying far away from the crack in the perturbation problem (Equations~\eqref{eqn:regularityConsequence} and ~\eqref{eqn:splitW}).

The mechanics governing the two configurations are, however, fundamentally different.
In the case of the axial crack, the beam is curved in its stress-free configuration and has a 
symmetric cross-section. 
In contrast, in the case of the circumferential crack, the beam is straight in its stress-free 
configuration and has an asymmetric cross-section.

In the first case, the curved beam subjected to free-clamped boundary conditions and loaded in 
compression at the free end tends to straighten, producing a rotation and an out-of-plane 
displacement away from its center of curvature. 
The induced rotations are equal and opposite on the two sides of the crack. 
Near the crack tips, however, these rotations are constrained by compatibility with the surrounding 
shell, which generates localized bending moments.
These moments provide an additional contribution to the crack-tip stress field relative to the flat 
case, thereby contributing to the larger stress intensity factor observed in 
Figure~\ref{fig:principalSIFS} for $\beta = 90^\circ$.

In the second case, the straight beam with asymmetric cross-section subjected to the same loading 
also develops a rotation, because the loading resultant does not pass through the neutral axis of 
the cross-section. 
As in the axial-crack configuration, compatibility with the surrounding shell constrains this 
rotation near the crack tips, generating localized bending moments and an additional contribution to 
the crack-tip stress field.
The magnitude of this contribution is, however, smaller than in the axial-crack configuration.
In fact, in the axial-crack case, the resisting moment scales with a lever arm set by the beam length,
and therefore by the shell size, whereas in the circumferential-crack case it scales with a lever
arm set by the crack length, which is much smaller than the shell size.
This leads to a weaker contribution to the crack-tip stress field and, consequently, to a lower
stress intensity factor for the circumferential crack than for the axial crack.

This mechanistic model explains why the same nominal curvature leads to a larger 
\sif\ when acting normal to the crack ($\beta = 90^\circ$) than when acting parallel to it
($\beta = 0^\circ$). 
While normal curvature acts over a characteristic length set by the shell size, parallel curvature 
acts only over the crack length. 
This analysis of the effects of normal and parallel curvature in isolation provides the basis for 
interpreting the behavior of doubly curved shells observed in Figure~\ref{fig:principalSIFS} as the 
superposition of the contributions associated with curvature normal and parallel to the crack.

We next examine how this interpretation explains the dependence of the mode I \sif\ on the curvature 
ratio $\kappa_X/\kappa_Y$.
For $\beta = 0^\circ$, varying $\kappa_X/\kappa_Y$ modifies the curvature normal 
to the crack while maintaining the curvature parallel to the crack constant and positive, 
since $\kappa_Y = 0.2$ throughout the analysis.
In the ellipsoidal branch, the two curvature contributions cooperate, leading to a monotonic 
increase of the \sif\ with the curvature ratio. 
In the hyperboloidal branch, the negative curvature normal to the crack counteracts the positive 
curvature parallel to the crack. 
Since the curvature normal to the crack has the stronger influence on the \sif, the competition 
between the two effects rapidly produces a minimum for a very small, negative curvature ratio. 
For more negative curvature ratios, the influence of the negative normal curvature dominates, 
although the slope remains weaker than in the ellipsoidal branch because the positive parallel 
curvature continues to oppose the response.

Conversely, for $\beta = 90^\circ$, varying the curvature ratio $\kappa_X/\kappa_Y$ modifies the 
curvature parallel to the crack while maintaining the curvature normal to the crack constant and 
positive.
As in the previous case, the two curvature contributions act cooperatively in the ellipsoidal 
branch, producing a monotonic increase of the \sif. 
However, because the curvature parallel to the crack has a weaker influence on the fracture 
response, the growth of the \sif\ with the curvature ratio is less pronounced than for 
$\beta = 0^\circ$. 
In the hyperboloidal branch, the negative curvature parallel to the crack competes with the positive 
curvature normal to the crack. 
Since the contribution of the parallel curvature is weaker, more negative curvature ratios are 
required before the competing effects balance, resulting in the minimum being shifted toward more 
negative values of $\kappa_X/\kappa_Y$. 
For even more negative curvature ratios, the influence of the negative parallel curvature 
gradually becomes dominant, although the slope remains less steep than in the ellipsoidal branch 
because the positive normal curvature continues to oppose the response.

The analysis above focused on principal cracks, for which the crack is aligned with one of the 
principal curvature directions. 
This setting allowed us to isolate the individual roles of curvature normal and parallel to the crack, 
and to interpret their combined effect in doubly curved shells. 
As shown in Figures~\ref{fig:tension-misurface} and \ref{fig:KandERRTension}, principal cracks 
($\beta = 0^\circ$ or $\beta = 90^\circ$) exhibit only the opening mode under membrane tension. 
We now move to the more general case of non-principal cracks, corresponding to intermediate values 
of $\beta$ between $0^\circ$ and $90^\circ$. 
In this case, under the same membrane tension loading, sliding and tearing modes appear alongside 
the opening mode in cylinders and hyperboloids, whereas they remain 
negligible in ellipsoids.

To understand why this mode coupling appears in some curvature profiles but not in others, we first
examine the curvature components experienced by the crack itself.
The prototypical shells are described in terms of the principal curvatures $\kappa_X$ and $\kappa_Y$, 
but the effect of this curvature on the cracked shell mechanics depends on the extent it acts normal 
to the crack, parallel to the crack, or in a mixed manner.
For this reason, it is useful to express the shell curvature in the crack-local frame $x$ and $y$, 
see Figure~\ref{fig:crack-reference}.
In this frame, the curvature tensor is obtained by rotating the principal curvature tensor by the 
crack inclination angle $\beta$, leading to
\begin{equation}
    \label{eqn:curvatureCrackReferenceMain}
    \begin{aligned}
       \bm{\kappa}(\beta) = 
       \begin{bmatrix}
            \kappa_x & \kappa_{xy}\\
            \kappa_{xy} & \kappa_y
        \end{bmatrix} = \begin{bmatrix} 
            \kappa_Y \sin^2 \beta + \kappa_X \cos^2 \beta & \left(\kappa_X - \kappa_Y\right) \sin \beta \cos \beta \\ 
            \left(\kappa_X - \kappa_Y\right) \sin \beta \cos \beta  & \kappa_Y \cos^2 \beta + \kappa_X \sin^2 \beta
        \end{bmatrix}, 
    \end{aligned}
\end{equation}
where $\kappa_x$ and $\kappa_y$ are the curvature components normal and parallel to the crack, 
respectively, and $\kappa_{xy}$ is the mixed curvature component in the crack-local frame.
For principal cracks, the crack-local axes coincide with the principal curvature directions, so 
$\kappa_{xy}=0$ and the curvature tensor remains diagonal. 
For non-principal cracks, instead, the rotation of the curvature tensor generally gives rise to a 
nonzero mixed curvature component. 
Geometrically, this term measures the extent to which the surface normal rotates about the crack 
tangent as one moves along the crack, and is therefore often interpreted as a torsional curvature
\citep{Novozhilov:1964}.

The torsional curvature component has two direct geometric consequences for the symmetry of the crack 
embedding. 
First, when $\kappa_{xy} \neq 0$ the crack plane is no longer a symmetry plane for the shell. 
In fact, in this case, the torsional curvature introduces a contribution to the local surface 
geometry that is antisymmetric with respect to the crack plane (Equation~\eqref{eqn:Paraboloid}).
Second, when $\kappa_{xy} \neq 0$ the normals to the shell along the crack twist out of the crack 
plane antisymmetrically with respect to the $xz$ plane, that is by equal angles in opposite 
directions for $y>0$ and $y<0$, respectively.

These losses of symmetry have direct implications for the fracture mechanics response. 
In fact, in a curved shell, membrane tension also induces out-of-plane displacements of the crack 
surfaces due to the coupling between membrane deformation and bending. 
For principal cracks, where $\kappa_{xy}=0$, these out-of-plane displacements remain compatible on 
the two sides of the crack, owing to the symmetry of both the in-plane response and the surface 
geometry \citep{Delale:1979a}.
Therefore membrane tension activates only the mode I, analogously to the classical center-cracked flat plate under mode I loading, where the 
symmetry of both the geometry and the loading results in in-plane displacement fields specular to 
the crack plane \citep{Westergaard:1939,Sih:1977} and in trivial compatibility at the crack tips.
For non-principal cracks, instead, the torsional curvature component breaks the mirror symmetry of the 
crack embedding within the shell. 
The curvature-induced out-of-plane displacements are then no longer equal on the two crack surfaces
and compatibility at the crack tips may engage antisymmetric modes, such as sliding (mode II) and tearing (mode III).

The different deformation regimes observed in Figure~\ref{fig:displacements-tension} can be 
interpreted in terms of the relative magnitudes of the crack-local curvature components. 
We first consider the sphere and ellipsoid. 
In the case of the sphere ($\kappa_X/\kappa_Y=1$), the curvature tensor is isotropic and remains 
diagonal in any crack-local frame. 
Consequently, $\kappa_{xy}=0$ for any crack orientation, the crack embedding remains symmetric, 
and only the mode I is activated. 
The ellipsoid ($\kappa_X/\kappa_Y=0.833$) represents a nearby case in which the principal curvatures 
have the same sign. 
Although an inclined crack may introduce a nonzero torsional curvature component, this contribution 
may become large only when one principal curvature is significantly larger than the other.
The crack-lip deformation therefore remains nearly symmetric, with compatible out-of-plane 
displacements on the two sides of the crack. 
This is consistent with Figures~\ref{fig:tension-misurface} and~\ref{fig:KandERRTension}, which show
a response governed by the mode I.

The hyperboloid and the minimal surface exhibit the opposite regime. 
Because the principal curvatures have opposite signs, rotating the curvature tensor into the 
crack-local frame produces a torsional curvature component that is larger than the normal and 
parallel curvatures. 
This causes a weaker opening and a stronger mode III and twisting mode II response, when compared 
with the ellipsoid with same principal curvature magnitudes.

The smaller values of $\kappa_x$ and $\kappa_y$ lead to smaller out-of-plane displacements than for 
principally embedded cracks. 
However, due to the antisymmetric component in the shell geometry introduced by $\kappa_{xy}$,
compatibility on the crack plane can only be maintained if the shell twists
antisymmetrically on both sides of the crack plane, producing the profiles observed in 
Figure~\ref{fig:displacements-tension}.n and Figure~\ref{fig:displacements-tension}.o.
The in-plane displacements still require a specular component about the crack plane because of the
symmetric loading. In addition to that symmetric component, the out-of-plane twisting of the shell 
induces an antisymmetric in-plane deformation of the crack lips, giving rise to the
response shown in Figure~\ref{fig:displacements-tension}.i and Figure~\ref{fig:displacements-tension}.j. 

The out-of-plane antisymmetric deformation reflects the mode III tearing response, while the 
in-plane antisymmetric deformation reflects the mode II twisting sliding response. Because the two 
deformation always occur simultaneously in the presence of $\kappa_{xy}$, the two fracture modes II and 
III must also occur simultaneously. 
This observation is also consistent with Hui and Zehnder \citep{Hui:1993}, who first demonstrated, 
by energetic arguments and under the assumption of separation of mode I and mode II, III, that 
twisting induced sliding and tearing responses always occur simultaneously in thin, Reissner plates.
Our mechanistic interpretation extends their observation to the more general case 
of arbitrarily aligned cracks in doubly-curved shells.

Since the coupling between membrane loading and bending mode I response is governed by the normal and 
parallel components of curvature in the crack-local frame, it is instructive to consider cases in 
which these components vanish. 
Equation~\eqref{eqn:curvatureCrackReferenceMain} shows that this can only happen when the principal 
curvatures have opposite signs. 
In this case, there exist critical crack orientations for which the normal and/or the parallel 
curvature component vanish. 
Specifically, one can define two critical angles
$\beta^{crit}_x = \arctan{\left(\sqrt{- \frac{\kappa_X}{\kappa_Y}}\right)}$ and 
$\beta^{crit}_y = \arctan{\left(\sqrt{- \frac{\kappa_Y}{\kappa_X}}\right)}$
at which the normal and parallel curvature components vanish, respectively.
This explains the reduction of the bending mode I response for inclined cracks in the hyperbolic 
shells shown in Figure~\ref{fig:KandERRTension}. 
In the limiting case of a minimal surface, where the two principal curvatures are equal in magnitude 
and opposite in sign, both components vanish simultaneously at $\beta=45^\circ$, leading to
vanishing bending mode I response.
In this case, the shell exhibits the exact out-of-plane twisting profile expected in flat plates 
subjected to twisting loads.

The cylinder represents the case with richest mixed-mode response among the geometries considered. 
Unlike the ellipsoid, where the normal and parallel curvature components dominate, and unlike the 
hyperboloid, where the torsional curvature dominates, in cylinders, normal, parallel and torsional 
curvatures have comparable magnitudes.
The symmetric opening promoted by the loading generates both the out-of-plane displacement typical 
of ellipsoids and the twisting response typical of hyperboloids. 
The interplay of these two contributions result in in-plane and twisting mode II, mode III, as well as 
bending mode I, see Figure~\ref{fig:tension-misurface}.

The governing mechanics remain the same for other loading types.
We now consider the response under in-plane shear loading $\tilde{N}^0_{xy}$.
Normal and parallel crack curvature couple in-plane displacements with out-of-plane displacements
and cross-section rotations; and the misalignment of the surface normals couples the opening response
with the sliding and tearing ones. 
The coupling-induced responses, however, differ substantially between geometries.

The sphere and ellipsoid exhibit very limited in-plane opening displacement, consistently with
their small torsional and large normal and parallel curvatures. 
They also display a perfectly antisymmetric out-of-plane profile, indicating a tearing
contribution, analogously to what we observed for membrane tension.
This deformation can be explained by the interplay between the membrane-bending kinematic coupling
and the twisting mode II and mode III coupling observed by \citep{Hui:1993}.
The large parallel curvature $\kappa_y$ induces a cross-section rotation, which manifests as the
large twisting mode II response shown in Figure~\ref{fig:KandERRShear}. 
Due to the need to maintain compatibility at the crack tips, the twisting mode II corresponds to a 
tearing motion of the crack tips, giving rise to the mode III response.
As the normal curvature $\kappa_x$ is comparable to $\kappa_y$ in magnitude, the neutral axis of
the shell with respect to rotations normal to the crack plane is shifted towards the lower edge.
Therefore, material points closer to the upper surface of the shell undergo larger displacements
than those closer to the bottom surface. 
The only way for the shell to maintain compatibility is for the crack tips to rotate in opposite
directions at the upper surface of the shell, producing the distinctive S-shape shown in 
Figure~\ref{fig:DisplacementsShear}, while the crack remains essentially straight at the lower
surface.
Additionally, modes II and III being the largest (see Figure~\ref{fig:KandERRShear}) for cracks
embedded at an angle $\beta = 0^\circ$ is consistent with the curvature aligned with the loading
direction having the largest influence on the increase of the stress concentration compared to the
flat plate case.
Indeed, the in-plane shear loading is applied parallel to the crack, and when $\beta = 0^\circ$, 
the parallel curvature is maximum for the geometries considered. 
This is analogous to the mode I response being the largest for cracks embedded at $\beta = 90^\circ$
when the shells are subjected to membrane tension.

The hyperboloid and minimal surface responses are completely different. Both sides of the crack
deform in the same out-of-plane direction, while the crack surfaces rotate about the shell 
midsurface in opposite directions, analogously to what is observed in flat plates subjected to
bending \citep{Jones:1975} (see Figure~\ref{fig:DisplacementsShear}).
As a result, an opening displacement is observed at the lower shell surface, whereas 
interpenetration occurs at the upper surface.
Although this response may appear counter-intuitive at first, it is consistent with the governing
mechanics we have unravelled. 
For principal crack embeddings, the in-plane and twisting mode II and mode III responses are 
non-zero in Figure~\ref{fig:KandERRShear}.
The mechanic principles involved are similar to those of the ellipsoids.
However, the opposite and competing shell principal curvatures reduce the magnitude of all three
responses. 
Moreover, the signs of both the twisting mode II and mode III responses are opposite between
$\beta = 0^\circ$ and $\beta = 90^\circ$, as the principal curvatures are equal and opposite,
causing the normal and parallel curvatures to switch sign between the two cases.
As we consider non-principal crack embeddings, the normal and parallel curvatures quickly become
negligible compared to the torsional term $\kappa_{xy}$ (see Equation~\ref{eqn:curvatureCrackReferenceMain}). 
Consequently, for the geometries shown in Figure~\ref{fig:DisplacementsShear}, when 
$\beta = 45^{\circ}$, the crack is effectively flat along its length and experiences nearly
antisymmetric shells on its two sides. 
The induced twisting mode II response therefore disappears, as does the mode III response.
The antisymmetric shear loading, together with the lack of a non-interpenetration constraint, then
causes the two antisymmetric halves of the shell to deform nearly symmetrically with respect to
the crack plane. 
The in-plane S-shape of the crack thus arises from the superposition of primary in-plane mode II
and coupled bending displacements.
In the limiting case of the minimal surface, for which the two halves are exactly antisymmetric,
the out-of-plane displacement field and the bending rotation become exactly symmetric, as shown in
the right-most plots of Figure~\ref{fig:DisplacementsShear}. 

Finally, cylinders display the most complex behavior, analogously to the membrane tension loading
case, due to the three curvature components $\kappa_x$, $\kappa_y$, and $\kappa_{xy}$ having
comparable magnitudes. 
Their dependence on the crack inclination angle $\beta$ is also the strongest.
For small $\beta$, the mode I response is purely in-plane, while both in-plane and twisting mode
II, and mode III responses are maximized. 
As the crack embedding angle increases, the parallel curvature $\kappa_y$ decreases while
remaining dominant, and the normal and torsional components progressively increase. 
Consequently, part of the response shifts towards membrane mode I, while the twisting mode II and
mode III responses decrease in magnitude.
As $\beta$ continues to increase, the twisting mode II and mode III responses become weaker, and a
bending mode I component appears for $\beta \approx 30^{\circ}$. 
This bending component reaches its maximum around $\beta \approx 60^{\circ}$, where the normal
curvature becomes dominant, while the parallel and torsional curvatures remain non-negligible.
This behavior is consistent with the ``curved beam'' effect discussed above.
Then, as we approach the axial crack case ($\beta = 90^{\circ}$), the membrane and bending mode I
components vanish, recovering the expected behavior for a principally embedded crack.

We can infer that the interpenetration observed in our results would result in crack closure,
thereby reducing the stress intensity factors and delaying the onset of crack propagation. 
In hyperboloids, the closure is induced by a bending mode I response.
Therefore, the mere presence of shear stress around the crack promotes closure at one of the shell
edges. 
It is therefore reasonable to expect that in-plane shear loading may have a beneficial effect in
suppressing failure of hyperboloidal shells.
This conclusion remains qualitative, however, since contact is not accounted for in the present analyses. 
For cylinders, the expected behavior is more complicated to infer. The closure is indeed created by a membrane mode I response. Thus, it depends on the combination of crack embedding and loading sign. When $\beta < 0$ and $\tilde{N}^0_{xy} > 0$ (or vice versa), our analyses indicate that the crack closes, whereas when $\beta < 0$ and $\tilde{N}^0_{xy} < 0$ (or vice versa), they indicate that the crack opens. Thus, it is possible that a given loading sign may promote or hinder crack growth in specific directions, but further analyses are needed.

When subjecting the shell to a bending load, the response is heavily dominated by the primary mode
I across all geometries (see Figure~\ref{fig:KandERRbending}).
The sliding response is strictly in-plane for cylinders and hyperboloids, while it is completely
absent in ellipsoids, as is the tearing response in all geometries.

The ellipsoid exhibits the weakest dependence on the crack embedding angle $\beta$. 
Ellipsoids have the largest normal and parallel curvatures, $\kappa_x$ and $\kappa_y$, and the
smallest $\kappa_{xy}$, and thus show the largest membrane-opening response 
(see Figure~\ref{fig:KandERRbending}). 
This is expected as a consequence of the known kinematic coupling between bending and membrane
responses in curved shells.
The normal and parallel curvatures having the same sign also shift the neutral axis of the
deformed crack towards the bottom shell edge, as observed in Figure~\ref{fig:DisplacementsBending}.a,b.

On the contrary, the hyperboloid has opposite principal curvatures, resulting in small normal and
parallel curvatures $\kappa_x$ and $\kappa_y$ when the cracks are inclined. 
Thus, its response lacks the membrane mode I contribution.
The whole load is therefore concentrated in generating a slightly larger bending mode I response
compared to the ellipsoids (see Figure~\ref{fig:KandERRbending}). 
We observe a small increase for inclined cracks.
While at first counterintuitive, we explain this by the lack of opposing normal and parallel
curvatures.
Indeed, inclined cracks experience a virtually flat crack along their length whereas principally
embedded cracks are affected by competing ``curved beam'' and ``asymmetric cross-section'' effects.
This observation is also corroborated by the lack of interpenetration at the midsurface of the
shell in Figure~\ref{fig:DisplacementsBending}.
The large torsional curvature $\kappa_{xy}$ also leads to the largest mode II contribution for
this loading type, as shown in Figure~\ref{fig:KandERRbending}.
As the crack is embedded with increasing $\beta$, the normals at its tip become increasingly
skewed with respect to the crack plane. 
This induces a sliding displacement parallel to the crack and antisymmetric with respect to the
crack plane, giving rise to the S-shape visible in Figure~\ref{fig:DisplacementsBending}.d,e.
This phenomenon produces a deformed profile analogous to the superposition of the response of a 
center-cracked, infinite flat plate in bending with the S-shape observed in 
Figure~\ref{fig:DisplacementsShear} for the in-plane shear loading.

Lastly, the cylinder once again displays an intermediate response between the ellipsoid and the hyperboloid. 
The stress response in Figure~\ref{fig:KandERRbending} shows the strongest dependence on the
embedding angle $\beta$. 
Circumferential cracks ($\beta = 0^\circ$) exhibit weaker bending and membrane responses. 
As $\beta$ increases, both responses become larger, reaching a maximum for axial cracks 
($\beta = 90^\circ$). 
This behavior is consistent with the observation that the curvature parallel to the load has a
larger effect on the stress concentration.
The sliding response is also present, but it is smaller than that of the hyperboloid.
It reaches a maximum for $\beta \approx 70^{\circ}$. 
At this angle, the normal and torsional curvatures $\kappa_x$ and $\kappa_{xy}$ become large
compared to the parallel curvature $\kappa_y$, leading to both strong membrane-bending and mode I
to mode II,III coupling.
This behavior is exactly specular to that observed for the opening response under in-plane shear
loading in Figure~\ref{fig:KandERRShear}, which also peaks for $\beta \approx 70^{\circ}$.
The intermediate nature of the cylinder response is also reflected in its in-plane deformation,
plotted in Figure~\ref{fig:DisplacementsBending}.c.
The neutral axis is shifted towards the lower surface, and both the elliptical opening
characteristic of a center-cracked flat plate under membrane tension and the antisymmetric S-shape
remain clearly recognizable.

Finally, we observe that the thin shells considered are significantly more compliant out-of-plane than in-plane. 
Note, in fact, that the displacement magnitudes in Figure~\ref{fig:DisplacementsBending} are larger than
those in Figures~\ref{fig:displacements-tension},\ref{fig:DisplacementsShear}, despite the application of the same stress magnitude in all cases. 

Our results qualitatively match the (limited) available literature on the effect of bending
loading on the fracture behavior of curved shells \citep{Liu:1999}.
The analysis of \citet{Liu:1999} for principally embedded cracks in ellipsoids and cylinders shows
that, the shell stress concentration is higher when the bending moment induces opening at the 
surface farther from the shell center of curvature than for the opposite bending moment, but
remains lower than in the case without a non-interpenetration constraint.
Although in our analyses we do not consider contact, resulting in the magnitude of the stress concentration not being affected by the sign of the applied moment, our results capture the difference in behavior at the two shell outer surfaces and the shift in the neutral axis due to curvature, and are therefore qualitatively consistent with Liu's findings.
Nonetheless, all available semi-analytical closure models assume a purely normal pressure 
distribution arising from crack closure. 
The presence of non-zero in-plane sliding in inclined cracks invalidates this assumption,
indicating that these models might not be sufficient for cylinders and hyperboloids.

\section{Conclusions}

In this study, we investigated the role of curvature in the fracture mechanics of thin,
shallow shells subjected to general loading conditions, with cracks embedded at  arbitrary orientations relative to the principal curvature directions.
We employed a semi-analytical methodology to compute stress intensity factors, energy
release rates, and crack-lip displacements for a broad range of geometries, including
spheres, ellipsoids, cylinders, hyperboloids, and minimal surfaces.

We showed that arbitrarily oriented cracks in curved shells can exhibit multiple fracture modes under a single loading component, in contrast to what observed in flat plates. 
We found that the secondary, or coupled, fracture modes are always weaker than the primary, or applied, fracture mode, which dominates the energy release rate. The strength of the coupled modes depends on both the shell curvature profile and on the crack embedding angle. The relative magnitudes of the curvature tensor components expressed in the crack reference frame are key to understanding the coupling between the symmetric mode I and the antisymmetric modes II and III. In ellipsoids, the normal and parallel curvatures dominate over the torsional curvature, resulting in weak coupling. In hyperboloids, the torsional curvature dominates, producing the strongest coupling observed.  In cylinders, all three curvature components are comparable, resulting in the most complex responses. Furthermore, we showed that, for arbitrarily oriented cracks subjected to twisting and transverse loading, curvature-induced coupling limits the applicability of shallow shell assumptions. Additionally, shear loading was found to induce mode I closure in hyperboloidal shells, suggesting a potential geometric pathway for enhancing damage resistance.

Overall, we extended the class of curved-shell fracture problems accessible to semi-analytical analysis to encompass geometries with positive and negative Gaussian curvature (alongside cylinders).
These results provide a foundation for understanding the fracture behavior of emerging architected materials with engineered curvature distributions, such as lattices based on triply periodic minimal surfaces and spinodoids.

The current framework is restricted to the shallow-shell regime and does not account for contact during crack closure. Future work will extend the framework to larger curvatures by relaxing the shallow-shell assumptions, enabling a more accurate analysis of twisting and transverse loading. Moreover, incorporating a contact formulation will enable explicit treatment of crack closure, improving the accuracy of predictions under loading conditions that lack sufficient membrane tension to maintain crack opening.

\section*{Acknowledgements}
This material is based upon work supported by the \emph{Army Research Laboratory} and the 
\emph{Army Research Office} under contract/grant number W911NF2310332.

R.B. thanks Dr. Santiago Garcia Espallargas for the several insightful discussions on the results 
and their implications.

R.B. thanks Dr. Claude Leibovici for his help in proving the boundedness of one of the 
Fredholm kernels. His derivation is given in Appendix~\ref{sssec:appSing}.

R.B. thanks Ne\v{z}a Jenko for her help with drafting and finalizing the illustrations in this work.

\section*{AI disclosure}

All mathematical content in this article was developed and written entirely by the authors, \textbf{without} the use of generative AI or AI-assisted technologies.  AI was used solely for minor grammatical and language corrections.

\appendix

\renewcommand{\theequation}{\thesection.\arabic{equation}}
\setcounter{equation}{0}
\renewcommand{\thetable}{\thesection.\arabic{table}}
\setcounter{table}{0}

\section{Appendix A: detailed derivations} \label{appendixA}

\subsection{From the equilibrium equations to the normalized field equations}
\label{app:EquilibriumToField}

In this appendix, we present the derivation of the shell field equations starting from the equilibrium equations.
The equilibrium equations in resultant form of the shallow, Reissner shell \citep{Naghdi:1957} in the principal
reference system $OXYZ$ are:

\begin{equation}
    \label{eqn:equilibriumA1}
    \begin{aligned}
        \partial_j& N_{ij}(X,Y) = 0\\
        \partial_i& V_{i}(X,Y) + \partial_j \left(\partial_i z N_{ij}(X,Y)\right) + q \left(X,Y\right)= 0\\
        \partial_j& M_{ij}(X,Y) - V_i(X,Y) = 0 
    \end{aligned}
\end{equation}

We introduce the Airy stress function, defined in terms of the membrane stress resultants, as:

\begin{equation}
    \label{eqn:AiryStressFunc}
    \begin{aligned}
        N_{ij}(x,y) &= \epsilon_{ik}\epsilon_{jl}\partial_{kl}F(x,y) 
    \end{aligned}
\end{equation}

Where the indices $i,j,k,l$ vary over $x,y$, and $\epsilon_{ij}$ denotes the two-dimensional permutation tensor:

\begin{equation}
    \label{eqn:PermutationTensor}
    \begin{aligned}
        \bm{\epsilon} &= 
        \begin{bmatrix}
            0 & 1 \\ -1 & 0
        \end{bmatrix}
    \end{aligned}
\end{equation}

We substitute the constitutive equations in stress resultant form (Equations~\eqref{eqn:ConstitutiveEquations}) and the
Airy stress function in the equilibrium equations (Equations~\eqref{eqn:equilibriumA1}) to obtain the field equations
for the shell in the principal reference system:

\begin{subequations}
    \label{eqn:fieldEqA1}
    \begin{align}
        \nabla^4 F &+ hE \left( \partial_{XX} z \partial_{YY} w_0 + \partial_{YY} z \partial_{XX} w_0 - 2 \partial_{XY} z \partial_{XY} w_0 \right) = 0 \label{eqn:ArbCurvInp} \\
        D \nabla^4 w_0& - \left(1 - \frac{D}{hB} \nabla^2 \right)\left( \partial_{XX} z \partial_{YY} + \partial_{YY} z \partial_{XX} - 2 \partial_{XY} z \partial_{XY} \right) F = \left(1 - \frac{D}{hB} \nabla^2 \right) q \label{eqn:ArbCurvOutp} \\
        \beta_X &+ \partial_X w_0 = \frac{D}{hB} \left(\nabla^2 \beta_X + \frac{1 + \nu}{2} \partial_Y \left(\partial_X \beta_Y - \partial_Y \beta_X \right)\right) \label{eqn:ArbCurvRot1} \\
        \beta_Y &+ \partial_Y w_0 = \frac{D}{hB} \left(\nabla^2 \beta_Y + \frac{1 + \nu}{2} \partial_X \left(\partial_Y \beta_X - \partial_X \beta_Y \right)\right) \label{eqn:ArbCurvRot2}
    \end{align}
\end{subequations}

For brevity, we omitted the arguments $(X,Y)$.
Here, $w_0$ denotes the midsurface out-of-plane displacement, $\beta_X,\beta_Y$ the cross-section rotations, 
$h$ the shell thickness, $E$ the Young's modulus, $\nu$ the Poisson's ratio, and:

\begin{equation}
    \label{eqn:ShellStiffnesses}
    \begin{aligned}
        D & = \frac{Eh^3}{12(1-\nu^2)} \\
        B & = \frac{5}{6} \frac{E}{2(1+\nu)}
    \end{aligned}
\end{equation}

Under the shallowness assumption, the shell midsurface $z(X,Y)$ is locally approximated by a paraboloid with 
curvature radii $R_X,R_Y$ \citep{Yahsi:1983}, so that $\partial_{ij} z(X,Y) \approx R_{ij}$.
We now transform the shell field Equations~\eqref{eqn:fieldEqA1} into the crack reference system $Oxyz$. 
We normalize all lengths in the problem by the crack halflength $a$:

\begin{equation}
    \label{eqn:NormalizedDisplacements}
    \begin{aligned}
        u& = \frac{u_0}{a}, \; \; v = \frac{v_0}{a}, \; \; w = \frac{w_0}{a} \;\; \beta_x = \beta_X \;\; \beta_y = \beta_Y
    \end{aligned}
\end{equation}

The field equations are then normalized by the Young's modulus $E$, introducing the dimensionless quantities:

\begin{equation}
    \label{eqn:NormalizedMaterialParams}
    \begin{aligned}
        \phi(x,y) &= \frac{F(x,y)}{h E a^2} \\
        \kappa &= \frac{E h^2}{12 B \left( 1 - \nu^2 \right) a^2} 
    \end{aligned}
\end{equation}

We further introduce the four fundamental dimensionless geometrical quantities 
$\lambda^2,\gamma_x,\gamma_y,\gamma_{xy}$:

\begin{equation}
    \label{eqn:normalizedGeometricParams}
    \begin{aligned}
        \lambda^2 &= \sqrt{12 \left( 1 - \nu^2 \right)} \frac{a}{h} \\
        \gamma_x &= \sqrt{12 \left( 1 - \nu^2 \right)} \frac{a^2}{h R_x} \\
        \gamma_y &= \sqrt{12 \left( 1 - \nu^2 \right)} \frac{a^2}{ h R_y} \\
        \gamma_{xy} &= \sqrt{12 \left( 1 - \nu^2 \right)} \frac{a^2}{h R_{xy}}
    \end{aligned}
\end{equation}

Finally, we introduce the auxiliary functions $\Omega(x,y), \psi(x,y)$:

\begin{equation}
    \label{eqn:normalizedAuxFuncs}
    \begin{aligned}
        \Omega \left(x,y \right) & = \partial_y \beta_x - \partial_x \beta_y  \\
        \psi \left(x,y \right) & = \kappa \left(\partial_x \beta_x + \partial_y \beta_y\right) - w 
    \end{aligned}
\end{equation}

Which can be inverted to obtain:

\begin{equation}
    \label{eqn:BetasAuxFuncs}
    \begin{aligned}
        \beta_x\left(x,y \right) & = \partial_x \psi + \frac{1 - \nu}{2} \kappa \partial_y \Omega \\
        \beta_y\left(x,y \right) & = \partial_y \psi - \frac{1 - \nu}{2} \kappa \partial_x \Omega
    \end{aligned}
\end{equation}

Without losing generality, we set the out-of-plane pressure load $q(X,Y)$ to vanish. 
Owing to the linearity of the problem, its effect can be recovered through the use of the superposition principle,
analogously to the far-field loading discussed in Section\ref{ssec:superposition} \citep{Sih:1977}. 
Substituting Eqs.~\ref{eqn:NormalizedMaterialParams},~\ref{eqn:normalizedGeometricParams},~\ref{eqn:BetasAuxFuncs} in
the transformed shell field equations yields the dimensionless governing equations:

\begin{subequations}
    \label{eqn:FinalEqA1}
    \begin{align}
        \nabla^4 \phi &+ \frac{1}{\lambda^2} \left( \gamma_x \partial_{yy} w + \gamma_y \partial_{xx} w - 2 \gamma_{xy} \partial_{xy} w \right) = 0 \label{eqn:FinalInp} \\
        \nabla^4 w& - \lambda^2 \left(1 - \kappa \nabla^2 \right)\left( \gamma_x \partial_{yy}  + \gamma_y \partial_{xx}  - 2 \gamma_{xy} \partial_{xy}  \right) \phi = 0 \label{eqn:FinalOutp} \\
        \kappa &\nabla^2 \psi - \psi - w = 0 \label{eqn:FinalRot1} \\
        \kappa &\frac{1 - \nu}{2} \nabla^2 \Omega - \Omega = 0 \label{eqn:FinalRot2}
    \end{align}
\end{subequations}

\subsection{Solution of the characteristic polynomial}
\label{app:CharacteristicPolynomials}

In this appendix, we discuss the properties of the roots of the characteristic polynomial. 
We repeat the polynomial here:

\begin{equation}
    \label{eqn:CharEqYahsiA2}
    \begin{aligned}
        p(m) = m^8 &- \left(4 \alpha^2 + \kappa \gamma_y^2\right) m^6 - 4 i  \kappa \gamma_{xy} \gamma_y \alpha m^5 + \\
        &+ \left(6 \alpha^4 + \gamma_y^2 + \kappa \alpha^2 \left(\gamma_y^2 + 4 \gamma_{xy}^2 + 2 \gamma_x \gamma_y \right) \right) m^4 + \\
        &- 4 i \alpha \gamma_{xy} \left( \gamma_y + \kappa \alpha^2 \left( \gamma_x + \gamma_y \right)\right) m^3 + \\
        &- \alpha^2 \left(4 \alpha^4 + 2 \left( 2 \gamma_{xy}^2 + \gamma_x \gamma_y\right) + \kappa \alpha^2 \left( \gamma_x^2 + 4 \gamma_{xy}^2 + 2 \gamma_x \gamma_y \right)\right) m^2 + \\
        &-  4 i \alpha^3 \gamma_x \gamma_{xy} \left(\kappa \alpha^2 + 1 \right) m + \alpha^4 \left(\alpha^4 + \gamma_x^2 \left(1 + \kappa \alpha^2\right)\right) = 0
    \end{aligned}
\end{equation}

This is an eighth order polynomial with complex coefficients. 
Consequently, no closed form solutions are available. 
We therefore compute the roots $m_j(\alpha)$ numerically ($j=1,\dots,8$). 
However, we can infer some of their properties \citep{marden:1949}:

\begin{itemize}
    \item the roots are, in general, complex-valued and depend on $\alpha$
    \item the coefficients are smooth functions of $\alpha$, and the roots depend smoothly on the polynomial coefficients. 
    Thus, the roots will be smooth functions of $\alpha$ \citep{Harris:1987}
\end{itemize}

Moreover, introducing the change of variable $m=is$, transforms the characteristic equation into a polynomial 
$p(s) = 0$ with real coefficients. 
Its roots $s$ therefore occur in complex conjugate pairs.
Consequently, the roots $m_j(\alpha)$, after a suitable ordering, satisfy:

\begin{equation}
    \label{eqn:OrderedRoots}
    \begin{aligned}
        \Re \left(m_{j}(\alpha)\right) = - \Re \left(m_{j+4}(\alpha)\right) < 0, \; j = 1, \dots,4, \; \forall \alpha
    \end{aligned}
\end{equation}

\subsubsection{Asymptotic expansion}

Since the roots are obtained numerically, an asymptotic expression for $m_j(\alpha)$ as $|\alpha| \to \infty$ is
required to carry out the asymptotic analysis leading to the singular integral equations' system.
To determine the leading-order behavior of the roots $m(\alpha)$ for $|\alpha|\to \infty$, the simplest approach is 
to substitute $m= n \alpha$ into Equation \eqref{eqn:CharEqYahsiA2}, divide the polynomial by $\alpha^8$, and introduce the small parameter $a = \frac{1}{\alpha}$. Assuming an integer inverse power series of the type $n = n_0 + \frac{n_1}{\alpha} + \dots$, the leading order equation yields $n_0 = \pm 1$, implying: 

$$m \approx \pm \alpha$$

Alternatively, inverting $p_j(\alpha) = m_j^2(\alpha) - \alpha^2$ with respect to $m_j$ and expanding the square root for large $|\alpha|$ \citep{Yahsi:1981,Joseph:1987} yields:

\begin{subequations}
    \label{eqn:largeAlphaYahsi}
    \begin{align}
        m_j(\alpha) &= 
        \begin{cases}- \sqrt{p_j + \alpha^2} \approx - |\alpha| \sum_{n=0}^{\infty} \binom{1/2}{n} (-1)^{n+1} \left( \frac{p_j}{\alpha^2} \right)^n & 1 \leq j \leq 4 \\
        \sqrt{p_j + \alpha^2} \approx |\alpha| \sum_{n=0}^{\infty} \binom{1/2}{n} (-1)^{n+1} \left( \frac{p_j}{\alpha^2} \right)^n & 5 \leq j \leq 8
        \end{cases} \label{eqn:largeAlphaMj}\\
        r_1(\alpha) &= - \sqrt{\frac{2}{\kappa(1-\nu)}+\alpha^2} \approx - |\alpha| \sum_{n=0}^{\infty} \binom{1/2}{n} (-1)^{n+1} \left( \frac{2}{\kappa(1-\nu)\alpha^2} \right)^n \label{eqn:largeAlphaR1}\\
        r_2(\alpha) &= \sqrt{\frac{2}{\kappa(1-\nu)}+\alpha^2} \approx |\alpha| \sum_{n=0}^{\infty} \binom{1/2}{n} (-1)^{n+1} \left( \frac{2}{\kappa(1-\nu)\alpha^2} \right)^n \label{eqn:largeAlphaR2}
    \end{align}
\end{subequations}

\subsubsection{Numerical computation}

We compute the characteristic roots with MATLAB's \texttt{roots} function on an interval $[\alpha_0,\alpha_u]$.
As a verification step, we evaluate the polynomial evaluation residual:
$$\rho_j (\alpha) = p(m_j(\alpha))$$
Whenever the residual exceeds a prescribed tolerance, i.e., $\rho_j > \varepsilon_0 = 1e-12$, the root is
refined with up to $20$ Newton-Rapson iterations. 
Numerical experiments indicate that the computation of the characteristic roots becomes increasingly sensitive
for very small ($\alpha_0 \lesssim 1e-12$) and large ($\alpha_u \gtrsim 300$) values of $\alpha$. 
In practice, however, the computation of the mapping functions $B_{jk}(\alpha),B_k(\alpha),C_k(\alpha)$
(Equation~\eqref{eqn:newToOldUnknownsMain}) requires values of $\alpha_0$ well above this lower limit. 
Similarly, choosing $\alpha_u < 300$ is sufficient, since all Fredholm kernels integrands have decayed to zero
at this point. 
Finally, the characteristic roots are ordered with MATLAB's \texttt{sort} to satisfy the ordering condition
of Equation~\eqref{eqn:OrderedRoots}. 
Since the sorting is performed only with respect to the real part of the roots, spurious discontinuities may
appear in $\Im(m_j(\alpha_i))$. 
These discontinuities do not affect the numerical solution because of the summation terms contained in the
$w_{lk}(\alpha)$ ($l,k = 1,\dots,5$) functions of the Fredholm kernels 
(Equations~\eqref{eqn:DerEffKerns},~\eqref{eqn:DerEffKernsK2},~\eqref{eqn:DerEffKernsK3},~\eqref{eqn:DerEffKernsK4},
\eqref{eqn:DerEffKernsK5}). 
Since these summations are evaluated before carrying out the integration with respect to $\alpha$, the
discontinuities cancel and therefore have no effect on the final result.

\subsection{Fourier expressions}
\label{app:FourierExpressions}

In this appendix we give the complete expressions for all the unknowns appearing in the governing equations in terms
of the functions $R_1(\alpha), \dots, R_8(\alpha),A_1(\alpha),A_2(\alpha)$.
We also derive the corresponding expressions of all the membrane, bending and transverse shear stress resultants. 
Using Equations~\eqref{eqn:Transforms},~\eqref{eqn:wHat} we obtain the expression for the dimensionless out-of-plane
displacement $w(x,y)$:

\begin{equation}
    \label{eqn:outOfPlaneDispFourier}
    \begin{aligned}
        w(x,y)& = \frac{1}{2 \pi} \begin{cases} \int_{- \infty}^{\infty} e^{-i y \alpha}\left( \sum_{j = 1}^4  R_j  e^{m_j x}\right)d \alpha & x >0 \\
        \int_{- \infty}^{\infty} e^{-i y \alpha}\left( \sum_{j = 5}^8  R_j  e^{m_j x}\right)d \alpha & x <0
        \end{cases}
    \end{aligned}
\end{equation}

For brevity, we have omitted the dependence on $\alpha$. We adopt the same convention throughout the remainder
of the appendix. 
Using Equations~\eqref{eqn:Transforms},~\eqref{eqn:phiHat},~\eqref{eqn:KjYahsi} we obtain the expression for the
dimensionless Airy stress function $\phi(x,y)$:

\begin{equation}
    \label{eqn:stressFunctionFourier}
    \begin{aligned}
        \phi(x,y)& = \frac{1}{2 \pi} \begin{cases} \int_{- \infty}^{\infty} e^{-i y \alpha}\left( \sum_{j = 1}^4  K_j R_j  e^{m_j x}\right)d \alpha & x >0 \\
        \int_{- \infty}^{\infty} e^{-i y \alpha}\left( \sum_{j = 5}^8  K_j R_j  e^{m_j x}\right)d \alpha & x <0
        \end{cases}
    \end{aligned}
\end{equation}

Combining Equations~\eqref{eqn:Transforms},~\eqref{eqn:psiHat},~\eqref{eqn:omegaHat},
~\eqref{eqn:rootsOmega} we obtain the expressions for the auxiliary function $\psi(x,y)$ and $\Omega(x,y)$:

\begin{subequations}
    \label{eqn:auxFunctionFourier}
    \begin{align}
        \psi(x,y)& = \frac{1}{2 \pi} \begin{cases} \int_{- \infty}^{\infty} e^{-i y \alpha}\left( \sum_{j = 1}^4  \frac{1}{\kappa p_j - 1} R_j  e^{m_j x}\right)d \alpha & x >0 \\
        \int_{- \infty}^{\infty} e^{-i y \alpha}\left( \sum_{j = 5}^8  \frac{1}{\kappa p_j - 1} R_j  e^{m_j x}\right)d \alpha & x <0
        \end{cases} \label{eqn:psiFourier}\\
        \Omega(x,y)& = \frac{1}{2 \pi} \begin{cases} \int_{- \infty}^{\infty} e^{-i y \alpha} A_1  e^{r_1 x} d \alpha & x >0 \\
        \int_{- \infty}^{\infty} e^{-i y \alpha} A_2  e^{r_2 x} d \alpha & x <0
        \end{cases} \label{eqn:OmegaFourier}
    \end{align}  
\end{subequations}

where:

\begin{equation}
    \label{eqn:pjYahsiMain}
    p_j = m_j^2 - \alpha^2
\end{equation}

We obtain the expressions of the membrane stress resultants using Equations~\eqref{eqn:stressFunctionFourier} and
Equations~\eqref{eqn:AiryStressFunc},~\eqref{eqn:NormalizedMaterialParams}:

\begin{subequations}
    \label{eqn:membraneResultantsFourier}
    \begin{align}
        N_{xx}(x,y) & = - \frac{1}{2 \pi} 
        \begin{cases}
            \int_{- \infty}^{\infty} \alpha^2 e^{-i y \alpha} \left( \sum_{j = 1}^4 K_j R_j e^{m_j x}\right)d \alpha & \text{ for } x >0 \\
            \int_{- \infty}^{\infty} \alpha^2 e^{-i y \alpha} \left( \sum_{j = 5}^8 K_j R_j e^{m_j x}\right)d \alpha & \text{ for } x < 0 \\
        \end{cases}\label{eqn:NxxFourierDomain} \\
        N_{yy}(x,y) & = \frac{1}{2 \pi} 
        \begin{cases}
            \int_{- \infty}^{\infty} e^{-i y \alpha} \left( \sum_{j = 1}^4 K_j R_j m_j^2 e^{m_j x}\right)d \alpha & \text{ for } x > 0 \\
            \int_{- \infty}^{\infty} e^{-i y \alpha} \left( \sum_{j = 5}^8 K_j R_j m_j^2 e^{m_j x}\right)d \alpha& \text{ for } x < 0 \\
        \end{cases} \label{eqn:NyyFourierDomain} \\
        N_{xy}(x,y) & = \frac{i}{2 \pi} 
        \begin{cases}
            \int_{- \infty}^{\infty} \alpha e^{-i y \alpha} \left( \sum_{j = 1}^4 K_j R_j m_j e^{m_j x}\right)d \alpha & \text{ for } x > 0 \\
            \int_{- \infty}^{\infty} \alpha e^{-i y \alpha} \left( \sum_{j = 5}^8 K_j R_j m_j e^{m_j x}\right)d \alpha& \text{ for } x < 0 \\
        \end{cases} \label{eqn:NxyFourierDomain}
    \end{align}
\end{subequations}

To derive the expressions of the bending stress resultants, we first substitute
Equations~\eqref{eqn:auxFunctionFourier} into Equations~\eqref{eqn:BetasAuxFuncs}, obtaining the cross-section rotation
expressions:

\begin{equation}
    \label{eqn:BetaiFourier}
    \begin{aligned}
        \beta_x(x,y) &= \frac{1}{2 \pi} \int_{-\infty}^{\infty} \left(\sum_{j=1}^4 \frac{m_j}{\kappa p_j -1} R_j e^{m_j x - i \alpha y}\right) d \alpha - \frac{\kappa (1 - \nu)}{4 \pi} \int_{-\infty}^{\infty} i \alpha A_1 e^{r_1 x - i \alpha y} d \alpha \\
        \beta_y(x,y) &= \frac{1}{2 \pi} \int_{-\infty}^{\infty} \left(\sum_{j=1}^4 \frac{- i \alpha}{\kappa p_j -1} R_j e^{m_j x - i \alpha y}\right) d \alpha - \frac{\kappa (1 - \nu)}{4 \pi} \int_{-\infty}^{\infty}  r_1 A_1 e^{r_1 x - i \alpha y} d \alpha 
    \end{aligned}
\end{equation} 

For brevity, we report only the expressions for $x>0$.
The corresponding expressions for $x<0$ follow by the same index change used previously.
We adopt the same convention for the remainder of the appendix. 
We substitute these equations into the constitutive relations~\eqref{eqn:MomResShallowShells}, and normalize the result
in the same manner as the shell field Equations~\eqref{eqn:FinalEqA1}, yielding:

\begin{subequations}
    \label{eqn:bendingResultantsFourier}
        \begin{align}
            M_{xx}(x,y)& = \frac{1}{2 \pi} \frac{a}{h \lambda^4} \int_{- \infty}^{\infty} e^{-i y \alpha} \left[\left( \sum_{j = 1}^4 \frac{p_j  + \alpha^2 \left(1 - \nu \right)}{\kappa p_j  - 1} R_j  e^{m_j x}\right) - \frac{\kappa i \left(1 - \nu \right)^2}{2} 
            A_1  \alpha r_1 e^{r_1 x} \right]d \alpha \label{eqn:MxxFourierDomain} \\
            M_{yy}(x,y)& = \frac{1}{2 \pi} \frac{a}{h \lambda^4} \int_{- \infty}^{\infty} e^{-i y \alpha} \left[\left( \sum_{j = 1}^4 \frac{p_j  - \alpha^2 \left(1 - \nu \right)}{\kappa p_j  - 1} R_j  e^{m_j x}\right)+ \frac{\kappa i \left(1 - \nu \right)^2}{2} A_1  \alpha r_1 e^{r_1 x} \right]d \alpha \label{eqn:MyyFourierDomain} \\
            M_{xy}(x,y)& = - \frac{1}{2 \pi} \frac{a \left(1 - \nu \right)}{h \lambda^4} \int_{- \infty}^{\infty} i \alpha e^{-i y \alpha} \left[\left( \sum_{j = 1}^4 \frac{m_j}{\kappa p_j  - 1} R_j  e^{m_j x}\right) + \frac{\kappa \left(1 - \nu \right)}{4}  A_1 \left(\alpha^2 + r_1^2 \right) e^{r_1 x} \right]d \alpha \label{eqn:MxyFourierDomain}
        \end{align}
\end{subequations}

We obtain the transverse shear stress resultants by substituting 
Equations~\eqref{eqn:BetaiFourier},~\eqref{eqn:outOfPlaneDispFourier} into the constitutive 
relations~\eqref{eqn:TransResShallowShells} and applying the same normalization used for the shell field 
Equations~\eqref{eqn:FinalEqA1}:

\begin{subequations}
    \label{eqn:transverseResultantsFourier}
    \begin{align}
        V_{x}(x,y)& = \frac{1}{2 \pi} \int_{- \infty}^{\infty} e^{-i y \alpha}\left[ \left( \sum_{j = 1}^4 \frac{ \kappa p_j  m_j}{\kappa p_j  - 1} R_j  e^{m_j x}\right)  - \frac{\kappa i \left(1 - \nu \right)}{2}   A_1  \alpha e^{r_1 x}\right]  d \alpha\label{eqn:VxFourierDomain} \\
        V_{y}(x,y)& = \frac{1}{2 \pi} i \int_{- \infty}^{\infty} \alpha e^{-i y \alpha}\left[ \left( \sum_{j = 1}^4 \frac{\kappa p_j }{\kappa p_j  - 1} R_j  e^{m_j x}\right)  + \frac{\kappa \left(1 - \nu \right)}{2}   A_1  r_1 e^{r_1 x} \right]d \alpha \label{eqn:VyFourierDomain}
    \end{align}
\end{subequations}

These expressions provide the Fourier transforms of the stress resultants:

\begin{subequations}
    \label{eqn:StressResultantsfourier}
    \begin{align}
        \hat{N}_{ij}(x,\alpha) &= \mathcal{F}(N_{ij}(x,y)) \\
        \hat{M}_{ij}(x,\alpha) &= \mathcal{F}(M_{ij}(x,y))  \\
        \hat{V}_i(x,\alpha) &= \mathcal{F}(V_i(x,y)) 
     \end{align}
\end{subequations}

Finally, we derive the Fourier expression for the in-plane midsurface displacements $u(x,y),v(x,y)$. 
We begin with the former. Using the kinematic relations of Equation~\eqref{eqn:Strains}, we obtain the expressions for 
$\partial_y \varepsilon_{xy}, \partial_x \varepsilon_{yy}$ :

\begin{subequations}
    \label{eqn:derivativesInplaneStrain}
    \begin{align}
        \partial_y \varepsilon_{xy}(x,y)& = \frac{1}{2} \left(\partial_{xy} v(x,y) + \partial_{yy} u(x,y) + \partial_{xy} z \partial_y w(x,y) + \partial_x z \partial_{yy} w(x,y) \right. + \nonumber \\
        &+ \left. \partial_{yy} z \partial_x w(x,y) + \partial_y z \partial_{xy} w(x,y) \right)\label{eqn:dyexy} \\
        \partial_x \varepsilon_{yy}(x,y) & = \partial_{xy} v(x,y) + \partial_{xy} z \partial_y w(x,y) + \partial_y z \partial_{xy} w \label{eqn:dxeyy}
    \end{align}
\end{subequations}

Substituting the first expression in the second and rearranging for $\partial_{yy} u(x,y)$ yields:

\begin{equation}
    \label{eqn:dyyuStrains}
    \begin{aligned}
        \partial_{yy} u(x,y)& = 2 \partial_y \varepsilon_{xy}(x,y) - \partial_x \varepsilon_{yy}(x,y) - \partial_x z \partial_{yy} w(x,y) - \partial_{yy} z \partial_x w(x,y)
    \end{aligned}
\end{equation}

Using the constitutive relations of Equation~\eqref{eqn:ConstitutiveEquations} yields:

\begin{equation}
    \label{eqn:dyyuResultants}
    \begin{aligned}
        \partial_{yy} u(x,y)& = 2 \left( 1 + \nu \right)\partial_y N_{xy}(x,y) - \partial_{x} N_{yy}(x,y) + \nu \partial_{x} N_{xx}(x,y) + \frac{\gamma_y}{\lambda^2} \partial_x w(x,y) -  \partial_{yy} w(x,y) \partial_x z
    \end{aligned}
\end{equation}

The shell midsurface $z(x,y)$ can be approximated by a paraboloid with constant curvature radii $R_x,R_y,R_{xy}$
\citep{Calladine:1983,Yahsi:1983}. Therefore, its dimensionless first derivative is approximated as: 

\begin{equation}
    \label{eqn:TayloExpCylinder}
    \begin{aligned}
        \partial_x z(x,y) \approx - \left( x \frac{\gamma_x}{\lambda^2} + y \frac{\gamma_{xy}}{\lambda^2} \right) 
    \end{aligned}
\end{equation}

We integrate once with respect to $y$ to obtain the derivative of the displacement:

\begin{equation}
    \label{eqn:CompStressResultantU}
    \begin{aligned}
        \partial_{y} u& = 2 \left( 1 + \nu \right) N_{xy} - \int \partial_{x} N_{yy} \;\text{d}y + \nu \int \partial_{x} N_{xx} \;\text{d}y + \frac{\gamma_y}{\lambda^2} \int \partial_x w \;\text{d}y + \int \left( x \frac{\gamma_x}{\lambda^2} + y \frac{\gamma_{xy}}{\lambda^2} \right) \partial_{yy} w\; \text{d}y
    \end{aligned}
\end{equation}

Integrating a second time yields the desired expression:

\begin{equation}
    \label{eqn:uResultants}
    \begin{aligned}
        u(x,y)& = \int 2 \left( 1 + \nu \right) N_{xy}(x,y) d y - \int \int \partial_{x} N_{yy}(x,y) \text{d}y \text{d}y + \nu \int \int \partial_{x} N_{xx}(x,y) \text{d}y \text{d}y + \\
        &+ \frac{\gamma_y}{\lambda^2} \int \int \partial_x w(x,y) \text{d}y \text{d}y +  \int \int \left( x \frac{\gamma_x}{\lambda^2} + y \frac{\gamma_{xy}}{\lambda^2} \right) \partial_{yy} w(x,y) \text{d}y \text{d}y
    \end{aligned}
\end{equation}

Finally, substituting Equations~\eqref{eqn:outOfPlaneDispFourier},~\eqref{eqn:membraneResultantsFourier} yields the Fourier expression for the in-plane displacement $u(x,y)$:

\begin{subequations}
    \label{eqn:uFourier}
    \begin{align}
        u(x,y) &= \frac{1}{2\pi} \int_{- \infty}^{\infty} e^{-i y \alpha} \left( \sum_{j = 1}^4 U_j(y) R_j e^{m_j x}\right)d \alpha \\
        U_j (y) & = \frac{K_j m_j^3}{\alpha^2} + \frac{\gamma_{xy}}{\lambda^2} \frac{y \alpha - 2 i}{\alpha} - m_j K_j(\nu +2) - \frac{\gamma_y}{\lambda^2} \frac{m_j}{\alpha^2}
    \end{align}
\end{subequations} 

The second in-plane displacement is obtained through an analogous procedure.
We again use the kinematic relations to derive an expression in terms of $\varepsilon_{yy}$ and approximate
$\partial_y z(x,y)$ as:

\begin{subequations}
    \label{eqn:dyvStrains}
    \begin{align}
        \partial_{y} v(x,y)& = \varepsilon_{yy} - \partial_y z \partial_y w \\
        \partial_y z(x,y) &\approx - \left( x \frac{\gamma_{xy}}{\lambda^2} + y \frac{\gamma_{y}}{\lambda^2} \right)
    \end{align}
\end{subequations}

Using the constitutive relations yields:

\begin{equation}
    \label{eqn:dyvFourier}
    \begin{aligned}
        \partial_y v(x,y) &= N_{yy}(x,y) - \nu N_{xx}(x,y) + \left( x \frac{\gamma_{xy}}{\lambda^2} + y \frac{\gamma_y}{\lambda^2} \right) \partial_y w(x,y)
    \end{aligned}
\end{equation}

We integrate once in $y$ to obtain the expression for $v(x,y)$:

\begin{equation}
    \label{eqn:vResultants}
    \begin{aligned}
        v(x,y) &= \int N_{yy}(x,y) \text{d}y - \nu \int N_{xx}(x,y) \text{d}y + x \frac{\gamma_{xy}}{\lambda^2} w(x,y) + \frac{\gamma_y}{\lambda^2} \int y \partial_y w(x,y) \text{d}y
    \end{aligned}
\end{equation}

Again, substituting Equations~\eqref{eqn:outOfPlaneDispFourier},~\eqref{eqn:membraneResultantsFourier} yields the Fourier
expression of the in-plane displacement $v(x,y)$:

\begin{equation}
    \label{eqn:vFourier}
    \begin{aligned}
        v(x,y) &= \frac{1}{2\pi} \int_{- \infty}^{\infty} e^{-i y \alpha} \left( \sum_{j = 1}^4 V_j(y) R_j e^{m_j x}\right)d \alpha \\
        V_j (y) & = i K_j \frac{m_j^2 + \nu \alpha}{\alpha} + \frac{\gamma_y}{\lambda^2} \frac{\alpha y - i}{\alpha}
    \end{aligned}
\end{equation}

\subsection{Fourier expressions of the displacement derivative jumps}
\label{app:FourierJumps}

In this appendix we provide the details of the derivation of
Equation~\eqref{eqn:newToOldUnknownsMain}, which maps the 
initial unknowns $R_1(\alpha), \dots, R_8(\alpha),A_1(\alpha),A_2(\alpha)$ to the displacement derivative jumps at the
crack plane:

\begin{subequations}
    \label{eqn:displacementJumpA4}
    \begin{align}
            g_1 \left( y \right) & = \bigg\llbracket \partial_y u (0,y) -  \frac{\gamma_{xy}}{\lambda^2} \int y \partial_{yy} w (0,y) \text{d}y \bigg\rrbracket \label{eqn:g1} \\
            g_2 \left( y \right) & =  \bigg\llbracket \partial_y \beta_x (0,y)  \bigg\rrbracket \label{eqn:g2} \\
            g_3 \left( y \right) & =  \bigg\llbracket \partial_y v (0,y) -  \frac{\gamma_y}{\lambda^2}  y \partial_{y} w (0,y)  \bigg\rrbracket \label{eqn:g3} \\
            g_4 \left( y \right) & =  \bigg\llbracket \partial_y \beta_y (0,y)  \bigg\rrbracket \label{eqn:g4} \\
            g_5 \left( y \right) & =  \bigg\llbracket \partial_y w (0,y)  \bigg\rrbracket \label{eqn:g5}
    \end{align}
\end{subequations}

Here:

$$\llbracket f(0) \rrbracket = \lim_{x\to 0^+}f(x) - \lim_{x\to 0^-}f(x)$$

The first step is to express the functions $g_k(y)$ in terms of the Fourier frequency $\alpha$. We use Equations~\eqref{eqn:uFourier} 
and~\eqref{eqn:outOfPlaneDispFourier} to obtain $g_1(y)$:

\begin{equation}
    \label{eqn:g1Fourier}  
    \begin{aligned}
        g_1 \left(y\right) &= \frac{1}{2 \pi} \left[ \int_{- \infty}^{\infty} i e^{-i \alpha y} \left[ \sum_{j = 1}^4 R_j \frac{m_j}{\alpha} \left(\frac{\gamma_y}{\lambda^2} - K_j  p_j \right) - \sum_{j = 5}^8 R_j \frac{m_j}{\alpha} \left(\frac{\gamma_y}{\lambda^2} - K_j  p_j \right)\right] d \alpha  \right. \\
        & + \left. i(1 + \nu)\int_{- \infty}^{\infty} e^{-i \alpha y} \alpha \left(\sum_{j = 1}^4 R_j m_j K_j - \sum_{j = 5}^8 R_j m_j K_j \right) d \alpha \right]
    \end{aligned} 
\end{equation}

For brevity, we omit the dependency on $\alpha$ of $R_j(\alpha), A_i(\alpha), p_j(\alpha), K_j(\alpha)$. 
The notation is used throughout the remainder of the appendix. 
Following the same procedure, Equations~\eqref{eqn:vFourier} and~\eqref{eqn:outOfPlaneDispFourier} give $g_3(y)$:

\begin{equation}
    \label{eqn:g3Fourier}  
    \begin{aligned}
        g_3 \left(y\right)& = \frac{1}{2 \pi} \int_{-\infty}^{\infty} e^{-i \alpha y} \left[ \sum_{j = 1}^4 R_j  K_j  \left( p_j + \left(1 + \nu \right) \alpha^2\right) - \sum_{j = 5}^8 R_j  K_j  \left( p_j + \left(1 + \nu \right) \alpha^2\right) \right] d \alpha
    \end{aligned} 
\end{equation}

The remaining functions $g_2(y),g_4(y)$ are obtained using Equations~\eqref{eqn:BetaiFourier} 
and~\eqref{eqn:outOfPlaneDispFourier}:

\begin{subequations}
    \begin{align}
        g_2 \left(y\right) &= -\frac{1}{2 \pi} \int_{- \infty}^{\infty} e^{-i \alpha y}\left[ i  \left( \sum_{j = 1}^4  \frac{R_j  m_j}{\kappa p_j - 1} - \sum_{j = 5}^8  \frac{R_j  m_j}{\kappa p_j - 1} \right) +  \frac{\alpha^2 \kappa (1 - \nu)}{2} \left( A_1  - A_2 \right)  \right] d \alpha \label{eqn:g2Fourier} \\
        g_4 \left(y\right) &= -\frac{1}{2 \pi} \int_{-\infty}^{\infty} e^{-i \alpha y}\left[ \alpha^2  \left( \sum_{j = 1}^4  \frac{R_j  }{\kappa p_j - 1} - \sum_{j = 5}^8  \frac{R_j }{\kappa p_j - 1} \right) -  i\frac{\alpha \kappa (1 - \nu)}{2} \left( r_1 A_1  - r_2 A_2 \right) \right] d \alpha \label{eqn:g4Fourier} 
    \end{align} 
\end{subequations}

Finally, Equation~\eqref{eqn:outOfPlaneDispFourier} gives $g_5(x,y)$:

\begin{equation}
    \label{eqn:g5Fourier}  
    \begin{aligned}
        g_5 \left(y\right)& = - \frac{i}{2 \pi} \int_{- \infty}^{\infty} \alpha e^{-i \alpha y} \left( \sum_{j = 1}^4  R_j  - \sum_{j = 5}^8 R_j  \right) d \alpha 
    \end{aligned} 
\end{equation}

We rearrange the boundary conditions (Equation~\eqref{eqn:boundaryConditionss}) to obtain one set of homogeneous and one set of mixed boundary conditions. The former prescribes the continuity of the stress resultants across the crack plane.

\begin{equation}
    \label{eqn:sBCContinuity}
    \begin{aligned}
        \left\{\begin{matrix}
                    &\lim_{x\to 0^+} N_{xx}(x,y) - \lim_{x\to 0^-} N_{xx}(x,y)= 0 \\
                    &\lim_{x\to 0^+} N_{xy}(x,y) - \lim_{x\to 0^-} N_{xy}(x,y)= 0 \\
                    &\lim_{x\to 0^+} M_{xx}(x,y) - \lim_{x\to 0^-} M_{xx}(x,y)= 0 \\
                    &\lim_{x\to 0^+} M_{xy}(x,y) - \lim_{x\to 0^-} M_{xy}(x,y)= 0 \\
                    &\lim_{x\to 0^+} V_x(x,y)    - \lim_{x\to 0^-} V_x(x,y)= 0 
                \end{matrix} \qquad \right. \;\;  \left| y \right| < \infty 
    \end{aligned}
\end{equation}

The latter prescribes the self-equilibrating loading applied to crack faces and the continuity of displacements outside of the crack length on the crack plane.

\begin{equation}
    \label{eqn:sBCLoading}
    \begin{aligned}
        &\begin{matrix}
            \begin{cases}
            \lim_{x\to 0^+} N_{xx}(x,y) + \lim_{x\to 0^-} N_{xx}(x,y)= 2 N^0_{xx}(y) \text{, }& \qquad \left| y \right| < 1\\
            \lim_{x\to 0^+} u(x,y) - \lim_{x\to 0^-} u(x,y)= 0 \text{, }& \qquad \left| y \right| \geq 1
            \end{cases}\\
            \begin{cases}
            \lim_{x\to 0^+} N_{xy}(x,y) + \lim_{x\to 0^-} N_{xy}(x,y)= 2 N^0_{xy}(y) \text{, }& \qquad \left| y \right| < 1\\
            \lim_{x\to 0^+} v(x,y) - \lim_{x\to 0^-} v(x,y)= 0 \text{, }&\qquad \left| y \right| \geq 1
            \end{cases} \\
            \begin{cases}
            \lim_{x\to 0^+} M_{xx}(x,y) + \lim_{x\to 0^-} M_{xx}(x,y)= 2M^0_{xx}(y) \text{, }& \qquad \left| y \right| < 1\\
            \lim_{x\to 0^+} \beta_x(x,y) - \lim_{x\to 0^-} \beta_x(x,y)= 0 \text{, }& \qquad \left| y \right| \geq 1
            \end{cases}\\
            \begin{cases}
            \lim_{x\to 0^+} M_{xy}(x,y) + \lim_{x\to 0^-} M_{xy}(x,y)= 2 M^0_{xy}(y) \text{, }& \qquad \left| y \right| < 1\\
            \lim_{x\to 0^+} \beta_y(x,y) - \lim_{x\to 0^-} \beta_y(x,y)= 0 \text{, }& \qquad \left| y \right| \geq 1
            \end{cases}\\
            \begin{cases}
            \lim_{x\to 0^+} V_{x}(x,y) + \lim_{x\to 0^-} V_{x}(x,y)= 2V^0_x(y) \text{, }& \qquad \left| y \right| < 1\\
            \lim_{x\to 0^+} w(x,y) - \lim_{x\to 0^-} w(x,y)= 0 \text{, }& \qquad \left| y \right| \geq 1
            \end{cases}
        \end{matrix}
    \end{aligned}
\end{equation}

Expressions~\eqref{eqn:g1Fourier},~\eqref{eqn:g2Fourier},~\eqref{eqn:g3Fourier},~\eqref{eqn:g4Fourier}, and~\eqref{eqn:g5Fourier} are the inverse Fourier transforms of five functions
$\hat{g}_1(\alpha), \dots, \hat{g}_5(\alpha)$, evaluated in the limit $x \to 0$. 
We therefore introduce the collocation variable $t$ and take the Fourier transform of $g_1(y), \dots, g_5(y)$. 
Owing to the displacement continuity conditions in Equation~\eqref{eqn:sBCLoading}, the integration interval reduces to the crack domain $|t| \leq 1$.

\begin{equation}
    \label{eqn:gHatDefinition}
    \begin{aligned}
        \hat{g}_k (\alpha) &= \mathcal{F}( g_k(0,t)) = \frac{1}{2\pi} \int_{-\infty}^\infty g_k(0,t) e^{i \alpha t} \text{d}t = \frac{1}{2\pi} \int_{-1}^{1} g_k(0,t) e^{i \alpha t} \text{d}t 
    \end{aligned}
\end{equation}

This yields five algebraic equations relating the unknown functions 
$R_1(\alpha), \dots, R_8(\alpha),A_1(\alpha),A_2(\alpha)$ to the Fourier transforms of the displacement derivative
jumps $\hat{g}_k (\alpha)$:

\begin{subequations}
    \label{eqn:gkHat}
    \begin{align}
        \hat{g}_1 (\alpha)& = \frac{\gamma_y}{\lambda^2} \frac{\hat{g}_2 (\alpha)}{\alpha^2} - \frac{i}{\alpha} \left(\sum_{j=1}^4 R_j  K_j  p_j  m_j- \sum_{j=5}^8 R_j  K_j  p_j  m_j \right) \label{eqn:ghat1} \\
        \hat{g}_2 (\alpha)& = \frac{A_1  - A_2 }{2} \label{eqn:ghat2} \\
        \hat{g}_3 (\alpha)& = \sum_{j=1}^4 R_j  K_j  p_j - \sum_{j=5}^8 R_j  K_j  p_j  \label{eqn:ghat3} \\
        \hat{g}_4 (\alpha)& = \frac{1}{ 1 - \nu } \left(\sum_{j=1}^4 \frac{p_j  R_j }{\kappa p_j  - 1 } - \sum_{j=5}^8 \frac{p_j  R_j }{\kappa p_j  - 1 } \right) \label{eqn:ghat4} \\
        \hat{g}_5 (\alpha)& = - i \alpha \left(\sum_{j=1}^4 R_j  - \sum_{j=5}^8 R_j   \right) \label{eqn:ghat5} 
    \end{align}
\end{subequations}

To complete the mapping, we substitute the Fourier expressions of the stress resultants derived in
Appendix~\ref{app:FourierExpressions} into the homogeneous boundary conditions \eqref{eqn:sBCContinuity}.
This yields a second set of algebraic equations in $R_1(\alpha), \dots, R_8(\alpha),A_1(\alpha),A_2(\alpha)$:

\begin{subequations}
    \label{eqn:HomBCMappingA5}
    \begin{align}
        \sum_{j=1}^4 & K_j R_j - \sum_{j=5}^8 K_j R_j = 0 \label{eqn:NxxHomBCA5} \\
        \sum_{j=1}^4 & m_j K_j R_j - \sum_{j=5}^8 m_j K_j R_j = 0 \label{eqn:NxyHomBCA5} \\
        \sum_{j=1}^4 & \frac{p_j  + \alpha^2 (1 - \nu)}{\kappa p_j  -1} R_j - \sum_{j=5}^8 \frac{p_j  + \alpha^2 (1 - \nu)}{\kappa p_j  -1} R_j - \frac{\kappa i \alpha \left(1 - \nu\right)^2}{2}  \left(r_1 A_1  - r_2 A_2 \right)= 0 \label{eqn:MxxHomBCA5} \\
        \sum_{j=1}^4 & \frac{m_j}{\kappa p_j  -1} R_j  - \sum_{j=5}^8 \frac{m_j}{\kappa p_j  -1} R_j + \frac{\kappa \left(1 - \nu\right)}{4} \left(\alpha^2 + r_1^2 \right)\left(A_1  - A_2 \right)= 0 \label{eqn:MxyHomBCA5} \\
        \sum_{j=1}^4 & \frac{m_j p_j }{\kappa p_j  -1}R_j  - \sum_{j=5}^8 \frac{m_j p_j }{\kappa p_j  -1}R_j + \frac{\kappa i \alpha \left(1 - \nu\right)}{2} \left(A_1  - A_2 \right)= 0 \label{eqn:VxHomBCA5}
    \end{align}
\end{subequations}

Collecting Equations~\eqref{eqn:gkHat},~\eqref{eqn:HomBCMappingA5}, we obtain a $10 \times 10$ linear system:

\begin{equation}
    \label{eqn:SystemInitial}
    \begin{aligned}
        \bm{D}(\alpha) \; \bm{r}(\alpha) = \bm{l}(\alpha) \Leftrightarrow \\
        \begin{bmatrix}
            K_1 p_1 m_1 & \cdots & - K_8 p_8 m_8 & 0 & 0 \\
            \vdots &  & & & \vdots \\
            \frac{m_1}{\kappa p_1 -1} & \cdots & - \frac{m_8}{\kappa p_8 -1} & \kappa (\alpha^2 + r_1^2) \frac{1 - \nu}{4} & - \kappa (\alpha^2 + r_1^2)\frac{1 - \nu}{4} \\
            \frac{m_1 p_1}{\kappa p_1 -1} & \cdots & - \frac{m_8 p_8}{\kappa p_8 -1} & i \alpha \kappa \frac{1 - \nu}{2} & -i \alpha \kappa \frac{1 - \nu}{2}
        \end{bmatrix}&
        \begin{bmatrix} 
            R_1 \\ \vdots \\ R_8 \\ A_1 \\ A_2
        \end{bmatrix} = 
        \begin{bmatrix} 
            i \left( \alpha \hat{g}_1(\alpha) - \frac{1}{\alpha} \frac{\gamma_y}{\lambda^2} \hat{g}_2(\alpha)\right) \\ \vdots \\ 0 \\ 0
        \end{bmatrix}
    \end{aligned}
\end{equation}

We decompose the loading vector $ \bm{l} (\alpha)$ into five contributions, one associated with each function $\hat{g}_k(\alpha)$. 
We decompose analogously the unknown vector $ \bm{r} (\alpha)$:

\begin{equation}
    \label{eqn:ImpsLinearity1}
    \begin{aligned}
       \mathbf{D}(\alpha) \left(\bm{r}_1(\alpha) + \dots + \bm{r}_5(\alpha)\right) & = \bm{l}_1(\alpha) \hat{g}_1(\alpha) + \dots +\bm{l}_5(\alpha) \hat{g}_5(\alpha)
    \end{aligned}
\end{equation}

The functions $\hat{g}_k(\alpha)$ are obtained by applying two linear operators (differentiation and Fourier transformation) to five independent kinematic quantities (namely, the displacements $u,v,w$ and cross-section rotations $\beta_x,\beta_y$). Consequently, they must be linearly independent and non-zero, when searching for non trivial solutions. 
Therefore, the system can be decomposed into five separated problems:

\begin{equation}
    \label{eqn:ImpsLinearity2}
    \begin{aligned}
        \begin{cases}  
            \bm{r}_1(\alpha) - \mathbf{D}^{-1}(\alpha) \; \bm{l}_1(\alpha) = \bm{0} \\ \vdots \\
            \bm{r}_5(\alpha) - \mathbf{D}^{-1}(\alpha) \; \bm{l}_5(\alpha)  = \bm{0} 
        \end{cases}
    \end{aligned}
\end{equation}

Each solution vector $\bm{r}_k(\alpha)$ provides eight functions $B_{jk}(\alpha)$, together with the functions $B_k(\alpha)$, $C_k(\alpha)$, thereby completing the mapping of Equation~\eqref{eqn:newToOldUnknownsMain}:

\begin{equation}
    \label{eqn:ImpsLinearity3}
    \begin{aligned}
            \bm{r}_k(\alpha) = \left[iB_{1k}(\alpha), \dots, iB_{8k}(\alpha), B_k(\alpha), C_k(\alpha) \right]^T
    \end{aligned}
\end{equation}

\subsubsection{Computation of the mapping function}

We compute the mapping functions numerically by solving the five linear systems of
Equation~\eqref{eqn:ImpsLinearity2} using the Singular Value Decomposition for each value of $\alpha$.
Specifically:

\begin{equation}
    \label{eqn:mappingSVD}
    \begin{aligned} 
            \mathbf{D}(\alpha) &= \bm{U}(\alpha) \bm{\Sigma}(\alpha) \bm{V}^T(\alpha)\\
            \bm{r}_k(\alpha) &=  \bm{V}(\alpha) \bm{\Sigma}^{-1}(\alpha) \bm{U}^T(\alpha) \bm{l}_k(\alpha)
    \end{aligned}
\end{equation}

We use the Singular Value Decomposition  because the system matrix $\bm{D}(\alpha)$ is ill-conditioned for both
ends of real line ($\alpha \in [0,\infty)$). 
In this way, the conditioning only reduces the significant digits retained in the mapping functions, according
to the heuristic:

\begin{equation}
    \label{eqn:relevantDigits}
    \begin{aligned} 
            t = m - |\log_{10}\kappa(\alpha_0)|
    \end{aligned}
\end{equation}

\begin{figure}[pos=ht]
    \centering
    \includegraphics[width =\linewidth]{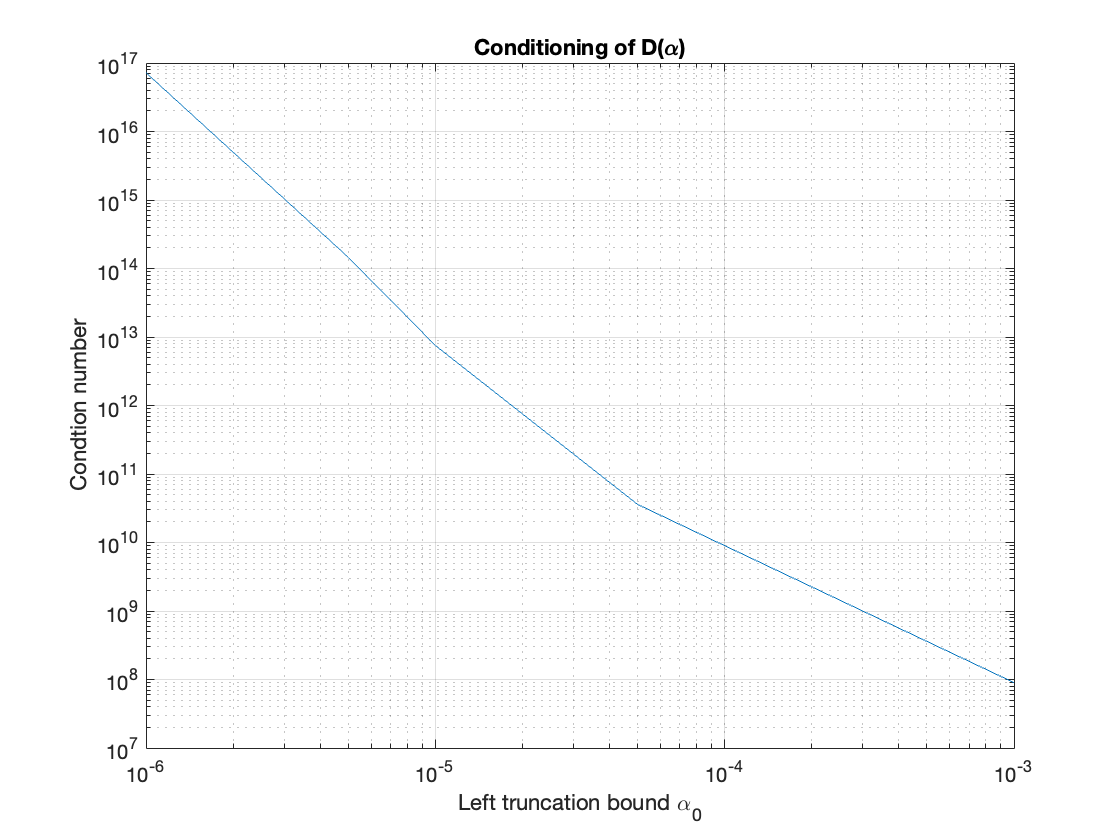}
    \caption{Condition number of $\mathbf{D}(\alpha = \alpha_0)$ as a function of the lower integration bound $\alpha_0$. 
    Both axes are shown in logarithmic scale. 
    Since the number of trustworthy digits is inversely related to the order of magnitude of the condition number, and 
    double precision provides approximately $16$ relevant digits, at least six trustworthy digits
    are retained $\alpha_0 \geq 1e-4$.}
    \label{fig:ExampleCondNummD}
\end{figure}

Where $t$ is the number of trustworthy numerical digits, $m = 16$ is the machine precision for standard 
double precision arithmetic, and $\kappa(\alpha_0)$ is the condition number of the matrix $\bm{D}(\alpha_0)$.
The Fredholm and displacement kernels (Equations~\eqref{eqn:DerEffKerns}, \eqref{eqn:DerEffKernsK2}, \eqref{eqn:DerEffKernsK3}, \eqref{eqn:DerEffKernsK4}, \eqref{eqn:DerEffKernsK5}, \eqref{eqn:discWp}, \eqref{eqn:BetaXpFull}, \eqref{eqn:BetaYpFull}, \eqref{eqn:UpFull}, \eqref{eqn:VpFull}) decay to numerical zero for $\alpha \gtrsim 300$. 
Consequently, $\alpha \to \infty$ is never reached in practice.
We investigated the behavior of the system matrix near $\alpha_0 \to 0$ with extensive numerical experiments.
By requiring at least six trustworthy digits in the solution of Equation~\eqref{eqn:ImpsLinearity2}, and using
Equation~\eqref{eqn:relevantDigits}, we obtain the lower bound for $\alpha_0 \approx 1e-4$.
We show in Figure~\ref{fig:ExampleCondNummD} and example of the condition number of the matrix $\bm{D}(\alpha)$.

\subsection{Expression of the kernels in the crack-loading conditions}
\label{app:Kernels}

In this appendix we provide the details of the derivation used to extract the singular part of the integral 
kernels in Equation~\eqref{eqn:SingIntegralsMixedLoading}. 

We substitute Equation~\eqref{eqn:gBackTranform} into~\eqref{eqn:newToOldUnknownsMain} to establish a relation between $g_k(y)$ and the initial unknowns $R_j(\alpha),A_k(\alpha)$. We then substitute this relations into the Fourier expressions for the stress resultants given in Appendix~\ref{app:FourierExpressions}. Finally, the resulting expressions are substituted into the loading boundary conditions (Equation~\eqref{eqn:sBCLoading}).

The integral kernels $H_{kj}^+\left(\alpha,x\right),H_{kj}^-\left(\alpha,x\right)$ of 
Equation~\eqref{eqn:SingIntegralsMixedLoading} are given by:

\begin{subequations}
    \label{eqn:kernelsWithSingularPart}
    \begin{align}
        H_{l1}^+(x,\alpha)&= - i\frac{\alpha^2}{2 \pi}\sum_{j=1}^4K_j(\alpha)B_{j1}(\alpha)e^{m_j(\alpha) x} \\
        H_{l1}^-(x,\alpha)&= - i\frac{\alpha^2}{2 \pi}\sum_{j=5}^8K_j(\alpha)B_{j1}(\alpha)e^{m_j(\alpha) x} \\
        H_{l2}^+(x,\alpha)&= i\frac{a}{2\pi \lambda^4 h} \left(\sum_{j=1}^4\frac{p_j(\alpha) + \alpha^2(1 - \nu)}{\kappa p_j(\alpha) -1}B_{j2}(\alpha)e^{m_j(\alpha) x} - \frac{\kappa (1- \nu)^2}{2} \alpha r_1(\alpha)e^{r_1(\alpha)x}B_l(\alpha)\right)\\
        H_{l2}^-(x,\alpha)&= i\frac{a}{2\pi \lambda^4 h} \left(\sum_{j=5}^8\frac{p_j(\alpha) + \alpha^2(1 - \nu)}{\kappa p_j(\alpha) -1}B_{j2}(\alpha)e^{m_j(\alpha) x} - \frac{\kappa (1- \nu)^2}{2} \alpha r_2(\alpha)e^{r_2(\alpha)x}C_l(\alpha)\right) \\
        H_{l3}^+(x,\alpha)&= - \frac{\alpha}{2 \pi}\sum_{j=1}^4K_j(\alpha)m_j(\alpha)B_{j3}(\alpha)e^{m_j(\alpha) x} \\
        H_{l3}^-(x,\alpha)&= - \frac{\alpha}{2 \pi}\sum_{j=5}^8K_j(\alpha)m_j(\alpha)B_{j3}(\alpha)e^{m_j(\alpha) x} \\
        H_{l4}^+(x,\alpha)&= \frac{a(1-\nu)}{2\pi \lambda^4 h} \alpha\left(\sum_{j=1}^4\frac{m_j(\alpha)}{\kappa p_j(\alpha) -1}B_{j4}(\alpha)e^{m_j(\alpha) x} - \frac{\kappa (1- \nu)}{2} (\alpha^2 + r_1^2(\alpha))e^{r_1(\alpha)x}B_l(\alpha)\right) \\
        H_{l4}^-(x,\alpha)&= \frac{a(1-\nu)}{2\pi \lambda^4 h} \alpha\left(\sum_{j=5}^8\frac{m_j(\alpha)}{\kappa p_j(\alpha) -1}B_{j4}(\alpha)e^{m_j(\alpha) x} - \frac{\kappa (1- \nu)}{2} (\alpha^2 + r_2^2(\alpha))e^{r_2(\alpha)x}C_l(\alpha)\right)\\
        H_{l5}^+(x,\alpha)&= i\frac{1}{2\pi} \left(\sum_{j=1}^4\frac{\kappa p_j(\alpha) m_j(\alpha)}{\kappa p_j(\alpha) -1}B_{j5}(\alpha)e^{m_j(\alpha) x} - \frac{\kappa (1- \nu)}{2} \alpha e^{r_1(\alpha)x}B_l(\alpha)\right) \\
        H_{l5}^-(x,\alpha)&= i\frac{1}{2\pi} \left(\sum_{j=5}^8\frac{\kappa p_j(\alpha) m_j(\alpha)}{\kappa p_j(\alpha) -1}B_{j5}(\alpha)e^{m_j(\alpha) x} - \frac{\kappa (1- \nu)}{2} \alpha e^{r_2(\alpha)x}C_l(\alpha)\right)
    \end{align}
\end{subequations}

\subsection{Extracting the singular terms}
\label{app:SingularTermExtraction}

The kernels $H_{kj}^{\pm}(\alpha,x)$ of Appendix \ref{app:Kernels} are available in closed form 
only in terms of the functions 
$B_{jk}(\alpha),B_k(\alpha),C_k(\alpha)$, which are obtained numerically. 
Therefore, when $t=y$ and $x =0$, an asymptotic analysis is required to verify that they remain bounded as $|\alpha| \to \infty$. 

In this appendix we provide the details of the asymptotic analysis, and the transformed expressions of Equation~\eqref{eqn:SingIntegralsMixedLoading}.

From previous studies \citep{Yuceoglu:1973,Delale:1979a}, we expect only the diagonal terms (i.e., 
$H_{lk}^+(x,\alpha),H_{lk}^-(x,\alpha)$ with $k=j$) to become unbounded for $t=y, x =0$. 
To extract their singularity, we need to study the behaviour of the kernels for large $\alpha$. 
The functions $K_j(\alpha),p_j(\alpha),B_{jk}(\alpha),B_k(\alpha),C_k(\alpha)$ depend on $\alpha$ either explicitly or through the characteristic roots $m_j(\alpha),r_1(\alpha),r_2(\alpha)$. 
Consequently, Equation~\eqref{eqn:largeAlphaYahsi} allows to compute all the relevant asymptotic expansions.

We begin with the case $k=1$, which is the only example discussed in \cite{Yahsi:1981}.
We substitute Equation~\ref{eqn:largeAlphaYahsi} and use Equation~\ref{eqn:newToOldUnknownsMain} to recover the dependence on $R_j(\alpha)$:

\begin{equation}
    \label{eqn:IntegEqK1}
    \begin{aligned}
        \lim_{x \to 0^+} &\int_{- \infty}^{\infty} \alpha^2 \sum_{j = 1}^4 K_j(\alpha) R_j(\alpha) e^{m_j x- i \alpha y} d \alpha - \lim_{x \to 0^-} \int_{- \infty}^{\infty} \alpha^2 \sum_{j = 5}^8 K_j(\alpha) R_j(\alpha) e^{m_j x- i \alpha y} d \alpha  = 4 \pi \frac{\tilde{N}_{xx}^0 (y)}{h} \\
        & \Rightarrow \lim_{x \to 0^+} - \int_{- \infty}^{\infty} \alpha^2 \left( \sum_{j = 1}^8 K_j(\alpha) R_j(\alpha) \right) e^{- \left| \alpha \right| x} e^{- i \alpha y} d \alpha + h.o.t. = 4 \pi \frac{\tilde{N}_{xx}^0 (y)}{h}
    \end{aligned}
\end{equation}

Where we have used that:

\begin{equation}
    \label{eqn:ExpIdentity}
    \begin{aligned}
        \lim_{x \to 0^+} e^x = \lim_{x \to 0^-} e^{- x}
    \end{aligned}
\end{equation}

It remains to define the behaviour of the summation term to determine the behaviour of the kernel. 
Using the in-plane shear boundary condition (Equation~\eqref{eqn:NxyHomBCA5}), together with Equation~\eqref{eqn:largeAlphaMj}, and rearranging gives:

\begin{subequations}
    \label{eqn:HomBC}
    \begin{align}
        \sum_{j=1}^4 & m_j(\alpha) K_j(\alpha) R_j(\alpha) - \sum_{j=5}^8 m_j(\alpha) K_j(\alpha) R_j(\alpha) = 0 \Leftrightarrow \sum_{j = 1}^8 \left(1+ \frac{p_j(\alpha)}{2\alpha^2}\right)K_j(\alpha) R_j(\alpha) + h.o.t. = 0 \\ 
        &\Leftrightarrow \sum_{j = 1}^8 K_j(\alpha) R_j(\alpha) \approx - \frac{1}{2\alpha^2}\sum_{j = 1}^8 p_j(\alpha) K_j(\alpha) R_j(\alpha) 
    \end{align}
\end{subequations}

We use Equation~\eqref{eqn:ghat1} together with Equation~\eqref{eqn:largeAlphaMj}, and substitute the result
in  Equation~\eqref{eqn:IntegEqK1} to obtain the asymptotic relation:

\begin{equation}
    \label{eqn:AsymptoticSummation11}
    \begin{aligned}
        \hat{g}_1 (\alpha)& = \frac{\gamma_y}{\lambda^2} \frac{\hat{g}_2 (\alpha)}{\alpha^2} - \frac{i}{\alpha} \left(\sum_{j=1}^4 R_j(\alpha)  K_j(\alpha)  p_j(\alpha)  m_j(\alpha)- \sum_{j=5}^8 R_j(\alpha)  K_j(\alpha)  p_j(\alpha)  m_j(\alpha) \right) \Leftrightarrow \\
        &\Leftrightarrow \hat{g}_1 (\alpha) \approx - \frac{i}{\alpha} |\alpha| \sum_{j=1}^8 R_j(\alpha) K_j(\alpha) p_j(\alpha) \Leftrightarrow \sum_{j = 1}^8 K_j R_j \sim - \frac{\sign (\alpha)}{2 i \alpha^2 } \hat{g}_1 ( \alpha )
    \end{aligned}
\end{equation}

Following the procedure of \citet{Muskhelishvili:1972}, we add and subtract the asymptotic contribution 
to eq. \ref{eqn:SingIntegralsMixedLoading} with $k=1$. 
Again, we use the identity~\ref{eqn:ExpIdentity} to obtain:

\begin{equation}
    \label{eqn:SingularIntermediateStep}
    \begin{aligned}
        \lim_{x \to 0^+}& - \int_{- \infty}^{\infty} \alpha^2 e^{- i \alpha y} \left[\left( \sum_{j = 1}^4 K_j(\alpha) R_j(\alpha) e^{m_j(\alpha) x} + \sum_{j = 5}^8 K_j(\alpha) R_j(\alpha) e^{- m_j(\alpha) x} \right) + \frac{\sign (\alpha)}{2 i \alpha^2 } \hat{g}_1 ( \alpha ) e^{- \left| \alpha \right| x} \right]d \alpha + \\
        &+ \lim_{x \to 0^+} \int_{- \infty}^{\infty} \frac{\sign (\alpha)}{2 i }\hat{g}_1(\alpha) e^{- \left| \alpha \right| x-i \alpha y} d \alpha = 4 \pi \frac{\tilde{N}_{xx}^0 (y)}{h}
    \end{aligned}
\end{equation}

We substitute the definition of $\hat{g}_1(\alpha)$ (Equation~\eqref{eqn:gHatDefinition}) into the second
integral and evaluate it in closed form by integration by parts:

\begin{equation}
    \label{eqn:IntegralByAlpha}
    \begin{aligned}
        \int_{- \infty}^{\infty}& \frac{\sign (\alpha)}{2 i } e^{- \left| \alpha \right| x} e^{i \alpha (t - y)} d \alpha = 2 \int_{0}^{\infty} \frac{1}{2} e^{- \alpha x} \sin \left(\alpha \left(t -y\right)\right) d \alpha =  \frac{t - y}{\left( t - y \right)^2 + x^2}
    \end{aligned}
\end{equation}

The first integrand is now uniformly bounded, so the limit and integration may be interchanged.
We apply the limit in $x$, express the result in terms of $g_j(t)$ using Equation~\eqref{eqn:newToOldUnknownsMain} and rearrange to obtain:

\begin{subequations}
    \label{eqn:DerEffKerns}
    \begin{align}
        \int_{-1}^{1} \frac{g_1 (t)} {t - y} \text{d}t &+ \sum_{j = 1}^5 \int_{-1}^{1} K_{1k}(t,y) g_k (t) \text{d}t = 4 \pi \frac{\tilde{N}_{xx}^0 (y)}{h} \\
        K_{11}(t,y) &= - \int_{- \infty}^{\infty} \alpha^2 i e^{i \alpha (t - y)} \left( \sum_{j = 1}^8 K_j(\alpha) B_{j1}(\alpha) - \frac{\sign (\alpha)}{2}\right) d \alpha \\
        K_{1k}(t,y) &= - \int_{- \infty}^{\infty} \alpha^2 i e^{i \alpha (t - y)} \left( \sum_{j = 1}^8 K_j(\alpha)B_{jk}(\alpha) \right)d \alpha 
    \end{align}
\end{subequations}

The singular contributions in the remaining four equations are extracted in an analogous manner. 
For brevity, we omit the dependency on $\alpha$ of $R_j,K_j,A_i,r_i,p_j,m_j$ in the remainder of the appendix. 

For $k = 2$, we begin with substituting Equations~\eqref{eqn:NxyFourierDomain},~\eqref{eqn:ExpIdentity} into
Equation~\eqref{eqn:sBCLoading}. 
After rearranging and using the Fourier representation of $\hat{\Omega}$ (Equation~\eqref{eqn:omegaHat}):

\begin{equation}
    \label{eqn:IntegEqK2Step1}
    \begin{aligned}
        \lim_{x \to 0^+} \frac{1}{\lambda^4}& \int_{- \infty}^{\infty} \left(\sum_{j = 1}^4 \frac{p_j^2 + (1-\nu) \alpha^2}{\kappa p_j -1} R_j e^{m_j x} + \sum_{j = 5}^8 \frac{p_j^2 + (1-\nu) \alpha^2}{\kappa p_j -1} R_j e^{- m_j x} \right. + \\
        -&\left. i \alpha \frac{\kappa(1-\nu)^2}{2} r_1 (A_1 - A_2)e^{r_1 x}\right) e^{- i \alpha y} d \alpha  = 4 \pi \frac{h}{a} \frac{\tilde{M}_{xx}^0(y)}{h}
    \end{aligned}
\end{equation}

We then substitute Equation~\eqref{eqn:largeAlphaYahsi} to obtain:

\begin{equation}
    \label{eqn:IntegEqK2Step2}
    \begin{aligned}
        \lim_{\alpha \to \infty} &\lim_{x \to 0^+} \frac{1}{\lambda^4} \int_{- \infty}^{\infty} \left(\sum_{j = 1}^8 \frac{p_j^2 + (1-\nu) \alpha^2}{\kappa p_j -1} R_j \right. + \left. i \alpha \frac{\kappa(1-\nu)^2}{2} |\alpha| (A_1 - A_2)\right) e^{- |\alpha| x} e^{- i \alpha y} d \alpha  = 4 \pi \frac{h}{a} \frac{\tilde{M}_{xx}^0(y)}{h}
    \end{aligned}
\end{equation}

Using Equation~\eqref{eqn:ghat2}, the term involving $A_1$ and $A_2$ can be expressed directly in terms of $\hat{g}_2(\alpha)$. 
To simplify the summation term involving $R_j$, we use Equations~\eqref{eqn:VxHomBCA5}, ~\eqref{eqn:largeAlphaMj} 
and rearrange:

\begin{equation}
    \label{eqn:IntegEqK2Step3}
    \begin{aligned}
        - |\alpha|\sum_{j=1}^8 & \frac{p_j }{\kappa p_j  -1}R_j + \frac{\kappa i \alpha \left(1 - \nu\right)}{2} \left(A_1  - A_2 \right)= 0 
        \to \sum_{j=1}^8 & \frac{p_j }{\kappa p_j  -1}R_j = - i \sign(\alpha) \frac{\left(1 - \nu\right)}{2} \left(A_1  - A_2 \right)
    \end{aligned}
\end{equation}

Then, substituting Equation~\eqref{eqn:largeAlphaMj} into Equation~\eqref{eqn:MxyHomBCA5} yields:

\begin{equation}
    \label{eqn:IntegEqK2Step4}
    \begin{aligned}
        i \alpha |\alpha| \sum_{j=1}^8 & \frac{1}{\kappa p_j  -1} R_j + 2 \alpha^2 \frac{\kappa \left(1 - \nu\right)}{4} \left(A_1  - A_2 \right)= 0 
        \to \sum_{j=1}^8 & \frac{1}{\kappa p_j  -1}R_j = - i \sign(\alpha) \frac{\kappa \left(1 - \nu\right)}{2} \left(A_1  - A_2 \right)
    \end{aligned}
\end{equation}

Finally, substituting these two relations into Equation~\eqref{eqn:IntegEqK2Step2} yields the asymptotic 
expression of the integrand. 
Defining it as $\rho_2(\alpha)$, we obtain:

\begin{equation}
    \label{eqn:IntegEqK2Step5}
    \begin{aligned}
        \rho_2(\alpha) = - i \sign(\alpha) (1 - \nu) \hat{g}_2(\alpha) e^{-(|\alpha| x + i \alpha y)}
    \end{aligned}
\end{equation}

We now extract the singular component of Equation~\eqref{eqn:IntegEqK2Step1}. 
We add and subtract additional scalar constants to obtain a consistent structure in the final singular integral system.

\begin{equation}
    \label{eqn:IntegEqK2Step6}
    \begin{aligned}
        \lim_{x \to 0^+} \frac{1}{\lambda^4} &\int_{- \infty}^{\infty} \left(\sum_{j = 1}^4 \frac{p_j^2 + (1-\nu) \alpha^2}{\kappa p_j -1} R_j e^{m_j x} + \sum_{j = 5}^8 \frac{p_j^2 + (1-\nu) \alpha^2}{\kappa p_j -1} R_j e^{- m_j x} \right. + \\
        - &i \alpha \frac{\kappa(1-\nu)^2}{2} r_1 (A_1 - A_2)e^{r_1 x} - \sign(\alpha) i(1 - \nu + \nu - \nu^2) \hat{g}_2(\alpha) e^{-|\alpha| x} + \\
        + &\left.  \sign(\alpha) i(1 - \nu + \nu - \nu^2) \hat{g}_2(\alpha) e^{-|\alpha| x}\right) e^{- i \alpha y} d \alpha  = 4 \pi \frac{h}{a} \frac{\tilde{M}^0_{xx}}{h}
    \end{aligned}
\end{equation}

We isolate the asymptotic term carrying the negative sign.
Rearranging, we obtain the same integral as Equation~\eqref{eqn:IntegralByAlpha}. 
The other integral is now uniformly convergent, and we can thus apply the limit in $x$. 
Finally, we again substitute Equation~\eqref{eqn:ghat2}:

\begin{equation}
    \label{eqn:IntegEqK2Step7}
    \begin{aligned}
        \frac{1 - \nu^2}{\lambda^4}& \int_{-1}^{1} \frac{g_2(t)}{t-y} \text{d}t +  \frac{1}{\lambda^4} \int_{- \infty}^{\infty} \left(\sum_{j = 1}^8 \frac{p_j^2 + (1-\nu) \alpha^2}{\kappa p_j -1} R_j - i \alpha \kappa(1-\nu)^2 r_1 \hat{g}_2(\alpha) \right. + \\
        &\left. + i \sign(\alpha) (1 - \nu^2) \hat{g}_2(\alpha) \right) e^{- i \alpha y} d \alpha  = 4 \pi \frac{h}{a} \frac{\tilde{M}_{xx}^0(y)}{h}
    \end{aligned}
\end{equation}

We then obtain the desired equation in Fredholm form:

\begin{subequations}
    \label{eqn:DerEffKernsK2}
    \begin{align}
        \frac{1 - \nu^2}{\lambda^4} \int_{-1}^{1} & \frac{g_2 (t)} {t - y} \text{d}t + \sum_{j = 1}^5 \int_{-1}^{1} K_{2k}(t,y) g_k (t) \text{d}t = 4 \pi \frac{h}{a} \frac{\tilde{M}_{xx}^0(y)}{h} \\
        K_{22}(t,y)  &= - \frac{1}{\lambda^4} \int_{-\infty}^{\infty} \alpha \left[ w_{2k}(\alpha) + (1 - \nu^2) - \kappa (1 - \nu)^2 \alpha r_1\right]e^{i\delta}  d \alpha \\
        K_{2k}(t,y) &= - \frac{1}{\lambda^4} \int_{-\infty}^{\infty} \alpha \; w_{2k}(\alpha) e^{i\delta} d \alpha 
        \text{ with }k \neq 2 \\
        w_{2k}(\alpha) &= \sum_{j = 1}^8 \frac{p_j^2 + (1-\nu) \alpha^2}{\kappa p_j -1} B_{jk} \text{ and } \delta = \alpha (t - y) \nonumber
    \end{align}
\end{subequations}

Regarding $k = 3$, we substitute Equations~\eqref{eqn:NxyHomBCA5},~\eqref{eqn:ExpIdentity} into \eqref{eqn:sBCLoading}:

\begin{equation}
    \label{eqn:IntegEqK3Step1}
    \begin{aligned}
        \lim_{x \to 0^+} \int_{- \infty}^{\infty} i \alpha^2 \left(\sum_{j = 1}^4 K_j R_j m_j e^{m_j x} + \sum_{j = 5}^8 K_j R_j m_j e^{- m_j x} \right) e^{- i \alpha y} d \alpha  = 4 \pi \frac{\tilde{N}_{xy}^0}{h}
    \end{aligned}
\end{equation}

We substitute Equation~\eqref{eqn:largeAlphaMj}, retaining terms up until $o(\alpha^{-4})$:

\begin{equation}
    \label{eqn:IntegEqK3Step2}
    \begin{aligned}
        \lim_{\alpha \to \infty}& \lim_{x \to 0^+} \int_{- \infty}^{\infty} i \alpha |\alpha| e^{- |\alpha| x} e^{- i \alpha y} \overset{= 0 \text{ from \eqref{eqn:NxxHomBCA5}}}{\overbrace{\left(- \sum_{j = 1}^4 K_j R_j + \sum_{j = 5}^8 K_j R_j  \right)}}  d \alpha  = 4 \pi \frac{\tilde{N}_{xy}^0}{h} \\
        \lim_{\alpha \to \infty} &\lim_{x \to 0^+} \int_{- \infty}^{\infty} i \frac{|\alpha|}{2 \alpha} e^{- |\alpha| x} e^{- i \alpha y} \underset{= - g_3(\alpha)}{\underbrace{\left(- \sum_{j = 1}^4 K_j p_j R_j + \sum_{j = 5}^8 K_j p_j R_j  \right)}}  d \alpha  = 4 \pi \frac{\tilde{N}_{xy}^0}{h} 
    \end{aligned}
\end{equation}

Using Equations~\eqref{eqn:IntegEqK3Step2} and~\eqref{eqn:ghat3}, we obtain the asymptotic behavior of the 
integrand $\rho_3(\alpha)$:

\begin{equation}
    \label{eqn:IntegEqK3Step3}
    \begin{aligned}
        \rho_3(\alpha) = - \sign(\alpha) \frac{i}{2} \hat{g}_3(\alpha) e^{-(|\alpha| x +i \alpha y)}
    \end{aligned}
\end{equation}

We add and subtract $\rho_3(\alpha)$ to the integrand of Equation~\eqref{eqn:IntegEqK3Step1}, isolating the term
carrying the negative sign. 
We obtain the usual integral, analogous to Equation~\eqref{eqn:IntegralByAlpha}. 
The remaining integrals are now uniformly convergent, and thus we can carry out the limiting process inside the
integral in $\alpha$.

\begin{equation}
    \label{eqn:IntegEqK3Step4}
    \begin{aligned}
        \lim_{x \to 0^+} &\int_{-1}^1 g_3(t) \frac{1}{2} i \int_{- \infty}^{\infty} \sign(\alpha) e^{- |\alpha| x} e^{i \alpha (t- y)} \text{d}\alpha \text{d}t + \\
        +\int_{- \infty}^{\infty}& i\left[ \alpha \left(\sum_{j = 1}^4 K_j m_j R_j e^{m_j x} + \sum_{j = 5}^8 K_j m_j R_j e^{- m_j x} \right) - \frac{1}{2} \sign(\alpha) \hat{g}_3(\alpha) e^{- |\alpha| x}\right] e^{- i \alpha y} d \alpha  = 4 \pi \frac{\tilde{N}_{xy}^0}{h}
    \end{aligned}
\end{equation}

Evaluating the first integral in closed form and taking the limit inside the second integral yields:

\begin{equation}
    \label{eqn:IntegEqK3Step5}
    \begin{aligned}
        \int_{-1}^1 & \frac{g_3(t)}{t-y} \text{d}t + 
        \int_{- \infty}^{\infty} i\left[ \alpha \sum_{j = 1}^8 K_j m_j R_j  - \frac{1}{2} \sign(\alpha) \hat{g}_3(\alpha)\right] e^{- i \alpha y} d \alpha  = 4 \pi \frac{\tilde{N}_{xy}^0}{h}
    \end{aligned}
\end{equation}

The remaining Fourier integral is bounded, yielding the Fredholm equation:

\begin{subequations}
    \label{eqn:DerEffKernsK3}
    \begin{align}
        \int_{-1}^{1} & \frac{g_3 (t)} {t - y} \text{d}t + \sum_{j = 1}^5 \int_{-1}^{1} K_{3k}(t,y) g_k (t) \text{d}t = 4 \pi \frac{\tilde{N}_{xy}^0}{h} \\
        K_{33}(t,y)  &= \int_{-\infty}^{\infty} \alpha \left(w_{3k}(\alpha) - \frac{1}{2}\right) e^{i\delta} d \alpha \\
        K_{3k}(t,y) &= \int_{-\infty}^{\infty} \alpha \;w_{3k}(\alpha) e^{i\delta}d \alpha 
        \text{ with }k \neq 3 \\
        w_{3k}(\alpha) &= \sum_{j = 1}^8 B_{kj} m_j  K_j \text{ and } \delta = \alpha (t - y) \nonumber
    \end{align}
\end{subequations}

For $k = 4$, we substitute Equations~\eqref{eqn:MxyFourierDomain},~\eqref{eqn:ExpIdentity} into \eqref{eqn:sBCLoading}. 
Rearranging and using Equation~\eqref{eqn:omegaHat}, we obtain:

\begin{equation}
    \label{eqn:IntegEqK4Step1}
    \begin{aligned}
        \lim_{x \to 0^+} \frac{1 - \nu}{\lambda^4}& \int_{- \infty}^{\infty} i \alpha \left(\sum_{j = 1}^4 \frac{m_j}{\kappa p_j -1} R_j e^{m_j x} + \sum_{j = 5}^8 \frac{m_j}{\kappa p_j -1} R_j e^{- m_j x} \right. + \\
        -&\left.  \frac{\kappa(1-\nu)}{4}(\alpha^2 + r_1^2) (A_1 + A_2) e^{r_1 x}\right) e^{- i \alpha y} d \alpha  = 4 \pi \frac{h}{a} \frac{\tilde{M}_{xy}^0}{h}
    \end{aligned}
\end{equation}

We again use Equation~\eqref{eqn:largeAlphaMj} and carry out the algebra:

\begin{equation}
    \label{eqn:IntegEqK4Step2}
    \begin{aligned}
        \lim_{x \to 0^+}& \frac{1 - \nu}{\lambda^4} \int_{- \infty}^{\infty} \left[ i \alpha |\alpha|\left(\sum_{j = 1}^4 \frac{1}{\kappa p_j -1} R_j - \sum_{j = 5}^8 \frac{1}{\kappa p_j -1} R_j\right) \right.+ \\
        -& \left.  \frac{\kappa(1-\nu)}{2} \alpha^2 (A_1 + A_2)\right] e^{- |\alpha| x} e^{- i \alpha y} d \alpha  = 4 \pi \frac{h}{a} \frac{\tilde{M}_{xy}^0}{h}
    \end{aligned}
\end{equation}

The term involving $(A_1 + A_2)$ has no explicit expressions in terms of $\hat{g}_k(\alpha)$. 
Thus, we use Equations~\eqref{eqn:MxxHomBCA5},\eqref{eqn:rootsOmega}:

\begin{equation}
    \label{eqn:MxxReworkedStart}
    \begin{aligned}
       i \frac{\kappa (1 - \nu)^2 \alpha r_1}{2}(A_1 + A_2) & = \sum_{j = 1}^4 \frac{p_j + (1- \nu) \alpha^2}{\kappa p_j - 1}R_j - \sum_{j = 5}^8 \frac{p_j + (1- \nu) \alpha^2}{\kappa p_j - 1}R_j
    \end{aligned}
\end{equation}

We then use the expansion of the characteristic root $r_1$ (Equation~\eqref{eqn:largeAlphaR1}s) and separate the
summation terms:

\begin{equation}
    \label{eqn:MxxReworkedAsymptotic}
    \begin{aligned}
        - i |\alpha|& \frac{\kappa (1 - \nu)^2 \alpha}{2}(A_1 + A_2) = \sum_{j = 1}^4 \frac{p_j}{\kappa p_j - 1}R_j - \sum_{j = 5}^8 \frac{p_j}{\kappa p_j - 1}R_j \\
        &+ (1- \nu) \alpha^2 \left(\sum_{j = 1}^4 \frac{1}{\kappa p_j - 1}R_j - \sum_{j = 5}^8 \frac{1}{\kappa p_j - 1}R_j \right) \\
        &\to - i |\alpha| \frac{\kappa (1 - \nu)^2 \alpha}{2}(A_1 + A_2) = \frac{1}{1-\nu}\hat{g}_4(\alpha) \\
        &+ (1- \nu) \alpha^2 \left(\sum_{j = 1}^4 \frac{1 + \kappa p_j - \kappa p_j}{\kappa p_j - 1}R_j - \sum_{j = 5}^8 \frac{1 + \kappa p_j - \kappa p_j}{\kappa p_j - 1}R_j \right) \\
        &\to - i |\alpha| \frac{\kappa (1 - \nu)^2 \alpha}{2}(A_1 + A_2) = \frac{1}{1-\nu} \hat{g}_4(\alpha) + (1- \nu) \alpha^2 \times \\
        &\left(- \sum_{j = 1}^4 R_j + \sum_{j = 5}^8 R_j + \kappa \sum_{j = 1}^4 \frac{p_j}{\kappa p_j - 1}R_j - \sum_{j = 5}^8 \frac{p_j}{\kappa p_j - 1}R_j\right) \\
        &\to - i |\alpha| \frac{\kappa (1 - \nu)^2 \alpha}{2}(A_1 + A_2) = \frac{1}{1-\nu} \hat{g}_4(\alpha) + (1- \nu) \alpha^2 \left(- \frac{\hat{g}_5(\alpha)}{i \alpha} + \kappa \hat{g}_4(\alpha)\right)
    \end{aligned}
\end{equation}

Where we have made repeated use of Equations~\eqref{eqn:ghat4},\eqref{eqn:ghat5}. 
Substituting the result back into Equation~\eqref{eqn:IntegEqK4Step2} gives the asymptotic behavior of the 
integrand $\rho_4(\alpha)$:

\begin{equation}
    \label{eqn:IntegEqK4Step3}
    \begin{aligned}
        \rho_4(\alpha) = - i \sign(\alpha) \hat{g}_4(\alpha) e^{-(|\alpha| x + i\alpha y)}
    \end{aligned}
\end{equation}

Moving on to the singularity extraction, we consider again Equation~\eqref{eqn:IntegEqK4Step1} and add and 
subtract the asymptotic integrand $\rho_4(\alpha)$:

\begin{equation}
    \label{eqn:IntegEqK4Step4}
    \begin{aligned}
        \lim_{x \to 0^+} \frac{1 - \nu}{\lambda^4}& \int_{- \infty}^{\infty} - i \alpha \left(\sum_{j = 1}^4 \frac{m_j}{\kappa p_j -1} R_j e^{m_j x} + \sum_{j = 5}^8 \frac{m_j}{\kappa p_j -1} R_j e^{- m_j x} \right. + \\
        -&\frac{\kappa(1-\nu)}{4}(\alpha^2 + r_1^2) (A_1 + A_2) e^{r_1 x} - \sign(\alpha) i \frac{1+\nu}{2} \hat{g}_4(\alpha) e^{-|\alpha| x} + \\
        +& \left.  \sign(\alpha) i \frac{1+\nu}{2} \hat{g}_4(\alpha) e^{-|\alpha| x} \right) e^{- i \alpha y} d \alpha  = 4 \pi \frac{h}{a} \frac{\tilde{M}_{xy}^0}{h}
    \end{aligned}
\end{equation}

Once again, we separate the asymptotic term carrying the negative sign and rearrange to obtain the singular
integrand of Equation~\eqref{eqn:IntegralByAlpha}. 
The remaining integral in $\alpha$ is uniformly convergent. 
Therefore, we can apply the limit inside of it:

\begin{equation}
    \label{eqn:IntegEqK4Step5}
    \begin{aligned}
        \frac{1 - \nu^2}{\lambda^4}& \int_{-1}^{1} \frac{g_4(t)}{t-y} \text{d}t - \frac{1 - \nu}{\lambda^4} \int_{- \infty}^{\infty} \left( i \alpha \sum_{j = 1}^8 \frac{m_j}{\kappa p_j -1} R_j \right. + \\
        +&\left. \frac{\kappa(1-\nu)}{4}(\alpha^2 + r_1^2) (A_1 + A_2) - \sign(\alpha) i \frac{1+\nu}{2} \hat{g}_4(\alpha) e^{-|\alpha| x} \right) e^{- i \alpha y} d \alpha  = 4 \pi \frac{h}{a} \frac{\tilde{M}_{xy}^0}{h}
    \end{aligned}
\end{equation}

Equation~\eqref{eqn:MxxReworkedAsymptotic} expresses $(A_1 + A_2)$ in terms of $\hat{g}_k(\alpha)$ only
asymptotically.
To complete the derivation, we require an exact expression. 
We therefore return to Equations~\eqref{eqn:MxxReworkedStart},~\eqref{eqn:ghat4}, and~\eqref{eqn:ghat5}:

\begin{equation}
    \label{eqn:MxxReworked}
    \begin{aligned}
        - i r_1 \frac{\kappa (1 - \nu)^2 \alpha}{2}(A_1 + A_2) & = \sum_{j = 1}^4 \frac{p_j}{\kappa p_j - 1}R_j - \sum_{j = 5}^8 \frac{p_j}{\kappa p_j - 1}R_j \\
        &+ (1- \nu) \alpha^2 \left(\sum_{j = 1}^4 \frac{1}{\kappa p_j - 1}R_j - \sum_{j = 5}^8 \frac{1}{\kappa p_j - 1}R_j \right) \\
        \to - i r_1 \frac{\kappa (1 - \nu)^2 \alpha}{2}(A_1 + A_2) & = \frac{1}{1-\nu}\hat{g}_4(\alpha) + (1- \nu) \alpha^2 \left(- \frac{\hat{g}_5(\alpha)}{i \alpha} + \kappa \hat{g}_4(\alpha)\right)
    \end{aligned}
\end{equation}

With further algebraic manipulation we obtain:

\begin{equation}
    \label{eqn:IntegEqK4Step6}
    \begin{aligned}
        \frac{1 - \nu^2}{\lambda^4}& \int_{-1}^{1} \frac{g_4(t)}{t-y} \text{d}t - \frac{1 - \nu}{\lambda^4} \int_{- \infty}^{\infty} \left[ i \left( \alpha \sum_{j = 1}^8 \frac{m_j}{\kappa p_j -1} R_j  + \frac{1 + \kappa \alpha^2 (1 - \nu)^2}{2 \alpha r_1}\times \right. \right. \\
        & \left.(\alpha^2 + r_1^2) \hat{g}_4(\alpha) - \left.\frac{1+\nu}{2} \sign(\alpha) \hat{g}_4(\alpha)\right) + \frac{\alpha^2 + r_1^2}{2r_1} \hat{g}_5(\alpha)\right]  e^{- i \alpha y} d \alpha  = 4 \pi \frac{h}{a} \frac{\tilde{M}_{xy}^0}{h}
    \end{aligned}
\end{equation}

We finally obtain the Fredholm form of the equation:

\begin{subequations}
    \label{eqn:DerEffKernsK4}
    \begin{align}
        \frac{1 - \nu^2}{\lambda^4}\int_{-1}^{1} & \frac{g_4 (t)} {t - y} \text{d}t + \sum_{k = 1}^5 \int_{-1}^{1} K_{4k}(t,y) g_k (t) \text{d}t = 4 \pi \frac{h}{a} \frac{\tilde{M}_{xy}^0}{h} \\
        K_{44}(t,y)  &= \frac{1-\nu}{\lambda^4} \int_{-\infty}^{\infty} \left[ \alpha \;w_{44}(\alpha) - i\frac{1 + \kappa (1 - \nu) \alpha^2}{2 r_1 \alpha}(\alpha^2 + r_1^2) - i\frac{1+\nu}{2} \right] d \alpha \\
        K_{45}(t,y)  &= \frac{1-\nu}{\lambda^4} \int_{-\infty}^{\infty} \left[ w_{45}(\alpha) + \frac{\alpha^2 + r_1^2}{r_1} \right] e^{i\delta} d \alpha\\
        K_{4k}(t,y) &= \frac{1-\nu}{\lambda^4} \int_{-\infty}^{\infty} \alpha \; w_{4k}(\alpha)e^{i\delta} d \alpha 
        \text{ with }k \neq 4,5 \\
        w_{4k}(\alpha) &= \sum_{j = 1}^8 \frac{m_j}{\kappa p_j -1}B_{jk} \text{ and } \delta = \alpha (t - y) \nonumber
    \end{align}
\end{subequations}

We obtain the last equation ($k = 5$ case) by substituting 
Equations~\eqref{eqn:VxHomBCA5},~\eqref{eqn:ExpIdentity},~\eqref{eqn:rootsOmega} into \eqref{eqn:sBCLoading}:

\begin{equation}
    \label{eqn:IntegEqK5Step1}
    \begin{aligned}
        \lim_{x \to 0^+}& \int_{- \infty}^{\infty} \left(\sum_{j = 1}^4\frac{\kappa m_j p_j}{\kappa p_j -1} R_j e^{m_j x} - \frac{\kappa(1-\nu)}{2}i \alpha A_1 e^{r_1 x} \right. + \\
        &+ \left.\sum_{j = 5}^8 \frac{\kappa m_j p_j}{\kappa p_j -1} R_j e^{- m_j x} - \frac{\kappa(1-\nu)}{2}i \alpha A_2 e^{r_1 x} \right) e^{- i \alpha y} d \alpha  = 4 \pi \frac{\tilde{V}_x^0(y)}{h}
    \end{aligned}
\end{equation}

We then substitute Equations~\eqref{eqn:largeAlphaMj},~\eqref{eqn:largeAlphaR1} to obtain:

\begin{equation}
    \label{eqn:IntegEqK5Step2}
    \begin{aligned}
        \lim_{x \to 0^+} \int_{- \infty}^{\infty}& e^{- i \alpha y} e^{- \left| \alpha \right| x} \left[- |\alpha| \kappa \left(\sum_{j = 1}^4\frac{p_j}{\kappa p_j -1} R_j - \sum_{j = 5}^8 \frac{p_j}{\kappa p_j -1} R_j\right) + \right. \\
        &- \left. \frac{\kappa(1-\nu)}{2}i \alpha \left(A_1 + A_2\right)\right]  d \alpha  = 4 \pi \frac{\tilde{V}_x^0(y)}{h}
    \end{aligned}
\end{equation}

We use Equation~\eqref{eqn:ghat4} for the first term between brackets. 
The second term is obtained from Equation~\eqref{eqn:MxxReworkedAsymptotic}.

\begin{equation}
    \label{eqn:IntegEqK5Step3}
    \begin{aligned}
        \rho_5(\alpha) &= \left[ - |\alpha| \kappa (1- \nu) \hat{g}_4(\alpha) + \frac{\sign(\alpha)}{\alpha} \times \right. \\
        &\left. \left(\hat{g}_4(\alpha) - i \alpha \hat{g}_5(\alpha) + \kappa (1-\nu) \alpha^2 \hat{g}_4(\alpha)\right) \right] e^{-(|\alpha| x + \alpha y)} \\
        \to \rho_5(\alpha) &= \left( - i \sign(\alpha) \hat{g}_5(\alpha) + \frac{\sign(\alpha)}{\alpha} \hat{g}_4(\alpha)\right) e^{-(|\alpha| x + \alpha y)}
    \end{aligned}
\end{equation}

We again add and subtract $\rho_5(\alpha)$ from the integrand of Equation~\eqref{eqn:IntegEqK5Step1}. 
We extract only the singular contribution associated with $\hat{g}_5(\alpha)$, while we let the terms involving
$\hat{g}_4(\alpha)$ cancel each other out. 
We then isolate the asymptotic term carrying the positive sign.

\begin{equation}
    \label{eqn:IntegEqK5Step4}
    \begin{aligned}
        \lim_{x \to 0^+} &\int_{-1}^{1} g_5(t) i \int_{- \infty}^{\infty} \sign(\alpha) e^{- |\alpha| x} e^{i \alpha (t- y)} \text{d}\alpha \text{d}t + \\
        + &\lim_{x \to 0^+} \int_{- \infty}^{\infty} \left[ \sum_{j = 1}^4 \frac{\kappa m_j p_j}{\kappa p_j -1} R_j e^{m_j x} + \sum_{j = 5}^8 \frac{\kappa m_j p_j}{\kappa p_j -1} R_j e^{- m_j x} \right. + \\
        - &\left. \frac{\kappa(1-\nu)}{2}i \alpha \left(A_1 + A_2\right) e^{r_1 x}  + i \sign(\alpha) \hat{g}_5(\alpha) e^{-|\alpha| x }\right] e^{- i \alpha y} d \alpha  = 4 \pi \frac{\tilde{V}_x^0(y)}{h}
    \end{aligned}
\end{equation}

Once again, the first integral has the same form of Equation~\eqref{eqn:IntegralByAlpha}. 
The latter is uniformly convergent, allowing the limit in $x$ to be taken inside the integral.

\begin{equation}
    \label{eqn:IntegEqK5Step5}
    \begin{aligned}
        2\int_{-1}^{1}& \frac{g_5(t)}{t-y} \text{d}t + \int_{- \infty}^{\infty}\left[ \sum_{j = 1}^8 \frac{\kappa m_j p_j}{\kappa p_j -1} R_j \right. + \\
        &- \left. \frac{\kappa(1-\nu)}{2}i \alpha \left(A_1 + A_2\right) + i \sign(\alpha) \hat{g}_5(\alpha)\right] e^{- i \alpha y} d \alpha  = 4 \pi \frac{\tilde{V}_x^0(y)}{h}
    \end{aligned}
\end{equation}

We again use Equation~\eqref{eqn:MxxReworked} to express the term involving $(A_1 + A_2)$ in terms of $\hat{g}_4(\alpha)$ and $\hat{g}_5(\alpha)$.

\begin{equation}
    \label{eqn:IntegEqK5Step6}
    \begin{aligned}
        2\int_{-1}^{1}& \frac{g_5(t)}{t-y} \text{d}t + \int_{- \infty}^{\infty}\left[ \sum_{j = 1}^8 \frac{\kappa m_j p_j}{\kappa p_j -1} R_j e^{m_j x} + i \sign(\alpha) \hat{g}_5(\alpha) + \right. \\
        &- \left. \frac{\hat{g}_4(\alpha)}{r_1} - \frac{\alpha^2}{r_1}\left(\kappa(1-\nu) \hat{g}_4(\alpha) - i \frac{\hat{g}_5(\alpha)}{\alpha}\right)\right] e^{- i \alpha y} d \alpha  = 4 \pi \frac{\tilde{V}_x^0(y)}{h}
    \end{aligned}
\end{equation}

We finally obtain the Fredholm form of the last singular integral equation:

\begin{subequations}
    \label{eqn:DerEffKernsK5}
    \begin{align}
        2\int_{-1}^{1} & \frac{g_5 (t)} {t - y} \text{d}t + \sum_{k = 1}^5 \int_{-1}^{1} K_{5k}(t,y) g_k (t) \text{d}t = 4 \pi \frac{\tilde{V}_x^0(y)}{h} \\
        K_{54}(t,y)  &= - \int_{-\infty}^{\infty} \left[ w_{54}(\alpha) + \frac{1 + \kappa (1 - \nu) \alpha^2}{r_1}\right] e^{i\delta}  d \alpha \\
        K_{55}(t,y)  &= - \int_{-\infty}^{\infty} \left[ \left( 1 + \frac{\alpha}{r_1}\right)i +w_{55}(\alpha)\right] e^{i\delta}  d \alpha\\
        K_{5k}(t,y) &= - \int_{-\infty}^{\infty} w_{5k}(\alpha) e^{i\delta} d \alpha 
        \text{ with }k \neq 4,5 \\
        w_{5k}(\alpha) &= \kappa \sum_{j = 1}^8 \frac{m_j p_j}{\kappa p_j -1}B_{jk} \text{ and } \delta = \alpha (t - y) \nonumber
    \end{align}
\end{subequations}

\subsection{Numerical solution of the singular integral equations}
\label{app:NumericalSingularIntegral}

In this appendix we provide the details of the discretization of the singular integral equation system (Equation~\eqref{eqn:SingIntegralsMixedLoading}).
The extraction of the singular terms of Equation~\eqref{eqn:SingIntegralsMixedLoading} shown in Appendix~\ref{app:SingularTermExtraction} leads to the final system of singular integral equations:

\begin{subequations}
    \label{eqn:singularIntegralsExtracted}
    \begin{align}
            \int_{-1}^{1}& \frac{g_k (t)} {t - y} \text{d}t + \sum_{j = 1}^5 \int_{-1}^{1} K_{kj}\left(t,y\right) g_j (t) \text{d}t =  4 \pi f_k (y) \text{   for } k = 1,3 \text{ and } \left| y \right| < 1 \label{eqn:SingInt1} \\
            \frac{1 - \nu^2}{\lambda^4} \int_{-1}^{1}& \frac{g_k (t)} {t - y} \text{d}t + \sum_{j = 1}^5 \int_{-1}^{1} K_{kj}\left(t,y\right) g_j (t) \text{d}t = 4 \pi \frac{h}{a} f_k (y) \text{   for } k = 2,4 \text{ and } \left| y \right| < 1 \label{eqn:SingInt2} \\
            \int_{-1}^{1}& \frac{g_5 (t)} {t - y} \text{d}t + \sum_{j = 1}^5 \int_{-1}^{1} K_{5j}\left(t,y\right) g_j (t) \text{d}t =  2 \pi f_5 (y) \text{   for } \left| y \right| < 1 \label{eqn:SingInt5} 
    \end{align}
\end{subequations}

The integral kernels $K_{kj}(t,y) = \int_{-\infty}^{\infty}k_{kj}(\alpha)e^{i(t-y)\alpha} \text{d}\alpha$ 
are bounded and commonly known as Fredholm kernels. 
The terms $\frac{g_k (t)} {t - y} $ are known as Cauchy-type singular integrals. 
The integrands are singular, but their integral is bounded.  
They can be numerically evaluated with a dedicated Gauss-Chebyshev quadrature
scheme \citep{Erdogan:1972a}. 
The change of variables introduced in Equations~\ref{eqn:dispJumps} requires five additional
integration constants. 
Thus, for a unique solution, five more conditions are required. 
We construct these by integrating the displacement derivative jumps defined by 
Equations~\eqref{eqn:dispJumps} and enforcing the displacement compatibility of 
Equations~\eqref{eqn:sBCLoading}:

\begin{equation}
    \label{eqn:SingleValuednessEq}
    \begin{aligned}
        \int_{-1}^1& \left( g_1 (t) + \frac{\gamma_{xy}}{\lambda^2} t g_5 (t)\right) \text{d}t - \frac{\gamma_{xy}}{\lambda^2} \int_{-1}^1 \int_{-1}^t g_5 (y) \text{d}y \text{d}t = 0 \\
        \int_{-1}^1& g_k (t)  \text{d}t = 0 \text{ for } k = 2,4,5\\
        \int_{-1}^1& \left( g_3 (t) + \frac{\gamma_y}{\lambda^2} t g_5 (t)\right) \text{d}t = 0 
    \end{aligned}
\end{equation}

The system is now complete and is in a form suitable for a numerical solution using the 
method of \citet{Erdogan:1972a,Yahsi:1981}. 
We decompose the solution into a bounded unknown solution $h_k(t)$ and a singular density function with Equation~\ref{eqn:shapeFuncDecomposition}.
We then discretize the singular integral system using Gauss-Chebyshev quadrature \citep{Abramowitz:1968}:

\begin{subequations}
    \label{eqn:GCQuad}
    \begin{align}
        t_j &= \cos \left(\pi \frac{j - 1}{N - 1}\right) \text{ } j = 1, \dots, N \text{ and } y_i = \cos \left(\pi \frac{2i - 1}{2(N - 1)}\right) \text{ } i = 1, \dots, N-1 \label{eqn:rowPointGC} \\
        w_j &= \frac{\pi}{N - 1} \text{ } j = 2, \dots, N - 1 \text{ and } w_1 = w_N = \frac{\pi}{2(N - 1)} \label{eqn:weight1nGC} 
    \end{align}
\end{subequations}

The discretization produces $5N$ algebraic equations, which we rearrange in a 
$5N \times 5N$ linear system. 
Solving this system yields the five discretized, bounded functions $h_k(\bm{t})$.

\begin{equation}
    \label{eqn:DiscsingEqSysArbMain}
    \begin{aligned}
        \begin{bmatrix} 
                                    \mathbf{K}_{11}(\bm{\Theta}) + \frac{1}{\bm{\Theta}}& & \cdots & & \mathbf{K}_{15}(\bm{\Theta}) \\
                                     & \mathbf{K}_{22}(\bm{\Theta}) + \frac{1 - \nu^2}{\lambda^4 \bm{\Theta}} &  & &  \\
                                    \vdots & & \mathbf{K}_{33}(\bm{\Theta}) + \frac{1}{\bm{\Theta}} & & \vdots \\
                                     & &  & \mathbf{K}_{44}(\bm{\Theta}) + \frac{1 - \nu^2}{\lambda^4 \bm{\Theta}} &  \\
                                    \mathbf{K}_{51}(\bm{\Theta}) & &  \cdots & & \mathbf{K}_{55}(\bm{\Theta}) + \frac{1}{\bm{\Theta}} \\
                                    [1,\dots,1] &  & \dots & & \frac{\gamma_{xy}}{\lambda^2} \underline{\tilde{t}}\\
                                    & [1,\dots,1] & & & \\
                                    \vdots & & [1,\dots,1] & & \frac{\gamma_y}{\lambda^2} [t_1, \dots,t_N]\\
                                    & & & [1,\dots,1] & \\
                                    [0,\dots,0]& & \dots & & [1,\dots,1]
        \end{bmatrix}\!\!
        \begin{bmatrix}
            h_1(t_1)\\
            \vdots \\
            h_1(t_N) \\
            h_2(t_1)\\
            \vdots \\
            h_4(t_N) \\
            h_5(t_1)\\
            \vdots \\
            h_5(t_N) \\
        \end{bmatrix} \!\!=\!
        \begin{bmatrix}
            \frac{\tilde{N}_{xx}^0(y_1)}{h}\\
            \vdots \\
            \frac{\tilde{N}_{xx}^0(y_{N-1})}{h}\\
            \vdots \\
            \frac{\tilde{V}_{x}^0(y_1)}{2h}\\
            \vdots \\
            \frac{\tilde{V}_{x}^0(y_{N-1})}{2h}\\
            0 \\
            \vdots \\
            0
        \end{bmatrix}
    \end{aligned}
\end{equation}

We define the matrix $\bm{\Theta}$ and the vector $\bm{\tilde{t}}$ as follows:

\begin{equation}
    \label{eqn:DiscSingEqDetails}
    \begin{aligned}
        \Theta_{ij} &= t_j - y_i; \;\; \tilde{t}_j = t_j - \sum_{k = j}^N w_k
    \end{aligned}
\end{equation}

The resulting linear system can be solved numerically using standard linear solvers. 
The most cumbersome numerical task is instead the computation of the Fredholm kernels $\mathbf{K}(\bm{\Theta})$. 
These are improper, oscillatory integrals with a complex-valued integrand. 
We provide more details of their numerical computation and convergence in Appendix~\ref{appendixB}.

\subsection{The displacement fields}
\label{app:DisplacementFields}

In this appendix we derive the expressions required to compute the displacements
$u(x,y),v(x,y),w(x,y),\beta_x(x,y),\beta_y(x,y)$.
These expression constitute our original contribution, since they are not present in literature. 
We discuss the numerical convergence of these expressions and the associated computational challenges in Appendix~\ref{ssec:displacementKernels}.

\subsubsection{Parity of numerically computed functions}

The displacement fields are computed in terms of Fourier integrals. 
Thus, we first establish the parity properties of the complex-valued functions $B_{jk}(\alpha)$,
$p_j(\alpha)$,$K_j(\alpha)$,$U_j(\alpha)$,$V_j(\alpha)$,$B_k(\alpha)$,$C_k(\alpha)$ and of the characteristic
roots $r_i(\alpha)$ and $m_j(\alpha)$. 
Since most of these functions are not known in closed form, we have found their properties through numerical inspection.

\begin{table}[pos=ht]
    \centering
    \begin{tabular}{|l|l|l|}
    \hline
    \textbf{Function}                   & \textbf{Real} & \textbf{Imaginary} \\ \hline
    $r_i(\alpha)$                       & Even          & -                  \\ \hline
    $m_j(\alpha)$                       & Even          & Odd                \\ \hline
    $p_j(\alpha),K_j(\alpha),U_j(\alpha),V_j(\alpha)$  & Even          & Odd                \\ \hline
    $B_{jk}(\alpha),B_k(\alpha),C_k(\alpha)$                    & Odd               & Even \\ \hline
    \end{tabular}
\caption{Collection of the odd/even properties of the various function of $\alpha$ needed to compute the displacements.}
\label{tab:oddEvenProps}
\end{table}

\subsubsection{Out-of-plane displacement field}

We start from the out-of-plane displacement $w(x,y)$. 
Substituting Equations~\eqref{eqn:newToOldUnknownsMain},~\eqref{eqn:gBackTranform} into Equation~\eqref{eqn:outOfPlaneDispFourier} yields:

\begin{equation}
    \label{eqn:wMMfuncGj}
    \begin{aligned}
        w(x^+,t) &= \frac{1}{2 \pi} \int_{-\infty}^{\infty}
        \sum_{j = 1}^4 \int_{-1}^1 \left( i \sum_k B_{jk}(\alpha) g_k(t) e^{m_j(\alpha) x + i \alpha (t-y)} \text{d}t \right) d \alpha \text{ for } x > 0
    \end{aligned}
\end{equation} 

We show only the $x>0$ expression, since that on the other side of the crack is trivially obtained by changing
the indices of the summation from $1,4$ to $5,8$. 
Noting that the characteristic roots $m_{j}(\alpha)$ are complex, with $\Re(m_j) < 0$ for $j \leq 4$, we
rearrange the expression by separating the exponential terms:

\begin{equation}
    \label{eqn:wpFull2ndStep}
    \begin{aligned}
        w(x^+,y) &= \frac{i}{2 \pi} \int_{-1}^1 \sum_k g_k(t) \underbrace{\int_{-\infty}^{\infty} \sum_{j = 1}^4 e^{\Re(m_j(\alpha)) x} B_{jk}(\alpha) e^{i(\Im(m_j(\alpha)) x + \alpha (t-y))}  d \alpha }_{I_k^w(x^+,y,t)}\text{d}t  
    \end{aligned}
\end{equation}

Using the parity properties listed in Table~\ref{tab:oddEvenProps}, the Fourier integral reduces to:

\begin{subequations}
    \label{eqn:reducedFourierW}
    \begin{align}
        I_k^w &= 2 i \int_{0}^{\infty} \left[\Im( \dots ) \cos (\dots) + \Re(\dots) \sin ( \dots ) \right] d \alpha 
    \end{align}
\end{subequations} 

We apply the Gauss-Chebyshev quadrature (Equation~\eqref{eqn:GCQuad}) and use 
Equation~\eqref{eqn:shapeFuncDecomposition} to obtain the algebraic representation of the displacement field:

\begin{subequations}
    \label{eqn:wpFullFinal}
    \begin{align}
        w(x^+,y) &= - \frac{1}{\pi} \left(\bm{w} \cdot \sum_{k = 1}^5 \mathbf{I}^w_k (t,y) \bm{h}_k (t) \right) \\
        \mathbf{I}^w_k(x^+,t,y) &= \int_{0}^{\infty} \left[\Im(f^w_k(x,\alpha)) \cos( \Im(m_j(\alpha)) + \alpha \bm{\Theta}) + \Re(f^w_k(x,\alpha)) \sin(\Im(m_j(\alpha)) +  \alpha \bm{\Theta}) \right] d \alpha \\
        f^w_k(x,\alpha) &= \sum_{j = 1}^4 e^{\Re(m_j(\alpha))x}  B_{jk}(\alpha)
    \end{align}
\end{subequations} 

Here, $\Theta_{ij} = t_j - y_i$. 
The crack lips displacements are obtained by taking the limit for $x \to 0^+$ of Equation \eqref{eqn:wpFullFinal}. 
These expressions are particularly convenient because the integral kernels contain only the linear oscillator
factor $e^{i\alpha \bm{\Theta}}$, instead of the non-linear oscillator of Equation~\eqref{eqn:wpFullFinal}. 
Consequently, they are significantly simpler to compute.

\begin{subequations}
    \label{eqn:discWp}
    \begin{align}
        w(0^+,y) &= - \frac{1}{\pi} \left(\bm{w} \cdot \sum_{k = 1}^5 \mathbf{I}^w_k (t,y) \bm{h}_k (t) \right) \label{eqn:discW}\\
        \mathbf{I}^w_k(t,y) &= \int_{0}^{\infty} \left( \Im(f^w_k(\alpha)) \cos( \alpha \bm{\Theta}) + \Re(f^w_k(\alpha)) \sin( \alpha \bm{\Theta}) \right) d \alpha \\
        f^w_k(\alpha) &= \sum_{j = 1}^4 B_{jk}(\alpha)
    \end{align}
\end{subequations} 

Here, $\bm{w}$ denotes the vector of Gauss-Chebyshev quadrature weights, while $\bm{h}_k(t)$ denotes the bounded
part of the unknown functions ${h}_k(t)$, evaluated at points $t_j$.

\subsubsection{Cross-section rotation fields}

The cross-section rotation fields are obtained by substituting Equations~\eqref{eqn:newToOldUnknownsMain} 
and~\eqref{eqn:gBackTranform} into Equation~\eqref{eqn:BetaiFourier}. 
The resulting expression is rearranged following the same procedure used for the out-of-plane displacement. 
Applying the parity properties listed in Table~\ref{tab:oddEvenProps} reduces the Fourier integrals to:

\begin{subequations}
    \label{eqn:reducedFourierBeta}
    \begin{align}
        I_k^x (t,y) &= 2 \int_0^\infty (\Re(\dots) \cos (\dots) - \Im(\dots) \sin (\dots)) d \alpha \\
        I_k^y (t,y) &= 2 \int_0^\infty (\Re(\dots) \cos (\dots) - \Im(\dots) \sin (\dots)) d \alpha
    \end{align}
\end{subequations} 

Applying the Gauss-Chebyshev quadrature rule (Equation~\eqref{eqn:GCQuad}) together with the decomposition of
$g_k(t)$ given by Equation~\eqref{eqn:shapeFuncDecomposition} yields:

\begin{subequations}
    \label{eqn:BetaXpFull}
    \begin{align}
        \beta_x(x^+,y) &= \frac{1}{\pi} \left( \bm{w} \cdot \left(\sum_{k=1}^5 \mathbf{I}^x_k(t,y) \bm{h}_k(t)\right)\right)\\
        \mathbf{I}^x_k(x^+,t,y) &= \int_{0}^{\infty} \left[ \Re(f^x_k(x^+,\alpha)) \cos(\Im(m_j(\alpha)) + \alpha \bm{\Theta}) \right. + \nonumber \\ 
        & \left. - \Im(f^x_k(x^+,\alpha)) \sin(\Im(m_j(\alpha)) + \alpha \bm{\Theta}) \right] d \alpha \\
        f^x_k(x,\alpha) &= i \left(\sum_{j=1}^4 e^{\Re(m_j(\alpha))x} \frac{m_j(\alpha)}{\kappa p_j(\alpha) -1} B_{jk}(\alpha) - \frac{\kappa(1-\nu)}{2} e^{r_1(\alpha) x} \alpha B_k(\alpha)\right)
    \end{align}
\end{subequations} 

\begin{subequations}
    \label{eqn:BetaYpFull}
    \begin{align}
        \beta_y(x^+,y) &= \frac{1}{\pi} \left( \bm{w} \cdot \left(\sum_{k=1}^5 \mathbf{I}^y_k(x^+,t,y) \bm{h}_k(t)\right)\right) \\
        f^y_k(x,\alpha) &= \sum_{j=1}^4 e^{\Re(m_j(\alpha))x} \frac{\alpha}{\kappa p_j(\alpha) -1} B_{jk}(\alpha) - \frac{\kappa(1-\nu)}{2} r_1(\alpha) e^{r_1(\alpha) x}B_k(\alpha)
    \end{align}
\end{subequations} 

The crack lips rotations are obtained by taking the limit $x \to 0^+$:

\begin{subequations}
    \label{eqn:discBetaXp}
    \begin{align}
        \beta_x(0^+,y) &= \frac{1}{\pi} \left( \bm{w} \cdot \left(\sum_{k=1}^5 \mathbf{I}^x_k(t,y) \bm{h}_k(t)\right)\right)\\
        \mathbf{I}^x_k(t,y) &= \int_{0}^{\infty} \left( \Re(f^x_k(\alpha)) \cos( \alpha \bm{\Theta}) - \Im(f^x_k(\alpha)) \sin( \alpha \bm{\Theta}) \right) d \alpha \\
        f_k^x(\alpha) &= i \left(\sum_{j=1}^4 \frac{m_j(\alpha)}{\kappa p_j(\alpha) -1} B_{jk}(\alpha) - \frac{\kappa(1-\nu)}{2} \alpha B_k(\alpha)\right)
    \end{align}
\end{subequations} 

\begin{subequations}
    \label{eqn:discBetaYp}
    \begin{align}
        \beta_y(0^+,y) &= \frac{1}{\pi} \left( \bm{w} \cdot \left(\sum_{k=1}^5 \mathbf{I}^y_k(t,y) \bm{h}_k(t)\right)\right) \\
        \mathbf{I}^y_k(t,y) &= \int_{0}^{\infty} \left( \Re(f^y_k(\alpha)) \cos( \alpha \bm{\Theta}) - \Im(f^y_k(\alpha)) \sin( \alpha \bm{\Theta}) \right) d \alpha \\
        f_k^y(\alpha) &= \sum_{j=1}^4 \frac{\alpha}{\kappa p_j(\alpha) -1} B_{jk}(\alpha) - \frac{\kappa(1-\nu)}{2} r_1(\alpha) B_{k}(\alpha)
    \end{align}
\end{subequations} 

\subsubsection{In-plane displacement fields}

We obtain the in-plane displacement $u(x,y)$ by substituting Equations~\eqref{eqn:newToOldUnknownsMain} 
and~\eqref{eqn:gBackTranform} into Equation~\eqref{eqn:uFourier}.
Applying the parity properties listed in Table~\ref{tab:oddEvenProps} we obtain:

\begin{equation}
    \label{eqn:parityU}
    \begin{aligned}
        I_k^u (t,y) &= 2 i \int_0^\infty (\Im(\dots) \cos (\dots) + \Re(\dots) \sin (\dots)) d \alpha 
    \end{aligned}
\end{equation} 

We therefore obtain the expression for $u(x,y)$:

\begin{subequations}
    \label{eqn:UpFull}
    \begin{align}
        u(x^+,y) =& - \frac{1}{\pi} \left(\bm{w} \cdot \sum_{k = 1}^5 \mathbf{I}^u_k (x^+,t,y) \bm{h}_k (t) \right) \\
        \mathbf{I}^u_k(x^+t,y) =& \int_{0}^{\infty} \left[ \Im(f^u_k(x^+,y,\alpha)) \cos(\Im(m_j(\alpha)) + \alpha \bm{\Theta}) \right. + \nonumber \\
        &+ \left. \Re(f^u_k(x^+,y,\alpha)) \sin(\Im(m_j(\alpha)) +  \alpha \bm{\Theta}) \right] d \alpha \\
        f^u_k(x,y,\alpha) =& \sum_{j = 1}^4 e^{\Re(m_j(\alpha))x} U_j(y,\alpha) B_{jk}(\alpha) \label{eqn:summU}
    \end{align}
\end{subequations} 

The $U_j (y,\alpha)$ term is given by Equation~\eqref{eqn:uFourier}. 
Taking the limit $x \to 0^+$ yields the crack-lip displacement:

\begin{subequations}
    \label{eqn:discUp}
    \begin{align}
        u(0^+,y) &= - \frac{1}{\pi} \left(\bm{w} \cdot \sum_{k = 1}^5 \mathbf{I}^u_k (t,y) \bm{h}_k (t) \right) \label{eqn:discU}\\
        \mathbf{I}^w_k(t,y) &= \int_{0}^{\infty} \left( \Im(f^u_k(y,\alpha)) \cos( \alpha \bm{\Theta}) + \Re(f^u_k(y,\alpha)) \sin( \alpha \bm{\Theta}) \right) d \alpha \\
        f^u_k(y,\alpha) &= \sum_{j = 1}^4 U_j(y,\alpha) B_{jk}(\alpha)
    \end{align}
\end{subequations} 

The second in-plane displacement $v(x,y)$, is obtain by substituting using 
Equations~\eqref{eqn:newToOldUnknownsMain} and~\eqref{eqn:gBackTranform} in Equation~\eqref{eqn:vFourier}. 
Applying the parity properties listed in Table~\ref{tab:oddEvenProps} gives:

\begin{equation}
    \label{eqn:parityV}
    \begin{aligned}
        I_k^v (t,y) &= 2 i \int_0^\infty (\Im(\dots) \cos (\dots) + \Re(\dots) \sin (\dots)) d \alpha 
    \end{aligned}
\end{equation} 

The final expression is:

\begin{subequations}
    \label{eqn:VpFull}
    \begin{align}
        v(x^+,y) =& - \frac{1}{\pi} \left(\bm{w} \cdot \sum_{k = 1}^5 \mathbf{I}^v_k (x^+,t,y) \bm{h}_k (t) \right) \label{eqn:discVFull}\\
        \mathbf{I}^v_k(t,y) =& \int_{0}^{\infty} \left[ \Im(f^v_k(x^+,y,\alpha)) \cos(\Im(m_j(\alpha)) + \alpha \bm{\Theta}) \right. + \nonumber \\
        & + \left. \Re(f^v_k(x^+,y,\alpha)) \sin(\Im(m_j(\alpha)) + \alpha \bm{\Theta}) \right] d \alpha \\
        f^v_k(x,y,\alpha) =& \sum_{j = 1}^4 e^{\Re(m_j(\alpha))x} V_j(y,\alpha)  B_{jk}(\alpha) 
    \end{align}
\end{subequations} 

Here, $V_j (y,\alpha)$ is given by Equation~\eqref{eqn:vFourier}. 
Taking the limit $x \to 0^+$ yields once again the crack-lip displacement:

\begin{subequations}
    \label{eqn:discVp}
    \begin{align}
        v(0^+,y) &= - \frac{1}{\pi} \left(\bm{w} \cdot \sum_{k = 1}^5 \mathbf{I}^v_k (t,y) \bm{h}_k (t) \right) \label{eqn:discV}\\
        \mathbf{I}^v_k(t,y) &= \int_{0}^{\infty} \left( \Im(f^v_k(y,\alpha)) \cos( \alpha \bm{\Theta}) + \Re(f^v_k(y,\alpha)) \sin( \alpha \bm{\Theta}) \right) d \alpha \\
        f^v_k(y,\alpha) &= \sum_{j = 1}^4 V_j(y,\alpha) B_{jk}(\alpha)
    \end{align}
\end{subequations} 

\subsection{The stress fields and stress intensity factors}
\label{app:StressFields}

In this appendix we derive the asymptotic crack tip stress fields and the corresponding stress intensity factors
as function of the $h_k(t)$ at $|t| = 1$.

\subsubsection{The (asymptotic) stress fields}

The derivation of the asymptotic crack tip fields is presented in multiple earlier works 
\citep{Yuceoglu:1973,Krenk:1978}.
We first transform the stress-stress resultant relations of Reissner's shell theory
(Equations~\eqref{eqn:stressFromResultants}) in the crack reference frame $Oxyz$.
We then use Equations~\eqref{eqn:membraneResultantsFourier},
~\eqref{eqn:bendingResultantsFourier},~\eqref{eqn:transverseResultantsFourier} and~\eqref{eqn:newToOldUnknownsMain} to
express the stress field in terms of the functions $h_k(t)$.
All resulting expressions contain singular, oscillating integrals of the form of Equation~\ref{eqn:singIntRes}.
Following Appendix B of \citet{Krenk:1978}, these can be expressed asymptotically as:

\begin{subequations}
    \label{eqn:asymResultantFourierInit}
    \begin{align}
        I(\alpha) &= \int_{-1}^1 \frac{h (\tau)}{\sqrt{1 - \tau^2}} e^{i \alpha \tau} d \tau  \label{eqn:singIntRes}\\
        I(\alpha) & \approx \sqrt{\frac{\pi}{2 \left| \alpha \right| }}  \left[ h(1) e^{i \left( \alpha - \frac{\pi \alpha}{4 \left| \alpha \right|}\right)} + h(-1) e^{ - i \left( \alpha - \frac{\pi \alpha}{4 \left| \alpha \right|}\right)} + o\left(\frac{1}{\left| \alpha \right|}\right) \right] \text{ for } \left| \alpha \right| \to \infty \label{eqn:krenkAsym}
    \end{align}
\end{subequations}

Following the procedure described by \citet{Yahsi:1981}, we use Equation~\ref{eqn:krenkAsym} and~\ref{eqn:largeAlphaMj} to obtain the leading term of the asymptotic integral expressions of the stress resultants in the crack reference frame:

\begin{subequations}
    \label{eqn:asymResultantFourier}
    \begin{align}
        N_{xx}(x,y) &=\frac{1}{4 \sqrt{2\pi}} \left(x h_3(1)I_2(x,y) + h_1(1) I_1(x,y) \right)\\
        N_{yy}(x,y) &=\frac{1}{4 \sqrt{2\pi}} \left( h_3(1)I_4(x,y) + h_1(1) I_3(x,y) \right) \\
        N_{xy}(x,y) &=\frac{1}{4 \sqrt{2\pi}} \left(x h_3(1)I_4(x,y) + x h_1(1) I_2(x,y) \right) \\
        M_{xx}(x,y) &=\frac{h}{12a}\frac{1}{4 \sqrt{2\pi}} \left( h_2(1)I_1(x,y) + x h_4(1) I_2(x,y) \right)\\
        M_{yy}(x,y) &=\frac{h}{12a}\frac{1}{4 \sqrt{2\pi}} \left( h_2(1)I_3(x,y) + h_4(1) I_4(x,y) \right)\\
        M_{xy}(x,y) &=\frac{h}{12a}\frac{1}{4 \sqrt{2\pi}} \left( |x|h_2(1)I_2(x,y) + x h_4(1) I_3(x,y) \right)\\
        V_{x}(x,y) &=\frac{1}{2 \sqrt{2\pi}} x h_5(1) I_5(x,y) \\
        V_{y}(x,y) &=\frac{1}{2 \sqrt{2\pi}} x h_5(1) I_4(x,y)
    \end{align}
\end{subequations}

Here, $I_k(x,y)$ ($k = 1,2,3,4,5$) are improper integrals given by:

\begin{subequations}
    \label{eqn:asymResultantIntegrals}
    \begin{align}
        \Upsilon & = \alpha(1 - y)- \frac{\pi}{4} \\
        I_1(x,y) &= \int_0^{\infty}\frac{1 + \alpha |x|}{\sqrt{\alpha}}e^{-\alpha |x|}\sin \left(\Upsilon \right) d \alpha \\
        I_2(x,y) &=\int_0^{\infty}\sqrt{\alpha}e^{-\alpha |x|}\cos \left(\Upsilon \right) d \alpha \\
        I_3(x,y) & =\int_0^{\infty}\frac{1 - \alpha |x|}{\sqrt{\alpha}}e^{-\alpha |x|}\sin \left(\Upsilon \right) d \alpha \\
        I_4(x,y) &=\int_0^{\infty}\frac{1}{\sqrt{\alpha}}e^{-\alpha |x|}\cos \left(\Upsilon \right) d \alpha \\
        I_5(x,y) &=\int_0^{\infty}\frac{1}{\sqrt{\alpha}}e^{-\alpha |x|}\sin \left(\Upsilon \right) d \alpha 
    \end{align}
\end{subequations}

At this point, we introduce the usual crack tip polar reference system:

\begin{equation}
    \label{eqn:referencePolar}
    \begin{aligned}
        x &= r \sin \theta \\
        y &= 1 + r \cos \theta \\
        z &=  z
    \end{aligned}
\end{equation}

We transform Equations~\ref{eqn:asymResultantIntegrals} into the crack tip polar reference system and use the
integral identities given in Equation 3.944 of  \citet{Gradshteyn:2014} to obtain the leading term of the
asymptotic crack tip resultants in polar coordinates:

\begin{subequations}
    \label{eqn:asymResultant}
    \begin{align}
        N_{xx}(r,\theta) &=- \frac{1}{4 \sqrt{2 \pi r}} \left(h_3(1)g_{xx}(\theta) + h_1(1) f_{xx}(\theta) \right)\\
        N_{yy}(r, \theta) &=- \frac{1}{4 \sqrt{2 \pi r}} \left( h_3(1)f_{yy}(\theta) + h_1(1) g_{yy}(\theta) \right) \\
        N_{xy}(r, \theta) &=- \frac{1}{4 \sqrt{2 \pi r}} \left(\ h_3(1)g_{yy}(\theta) + h_1(1) f_{xx}(\theta) \right) \\
        M_{xx}(r, \theta) &=- \frac{h}{12a}\frac{1}{4 \sqrt{2 \pi r}} \left( h_2(1)f_{xx}(\theta) + h_4(1) g_{xx}(\theta) \right)\\
        M_{yy}(r, \theta) &=- \frac{h}{12a}\frac{1}{4 \sqrt{2 \pi r}} \left( h_2(1)g_{yy}(\theta) + h_4(1) f_{yy}(\theta) \right)\\
        M_{xy}(r, \theta) &=- \frac{h}{12a}\frac{1}{4 \sqrt{2 \pi r}} \left( h_2(1)f_{xx}(\theta) + h_4(1) g_{yy}(\theta) \right)\\
        V_{x}(r, \theta) &=\frac{1}{2 \sqrt{2 \pi r}} h_5(1) \cos \frac{\theta}{2} \\
        V_{y}(r, \theta) &=\frac{1}{2 \sqrt{2 \pi r}} h_5(1) \sin \frac{\theta}{2}
    \end{align}
\end{subequations}

Here, the angular functions $f_{ii}(\theta)$, $g_{ii}(\theta)$ are defined as:

\begin{subequations}
    \label{eqn:asymResultantAngFuncs}
    \begin{align}
        f_{xx}(\theta) &= - \frac{1}{4} \sin \frac{\theta}{2} + \frac{1}{4} \sin \frac{5\theta}{2} \\
        g_{xx}(\theta) &= \frac{5}{4} \cos \frac{\theta}{2} - \frac{1}{4} \cos \frac{5\theta}{2} \\
        f_{yy}(\theta) &= - \frac{7}{4} \sin \frac{\theta}{2} - \frac{1}{4} \sin \frac{5\theta}{2} \\
        g_{yy}(\theta) &= \frac{3}{4} \cos \frac{\theta}{2} + \frac{1}{4} \cos \frac{5\theta}{2} 
    \end{align}
\end{subequations}

Finally, using Equation~\ref{eqn:stressFromResultants}, one can show that these expressions possess exactly the
same angular dependence as the classical plane stress crack tip fields in isotropic, flat plates \citep{anderson2005}.

\subsubsection{Stress intensity factors}

We compute the stress intensity factors as functions of $h_k(t)$ by assuming self-similar crack growth and using
the standard definitions \citep{anderson2005}. 

\begin{subequations}
    \label{eqn:SIFSasLimit}
    \begin{align}
        K_{I}(z) & = \lim_{r \to 0} \sqrt{2 \pi r} \; \sigma_{xx}(r, 0, z )\label{eqn:sifI} \\
        K_{II}(z) & = \lim_{r \to 0} \sqrt{2 \pi r} \; \sigma_{xy}(r, 0, z)\label{eqn:sifII} \\
        K_{III}(z) & = \lim_{r \to 0}\sqrt{2 \pi r} \; \sigma_{xz}(r, 0, z )\label{eqn:sifIII}
    \end{align}
\end{subequations}

We then use Equations~\eqref{eqn:stressFromResultants},~\eqref{eqn:asymResultant} to obtain 
Equation~\eqref{eqn:SIFSfromhk}.
We next consider the integral system from 
Equations~\eqref{eqn:singularIntegralsExtracted},~\eqref{eqn:SingleValuednessEq} and~\eqref{eqn:DiscsingEqSysArbMain}.
Writing the system in matrix form, it becomes evident that the loading vector can be decomposed in five
independent contributions, one for each stress resultant:

\begin{equation}
    \label{eqn:linearizedFredholmLoads}
    \begin{aligned}
        \bm{K}(t,y) & \bm{h}(t) = \bm{f} =  \bm{f}_m + \dots + \bm{f}_v \\
        \bm{f}_m & = \tilde{N}_{xx}^0(y) \neq 0; \; \tilde{M}_{xx}^0(y), \dots, \tilde{V}_{x}^0(y) = 0  \\
        \dots& \\
        \bm{f}_v & = \tilde{V}_{x}^0(y) \neq 0; \; \tilde{N}_{xx}^0(y), \dots, \tilde{M}_{xy}^0(y) = 0
    \end{aligned}
\end{equation}

By linearity, the superposition principle allows the functions $h_k(t)$ to be decomposed into five corresponding
contributions, yielding five independent systems.

\begin{equation}
    \label{eqn:linearizedFredholmSystems}
    \begin{aligned}
        \bm{K}(t,y) & (\bm{h}_m(t) + \dots + \bm{h}_v(t)) = \bm{f}_m + \dots + \bm{f}_v \\
        \bm{K}(t,y) & \bm{h}_m(t) = \bm{f}_m(y) \\
        \dots& \\
        \bm{K}(t,y) & \bm{h}_v(t) = \bm{f}_v(y)
    \end{aligned}
\end{equation}

Using Equations~\eqref{eqn:referenceSIFS} and~\eqref{eqn:referenceSIFS}, we obtain the contribution of each 
loading component. 
The latter equation introduces the assumption of uniformly distributed stresses along the crack length. 
Superposing the five contributions yields Equation~\eqref{eqn:SIFSTradMain}.

\subsection{Energy release rate}
\label{app:ERR}

In this appendix, we derive the expressions for the energy release rate. 
By definition, the energy release rate is given by:

$$G = d(U-V)/dA$$

Here, $U$ is the potential energy, $V$ the work done by external loads and $dA\approx 2h da$ is the increase in
crack surface area. 
We assume self-similar crack growth and sufficient membrane tension to prevent crack closure. 
Considering an infinitesimal increase in crack length $da \ll 1$, the external work associated with crack growth
can be expressed as:

\begin{equation}
    \label{eqn:extWorkEnrgRelrate}
    \begin{aligned}
        dV = &-  \frac{1}{2} \int_{- \frac{h}{2}}^{\frac{h}{2}} \int_0^{da/a}\sum_{j=x,y,z} \sigma_{xj}(0,y,z) \llbracket u_j(0,y - da/a,z) \rrbracket \;\text{d}y \; \text{d}z
   \end{aligned}
\end{equation}

where the displacement jump is defined as:

$$\llbracket u_j(0,y - da/a,z) \rrbracket = \lim_{x \to 0^+} u_j(x,y - da/a,z) - \lim_{x \to 0^-} u_j(x,y - da/a,z)$$

Only the stress components associated with the stress intensity factors perform work. 
Using the standard asymptotic expressions for the crack tip displacements \citep{anderson2005}, we obtain:

\begin{subequations}
    \begin{align}
        u_{x}^{I}(r,\theta,z)
        & = \sqrt{\dfrac{r}{2\pi}}
            \sin\left(\frac{\theta}{2}\right) \frac{K_{I}(z)}{2\mu}
            \left[
                -2\cos^{2}\left(\frac{\theta}{2}\right)
                +\frac{3-\nu}{\nu+1}+1
            \right] = \nonumber \\
        & \Rightarrow u_x^{I}(0^{\pm}, y - da/a,z) \approx \pm 2 \frac{K_I(z)}{E}\sqrt{2\pi(1 - y + da/a)} \\
        u_{y}^{I}(r,\theta,z)
        &= \sqrt{\dfrac{r}{2\pi}}
            \cos\left(\frac{\theta}{2}\right)\frac{K_{I}(z)}{2\mu}
            \left[
                2\sin^{2}\left(\frac{\theta}{2}\right)
                +\frac{3-\nu}{\nu+1}-1
            \right]\nonumber \\
        & \Rightarrow u_y^{I}(0^{\pm}, y - da/a,z) \approx 0 \\
        u_{x}^{II}(r,\theta,z)
        &= -\sqrt{\dfrac{r}{2\pi}}
            \cos\left(\frac{\theta}{2}\right)\frac{K_{II}(z)}{2\mu}
            \left[
                -2\sin^{2}\left(\frac{\theta}{2}\right)
                +\frac{3-\nu}{\nu+1}-1
            \right] \nonumber \\
        & \Rightarrow u_x^{II}(0^{\pm}, y - da/a,z) \approx 0 \\
        u_{y}^{II}(r,\theta,z)
        &= \sqrt{\dfrac{r}{2\pi}}
            \sin\left(\frac{\theta}{2}\right)\frac{K_{II}(z)}{2\mu}
            \left[
                2\cos^{2}\left(\frac{\theta}{2}\right)
                +\frac{3-\nu}{\nu+1}+1
            \right]\nonumber \\
        & \Rightarrow u_y^{II}(0^{\pm}, y - da/a,z) = \pm 2 \frac{K_{II}(z)}{E}\sqrt{2\pi(1 - y + da/a)} \\
            u_{z}^{III}(r,\theta,z)
        &= 2\sqrt{\dfrac{r}{2\pi}}
            \sin\left(\frac{\theta}{2}\right)\frac{K_{III}(z)}{\mu} \nonumber \\
        & \Rightarrow u_z^{III}(0^{\pm}, y - da/a,z) = \pm 2 \frac{K_{III}(z)}{\mu}\sqrt{2\pi(1 - y + da/a)} 
    \end{align}
\end{subequations}

These expressions show that the crack tip displacements are equal in magnitude and opposite in sign across the
crack plane:

$$u_i(0^+,y,z) \approx - u_i(0^-,y,z)$$

Using the definitions of stress intensity factors (Equation~\eqref{eqn:SIFSasLimit}) and the displacement jumps, 
we obtain:

\begin{subequations}
    \label{eqn:EnrgInteg}
    \begin{align}
        (\sigma_{xx},\sigma_{xy},\sigma_{xz})(0,y\;a/\sqrt{c},z) &= \frac{(K_I,K_{II},K_{III})(z)}{\sqrt{2\pi(y - 1)}} \label{eqn:CrackSSurfDispStress} \\
        \llbracket u_i(0,y - da/a,z) \rrbracket &=  \frac{4 K_j(z)}{E} \sqrt{2\pi(1 - y + da/a)} \text{ with } i,j \in [(x,I),(y,II)] \label{eqn:CrackSSurfDispPlane} \\
        \llbracket u_z(0^+,y - da/a,z) \rrbracket &= \frac{K_{III}(z)}{\mu} \sqrt{2 \pi(1 - y + da/a)} \label{eqn:CrackSSurfDispOutp} 
    \end{align}
\end{subequations}

Where $\mu = \frac{E}{2(1+\nu)}$ is the shear modulus of the material.
We substitute Equations~\eqref{eqn:SIFSTradMain} into Equation~\eqref{eqn:EnrgInteg}, and then the resulting
expression into Equation~\eqref{eqn:extWorkEnrgRelrate}.

\begin{equation}
    \label{eqn:dvYahsi}
    \begin{aligned}
        dV &= - \frac{\pi}{E} \frac{1}{2} \int_0^{da/a}\int_{- \frac{h}{2}}^{\frac{h}{2}}  \left(4 K_I^2(z) + 4 K_{II}^2(z) + 2 (1 + \nu) K_{III}^2(z)\right) \;\text{d}y \text{d}z
    \end{aligned}
\end{equation}

We perform the integrations in $y,z$ to obtain:

\begin{equation}
    \label{eqn:GYahsi}
    \begin{aligned}
        dV &= - \frac{2\pi}{E} da/a 
            \left[
                \left( \dots \right)_1 h + \left( \dots \right)_2 \frac{h}{3} +
                \left( \dots \right)_3 h + \left( \dots \right)_4 \frac{h}{3} +
                (1+\nu) \left( \dots \right)_5 \frac{4}{15}h
            \right]
    \end{aligned}
\end{equation}

Each quantity $(\dots)_k$ is a lengthy expression depending on $h_k(1)$, the reference stress intensity factors
(Equation~\eqref{eqn:SIFSTradMain}), the normalized crack surface loads, the parameter $\gamma$, and the Poisson's
ratio $\nu$. 
Each expression contains coupling terms arising from the decomposition $h_k(1) = h^m_1(1) + \dots + h^v_1(1)$. 
For an infinitesimal crack extension, the newly created crack surface is $\approx 2h \; da$.
Assuming that only one loading component is acting at the time, Equation~\eqref{eqn:GYahsi} reduces to
Equation~\eqref{eqn:ERRsingleLoading}.

\renewcommand{\theequation}{\thesection.\arabic{equation}}
\setcounter{equation}{0}
\renewcommand{\thetable}{\thesection.\arabic{table}}
\setcounter{table}{0}

\section{Numerical implementation} \label{appendixB}
\setcounter{equation}{0}

In this appendix we present the numerical calculation of the Fredholm Kernel $K_{ij}(t,y)$
(Equations~\eqref{eqn:DerEffKerns},~\eqref{eqn:DerEffKernsK2},~\eqref{eqn:DerEffKernsK3},
\eqref{eqn:DerEffKernsK4},
~\eqref{eqn:DerEffKernsK5}) and study their convergence properties in Section~\ref{ssec:fredholmKernels}.
In Section~\ref{ssec:solvingSing} we provide the details about the convergence study we performed on the solution $h_k(t)$ the
singular integral system (Equation~\eqref{eqn:singularIntegralsExtracted}). 
Finally, in Section~\ref{ssec:displacementKernels} we describe the computation of the crack lip displacements (Equations~\eqref{eqn:discWp},
 \eqref{eqn:BetaXpFull}, \eqref{eqn:BetaYpFull}, \eqref{eqn:UpFull}, \eqref{eqn:VpFull}).
We have implemented the methodology in a \texttt{MATLAB 2024a} code. 
We followed the blueprint of \citet{Chan:2020} for the Matlab implementation.

\subsection{Numerical computation and convergence study of the Fredholm kernels} \label{ssec:fredholmKernels}

We consider a Fredholm kernel from Equation~\eqref{eqn:singularIntegralsExtracted}:

$$K_{ij}(t,y) =  \int_{-\infty}^{\infty} k(\alpha) e^{i\alpha(\tau-\eta)} \text{d}\alpha$$

Each integrand $k(\alpha)$ is a complex valued function whose real and imaginary parts always possess opposite
parity, as observed through extensive numerical experiments.
The integrands are known in closed form only up to the mapping functions $B_{jk}(\alpha)$,$B_k(\alpha)$,
$C_k(\alpha)$ and the characteristic roots $m_j(\alpha)$.
Exploiting this parity properties allows the kernel to be decomposed into two real valued contributions, while
halving the integration interval: 

\begin{equation}
    \label{eqn:fredholmHalfInterval}
    \begin{aligned}
        K(t,y) & = \int_{-\infty}^{\infty} k(\alpha) e^{i\alpha(t-y)} \text{d}\alpha\\
        & \to 2 \int_0^{\infty} \left(k^e(\alpha) \cos(\alpha(t-y)) + k^o(\alpha) \sin(\alpha(t-y)) \right) \text{d}\alpha
    \end{aligned}
\end{equation}

We discretize the kernels using Gauss-Chebyshev quadrature (Equation~\eqref{eqn:GCQuad}).
After discretization, each kernel is evaluated for all values of $(t-y) \in (-2,2)$.
We exploit the vectorization capabilities of \texttt{MATLAB} by implementing each kernel as a matrix valued
function $\bm{K}_{ij}(\bm{\Theta})$, with $\Theta_{ij} = t_j - y_i$
(Equations~\ref{eqn:DiscsingEqSysArbMain},~\ref{eqn:DiscSingEqDetails}), following the example of \citet{Chan:2020}. 
To facilitate the numerical evaluation, we split the infinite integration interval and treat each
contribution separately:

\begin{equation}
    \label{eqn:FredholmDecomp}
    \begin{aligned}
        \mathbf{K}(\bm{\Theta}) &= \int_{0}^{\alpha_0} \mathbf{k}(\alpha,\bm{\Theta}) d \alpha + \int_{\alpha_0}^{\alpha_u} \mathbf{k}(\alpha,\bm{\Theta}) d \alpha + \int_{\alpha_u}^{+\infty} \mathbf{k}(\alpha,\bm{\Theta}) d \alpha =\nonumber \\
        &= \mathbf{K}_s(\bm{\Theta}) + \mathbf{K}_n(\bm{\Theta}) + \mathbf{K}_t(\bm{\Theta}) 
    \end{aligned}
\end{equation}

The right-most contribution $\mathbf{K}_t(\bm{\Theta})$ is evaluated analytically using the
asymptotic analysis presented in appendix III of \citet{Yahsi:1981}.
Consequently, the integration bound $\alpha_u$ has no effect on the numerical convergence of the kernels.

We evaluate numerically the central contribution $\mathbf{K}_n(\bm{\Theta})$.
Because of the oscillatory nature of the kernels, \texttt{MATLAB}'s standard Gauss-Kronrod adaptive quadrature
becomes computationally expensive. 
Since the oscillatory terms of the kernels are purely trigonometric (Equation~\ref{eqn:fredholmHalfInterval}),
we instead employ Filon's method \citep{Abramowitz:1968}. 
Although \citet{Yahsi:1981} do not explicitly state the numerical integration technique they used,
similar works have employed Filon's method \citep{Krenk:1978,Yuceoglu:1973}.
Moreover, the fixed integration grid allows the integrands $k(\alpha)$ to be computed only once and the 
subsequent integration to be fully vectorized, considerably reducing the integration computational cost.
We follow the vectorized implementation of \href{https://github.com/johannesgerer/jburkardt-m/blob/master/filon/}{Johannes Gerer}.

We choose the lower integration bound $\alpha_0$ so that the left-most contribution $\mathbf{K}_s(\bm{\Theta})$
becomes negligible. 
The kernels $\mathbf{K}_{22}$ and $\mathbf{K}_{44}$ require additional care because their numerical 
implementation displays apparent singularities as $\alpha \to 0$.

\subsubsection{Apparent singularities in the Fredholm kernel integrands}
\label{sssec:appSing}

From numerical experiments, we observed that $\Re(k_{22}(t,y)),\Im(k_{44}(t,y))$ show mild numerical 
singularities as $\alpha \to 0$.
In the former kernel, the apparent singularity originates from the summation term in $B_{j2}(\alpha)$ 
(Equation~\eqref{eqn:DerEffKernsK2}).
A rigorous proof that this singularity is spurious would require an asymptotic closed form solution of the
system in Equation~\ref{eqn:SystemInitial}, which we were unable to derive.
Nevertheless, following appendix VIII of \citet{Hagendorf:1972}, we show that the integral $K^s_{22}(t,y)$
exists and is bounded by applying the mean value theorem.

\begin{equation}
    \label{eqn:k22re}
    \begin{aligned}
        K^s_{22}(\bm{\Theta}) &= \int_0^{\alpha_0} \Re(w_{22}(\alpha)) \cos(\alpha \bm{\Theta}) d \alpha =\\
        &= \Re(w_{22}(\frac{\alpha_0}{2})) \int_0^{\alpha_0} \cos(\bm{\Theta} \alpha) d \alpha = \Re(w_{22}(\frac{\alpha_0}{2})) \frac{\sin(\alpha_0 \bm{\Theta})}{\bm{\Theta}}
    \end{aligned}
\end{equation}

Since $|\Theta_{ij}| \leq 2$, and $\alpha_0 \ll 1$, the sine function can be expanded in Taylor series.
Therefore, if $\Re(w_{2k}(\alpha)) \propto C \alpha^n $, as long as $n > -2$, the integral $K^s_{22}$ is bounded. 
Here, $C$ a constant. 
Extensive numerical experiments over a wide range of shell geometries consistently satisfied this condition.
Together with the convergence displayed in Figure~\ref{fig:LeftConvg} and the agreement with the published 
benchmark results presented in Appendix~\ref{appendixC}, this provides confidence in the numerical 
implementation.

The apparent singularity in $\Im(k_{44}(t,y))$, instead, originates from a term available in closed form
(Equation~\eqref{eqn:DerEffKernsK4}).
We isolate the singular contribution by rewriting the kernel as:

\begin{equation}
    \label{eqn:imagK44}
    \begin{aligned}
        \Im(k_{44}(\alpha)) &= \Im(k_{44}^n(\alpha)) + \Im(k_{44}^c(\alpha)) \\
        \Im(k_{44}^n(\alpha)) &= \Im(w_{44}(\alpha))  - \frac{1+\nu}{2} - \frac{\kappa (1 - \nu)\alpha}{2 r_1}(\alpha^2 + r_1^2) - \frac{\alpha}{r_1}\\
        \Im(k_{44}^c(\alpha)) &= - \frac{1}{\kappa(1-\nu) r_1 \alpha}
    \end{aligned}
\end{equation}

To evaluate the singular contribution analytically, we employ a result due to Claude Lebovici, published on 
\href{https://math.stackexchange.com/questions/5122079/oscillatory-improper-integral/5122191?noredirect=1#comment11040724_5122191}{MathStackExchange}. 
Indeed, the integral of $k_{44}^c(\alpha)$ can be computed in closed form.

\begin{equation}
    \label{eqn:K44rearrange}
    \begin{aligned}
        k_{44}^c(\alpha,\Theta_{ij}) &= - \frac{1}{\kappa(1-\nu)} \int_0^{\infty}\frac{\sin(\Theta_{ij} \alpha)}{r_1 \alpha} d \alpha = - \frac{1}{\kappa(1-\nu)} \int_0^{\infty}\frac{\sin(\Theta_{ij} \alpha)}{\sqrt{\alpha^2 + \frac{2}{\kappa(1-\nu)}}\alpha} d \alpha\\
        & \text{ use } t = \Theta_{ij} \alpha, a = \sqrt{\frac{2}{\kappa(1-\nu)}} \Theta_{ij} \Rightarrow k_{44}^c(\alpha,\Theta_{ij}) = 2 \Theta_{ij} \int_0^{\infty} \frac{\sin(t)}{t\sqrt{t^2 + a^2}} d t
    \end{aligned}
\end{equation}

This integral admits a closed form representation in terms of modified Bessel functions of the second kind $K_o(a),K_1(a)$ and Struve functions $L_{-1}(a),L(a)$:

\begin{equation}
    \label{eqn:Bessel}
    \begin{aligned}
        J(b) &= \int_0^{\infty} \frac{\sin(bt)}{t \sqrt{t^2 + a^2}} \text{d}t  \text{ and } d J(b) / db = \int_0^{\infty} \frac{\cos(bt)}{\sqrt{t^2 + a^2}} \text{d}t = K_0(ab) \\
        \Rightarrow J = &\int_0^1 K_0(ab) db = \frac{\pi}{2} \left(L_{-1}(a) K_0(a) + L_0(a) K_{1}(a)\right) \\
        \Rightarrow K_{44}^c(\alpha,\Theta_{ij}) &= \pi \Theta_{ij} \left( L_{-1}(c \Theta_{ij}) K_0(c \Theta_{ij}) + L_0(c \Theta_{ij}) K_{1}(c \Theta_{ij})\right) \text{ with } c = \sqrt{\frac{2}{\kappa(1-\nu)}}
    \end{aligned}
\end{equation}

The kernel is therefore bounded.

\subsubsection{Convergence study}

\begin{figure}
    \centering
    \includegraphics[width =\linewidth]{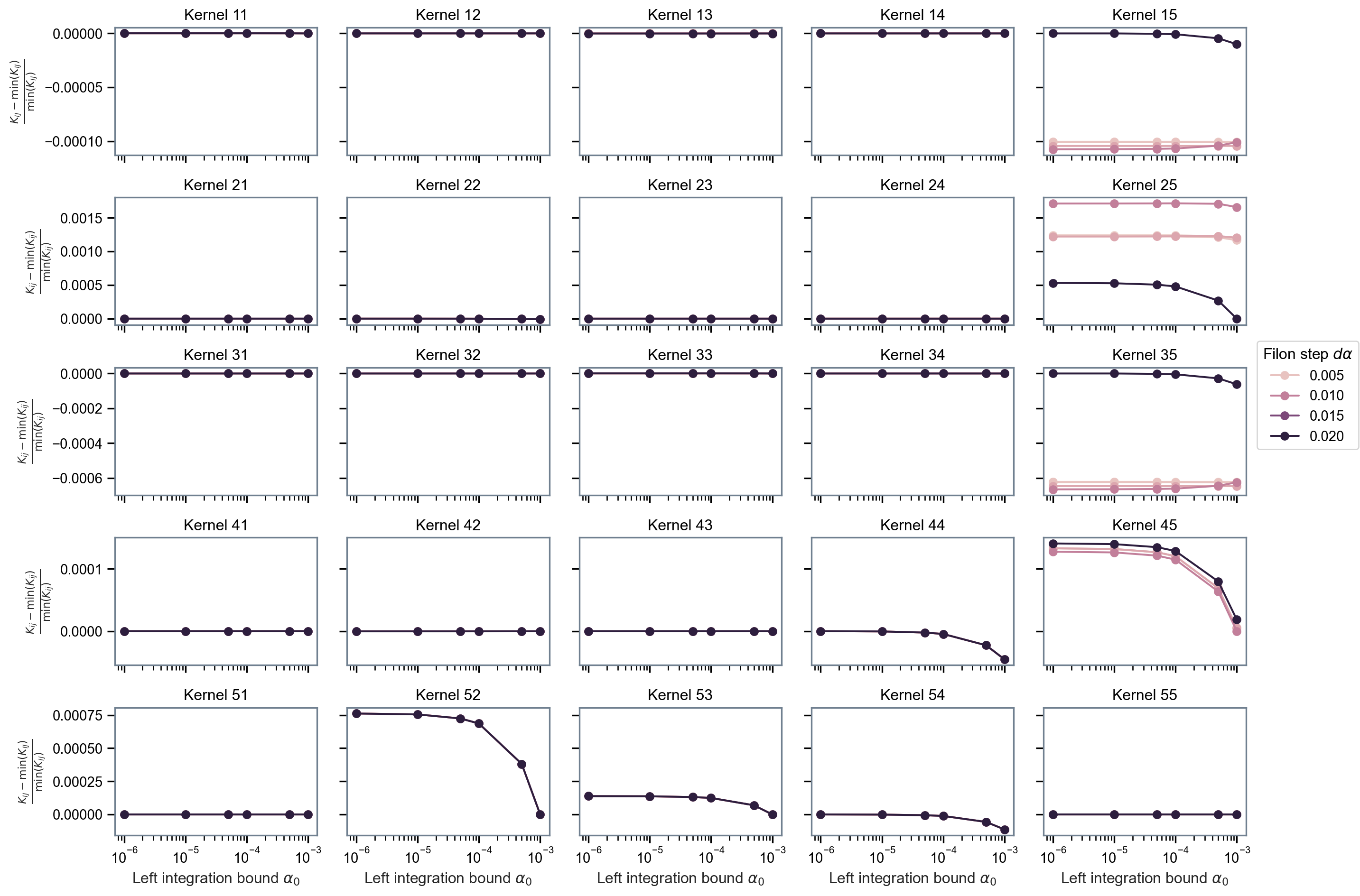}
    \caption{
    Convergence of the Fredholm kernels $K_{ij}$ with respect to left integration bound $\alpha_0$ and the Filon
    discretization step $\Delta \alpha$. 
    The horizontal axis is shown on a logarithmic scale. 
    Darker curves correspond to larger discretization steps (smaller values of $n_f$).
    Convergence of all kernels requires the finest discretization step considered, corresponding to $\alpha_u =100$ and $n_F = 28001$. 
    Most kernels are already converged for $\alpha_0 \approx 1e-4$. 
    }
    \label{fig:LeftConvg}
\end{figure}

We performed an extensive numerical convergence study to determine the influence of the lower integration bound
$\alpha_0$ and the number of Filon nodes $n_F$ on the Fredholm kernels.
We show the results in Figure~\ref{fig:LeftConvg}. 
We define the relative kernel error as $\Delta K_{ij} = \frac{K_{ij} - K^{ref}_{ij}}{K^{ref}_{ij}}$.
Here, $K^{ref}_{ij}$ is the value obtained for the minimum number of Filon nodes and the largest value of
$\alpha_0$.
We plot the relative kernel error as a function of $\alpha_0$ and of the Filon step 
$\Delta \alpha = \frac{\alpha_u - \alpha_0}{n - 1}$. 
Owing to the oscillatory nature of the kernels, the slowest convergence is expected for the smallest value of
$|\Theta|$, and only those results are reported.

\subsection{Numerical solution of the singular integral equation system} \label{ssec:solvingSing}

\begin{figure}
    \centering
    \includegraphics[width =\linewidth]{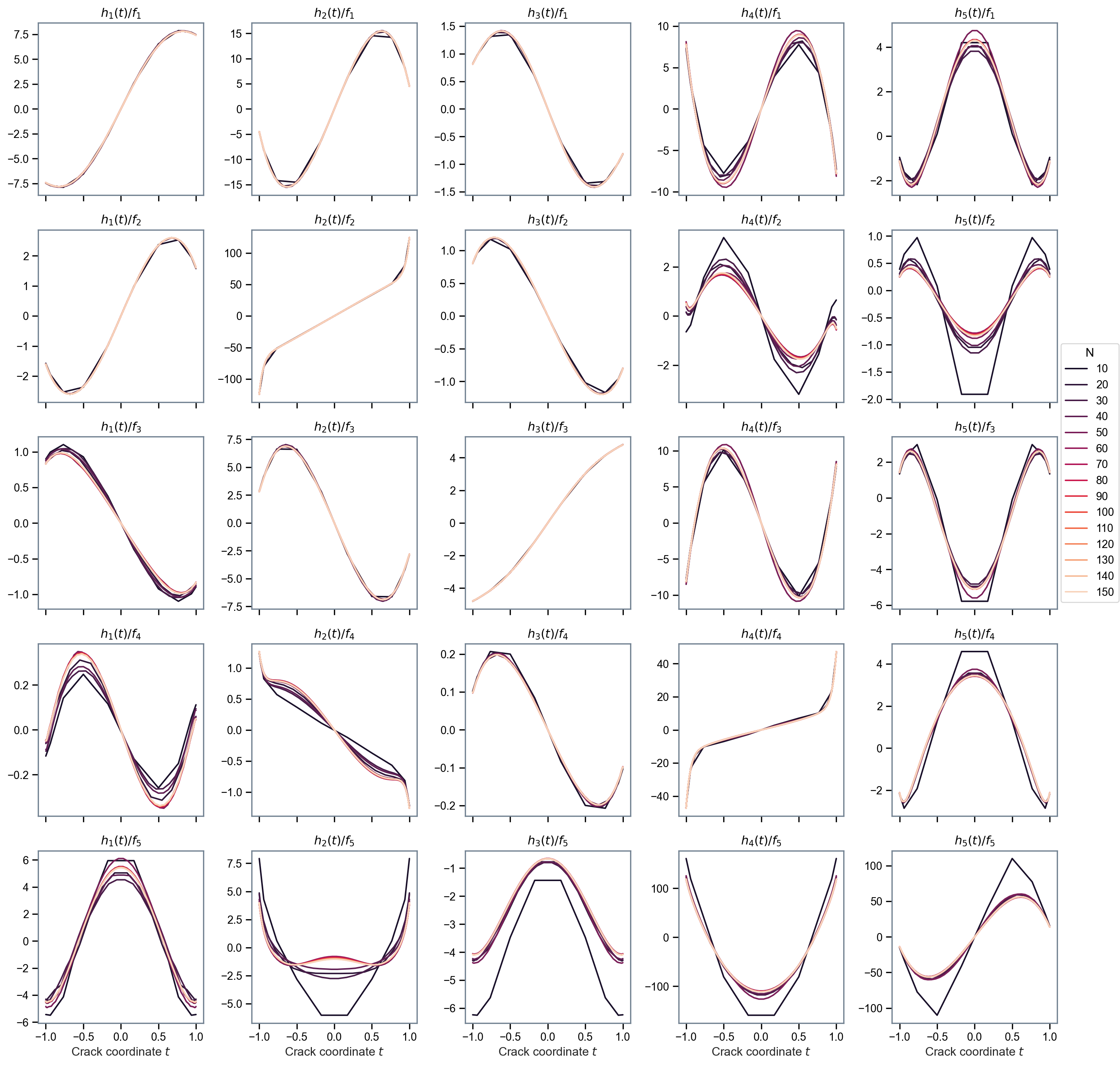}
    \caption{
    Normalized target functions $h_k(t)$, for an increasing number of Gauss-Chevyshev quadrature nodes $N$.
    Each row corresponds to one of the five loading cases.
    The endpoint values $h_k(\pm1)$, which determine the stress intensity factors, converge with relatively few 
    quadrature nodes. 
    In contrast, evaluating the crack lip displacement fields require accurate values of $h_k(t))$ over the 
    entire interval, and therefore demands a substantially finer discretization.
    }
    \label{fig:outputGCCOnvergence}
\end{figure}

We solve the linear system (Equation~\eqref{eqn:DiscsingEqSysArbMain}) using \texttt{MATLAB}'s \texttt{mldivide}
routine. 
We then perform an extensive numerical study to assess the effect of the number of Gauss-Chebyshev quadrature
nodes $N$ on the unknown functions $h_k(t)$.
We plot the results in Figure~\ref{fig:outputGCCOnvergence} for five unitary, uniformly distributed loading 
cases, each corresponding to $f_i(y) = f_i\neq 0,f_j(y) = 0$ for $i\neq j$ and $i,j = 1,\dots5$
(Equation~\eqref{eqn:SingIntegralsMixedLoading}).

We further investigate the convergence of the functions $h_k(t)$ with respect to the number of Gauss-Chebyshev
nodes $N$ by comparing our results with those reported by \citet{Yahsi:1981,Yahsi:1983}. 
Since the authors do not specify the number of quadrature nodes used in their computations, we consider the
same benchmark problem: a cylindrical shell with $a/h = 10$, $h/R =0.2$ and $\nu = 0.3$.
We vary both the number of Gauss-Chebyshev quadrature points $N$ and the crack orientation $\beta$, and 
following \citet{Yahsi:1981,Yahsi:1983}, plot the normalized stress intensity factors $\tilde{K}_{i}$.

\begin{figure}
    \centering
    \includegraphics[width =\linewidth]{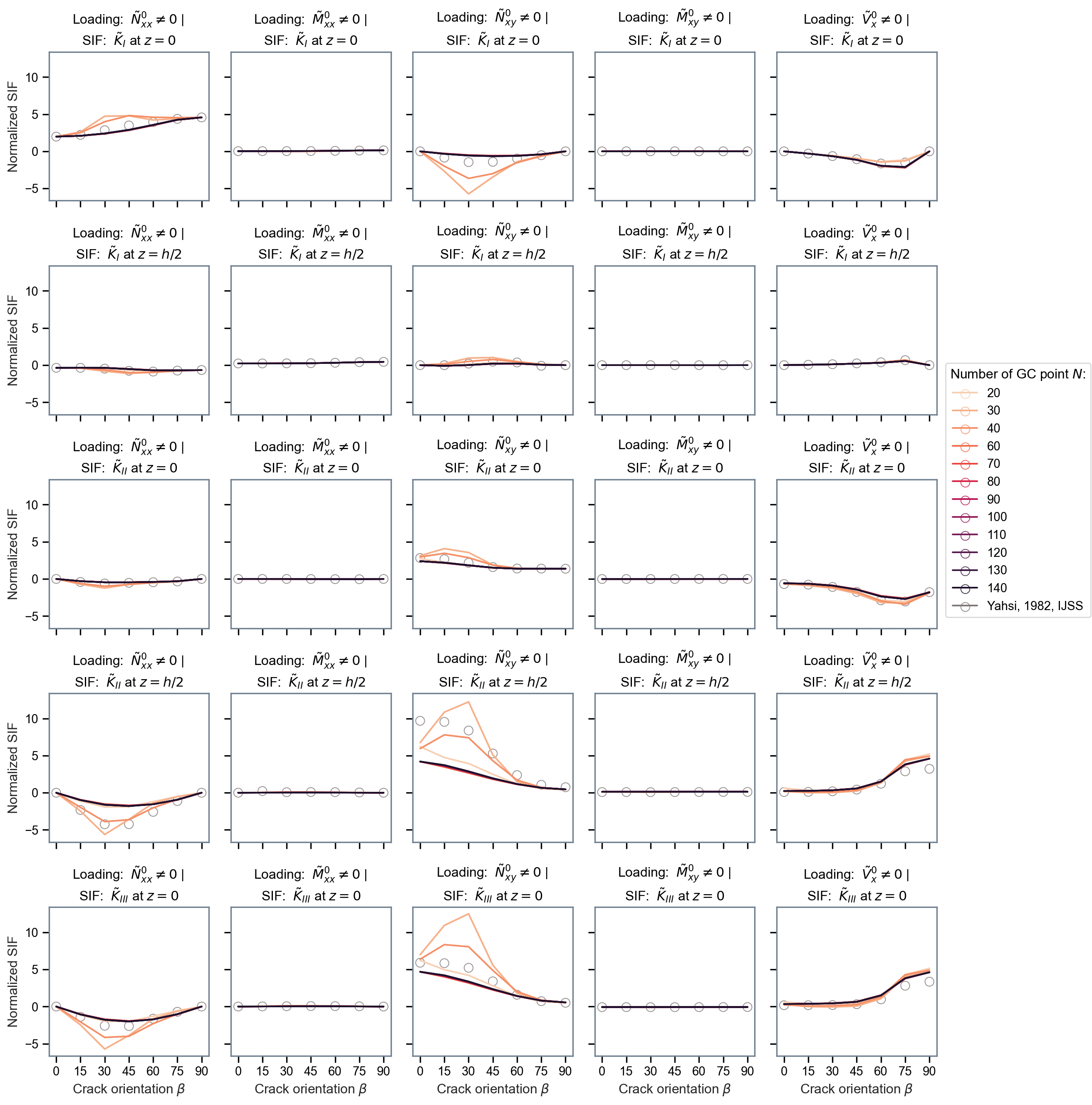}
    \caption{
    Normalized stress intensity factors $\tilde{K}_{i}$ for an inclined crack in a cylindrical shell as
    functions of the crack orientation $\beta$ and of the number of Gauss-Chebyshev points $N$.
    Most discrepancies with the results of \citet{Yahsi:1981,Yahsi:1983} can be attributed to an
    insufficient numerical convergence. 
    The remaining differences are likely due to implementation details and differences in the computational
    hardware used.
    }
    \label{fig:convSlantedIso}
\end{figure} 

We observe convergence of the normalized stress intensity factors for $N \geq 70$.
Nevertheless, the results reported by \citet{Yahsi:1981,Yahsi:1983} almost always lie in the envelope 
defined by our computations using the minimum and maximum numbers of quadrature nodes. 
This suggests that, owing to the computational limitations of the hardware available more than four decades ago,
the authors employed fewer quadrature nodes than required for full convergence. 
Such an interpretation provides a plausible explanation for the remaining discrepancies between the two sets of
results. 

\subsection{Computing the crack lip displacement fields}
\label{ssec:displacementKernels}

To the best of the authors' knowledge, closed-form expressions and numerical solution procedures
for the displacement fields are not available in the literature. 
All displacement fields share the same mathematical structure:

\begin{equation}
    \label{eqn:displacementFields}
    \begin{aligned}
        (w,u,v,\beta_x,\beta_y)(x^{\pm},y) &= \frac{i}{2 \pi} \int_{-1}^1 \sum_{k = 1}^5 g_k(t) (I_k^w,I_k^u,I_k^v,I_k^x,I_k^y) (x^{\pm},y,t) d t  \\
       I_k^l (x^+,y,t)& =\int_{-\infty}^{\infty} \sum_{j = 1}^4 e^{\Re(m_j(\alpha)) x^+} f_{jk}(\alpha,y) e^{i(\Im(m_j(\alpha)) x + \alpha (t-y))}  d \alpha \\
       I_k^l (x^-,y,t)& =\int_{-\infty}^{\infty} \sum_{j = 5}^8 e^{\Re(m_j(\alpha)) x^-} f_{jk}(\alpha,y) e^{i(\Im(m_j(\alpha)) x + \alpha (t-y))}  d \alpha
    \end{aligned}
\end{equation}

The functions $f_{jk}(\alpha,\eta)$ are defined in 
Equations~\eqref{eqn:wpFullFinal},~\eqref{eqn:BetaXpFull},
~\eqref{eqn:BetaYpFull},~\eqref{eqn:UpFull} and~\eqref{eqn:VpFull}. 

The oscillator's phase is no longer trigonometric, but depends on the imaginary part of the 
characteristic roots $m_j(\alpha)$. 
Consequently, Filon's method is no longer applicable \citep{Huybrechs:2009}, since numerical
experiments show that $\Im(m_j(\alpha))$ is a non-linear function of $\alpha$.
Furthermore, additional numerical experiments indicate that the displacement kernels integrands 
exhibit strong singularities for $\alpha \to 0$.
As a result, computing the complete displacement fields becomes computationally demanding.
For this reason, we restrict the numerical computation to the crack lips displacements
(Equation~\eqref{eqn:discWp},~\eqref{eqn:discBetaXp},~\eqref{eqn:discBetaYp},~\eqref{eqn:discUp} and
~\eqref{eqn:discVp}), which possess trigonometric oscillating terms and thus allow to use Filon's 
quadrature.
Exploiting parity properties, analogously to the treatment of the Fredholm kernels, yields:

\begin{equation}
    \label{eqn:SingIntDispDecompLip}
    \begin{aligned}
        I_k^{l}(0^+,\bm{\Theta}) &= 2 \left[ \int_{0}^{\infty} \sum_{j=1}^4 \Re(f_{jk}(\alpha,\eta))  \cos(\alpha\bm{\Theta}) d \alpha + \int_{0}^{\infty} \sum_{j=1}^4 \Im(f_{jk}(\alpha,\eta)) \sin(\alpha\bm{\Theta}) d \alpha  \right] \\ 
        I_k^{l}(0^-,\bm{\Theta}) &= 2 \left[ \int_{0}^{\infty} \sum_{j=5}^8 \Re(f_{jk}(\alpha,\eta))  \cos(\alpha\bm{\Theta}) d \alpha + \int_{0}^{\infty} \sum_{j=5}^8 \Im(f_{jk}(\alpha,\eta)) \sin(\alpha\bm{\Theta}) d \alpha  \right]
    \end{aligned}
\end{equation}

We decompose each displacement kernels into three intervals, analogously to the Fredholm kernels:

\begin{equation}
    \label{eqn:DisplacementTransform}
    \begin{aligned}
        \mathbf{I}_{l}^k &= \mathbf{L}_{l}^k + \mathbf{R}_{l}^k + \mathbf{T}_{l}^k 
    \end{aligned}
\end{equation}

The convergence requires additional care. The contribution $\mathbf{L}_{l}^k$ contains strong
singularities as $\alpha \to 0^{\pm}$, while no asymptotic expressions are available for the
tail contribution $\mathbf{T}_{l}^k $. 
The contribution $\mathbf{R}_{l}^k$ is evaluated numerically using vectorized Filon quadrature,
following the same procedure adopted for the Fredholm kernels.

Nevertheless, numerical experiments consistently show that all integrands $f_{jk}(\alpha,\eta)$
decay to numerical zero for $\alpha \geq 100$. 
Therefore, the convergence of the crack lip displacements is governed by the behavior of the
contribution $\mathbf{L}_{l}^k$. 

\begin{figure}
    \centering
    \includegraphics[width =\linewidth]{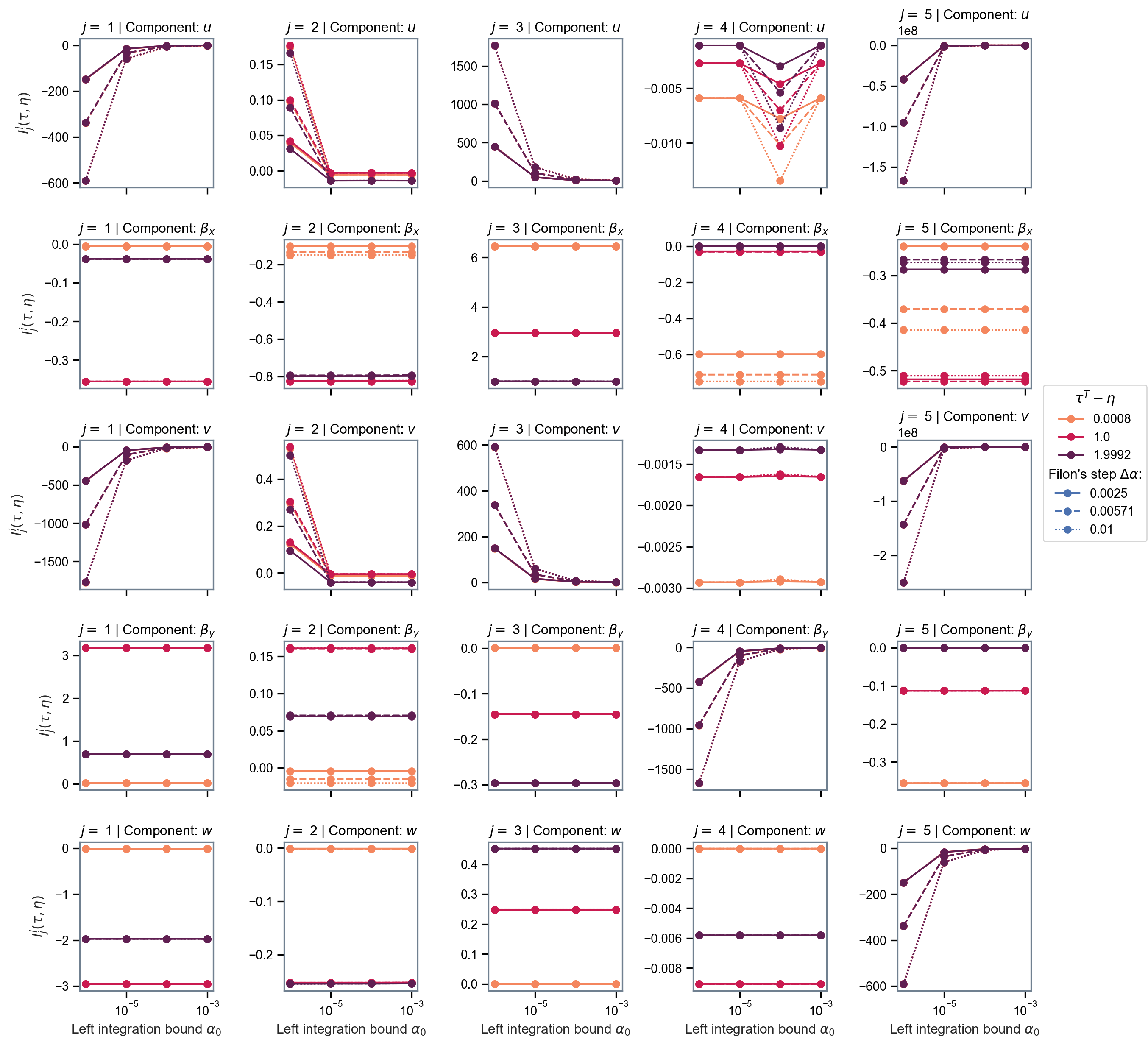}
    \caption{
    Relative error $\Delta I^k_{i}$ of the displacement kernel for an ellipsoidal shell as
    functions of the lower integration bound $\alpha_0$ and the Filon step $\Delta \alpha$.
    The kernels exhibit a strong sensitivity to both parameters, preventing a clear 
    identification of numerical convergence.  
    }
    \label{fig:dispTransfEllip}
\end{figure}

\begin{figure}
    \centering
    \includegraphics[width =\linewidth]{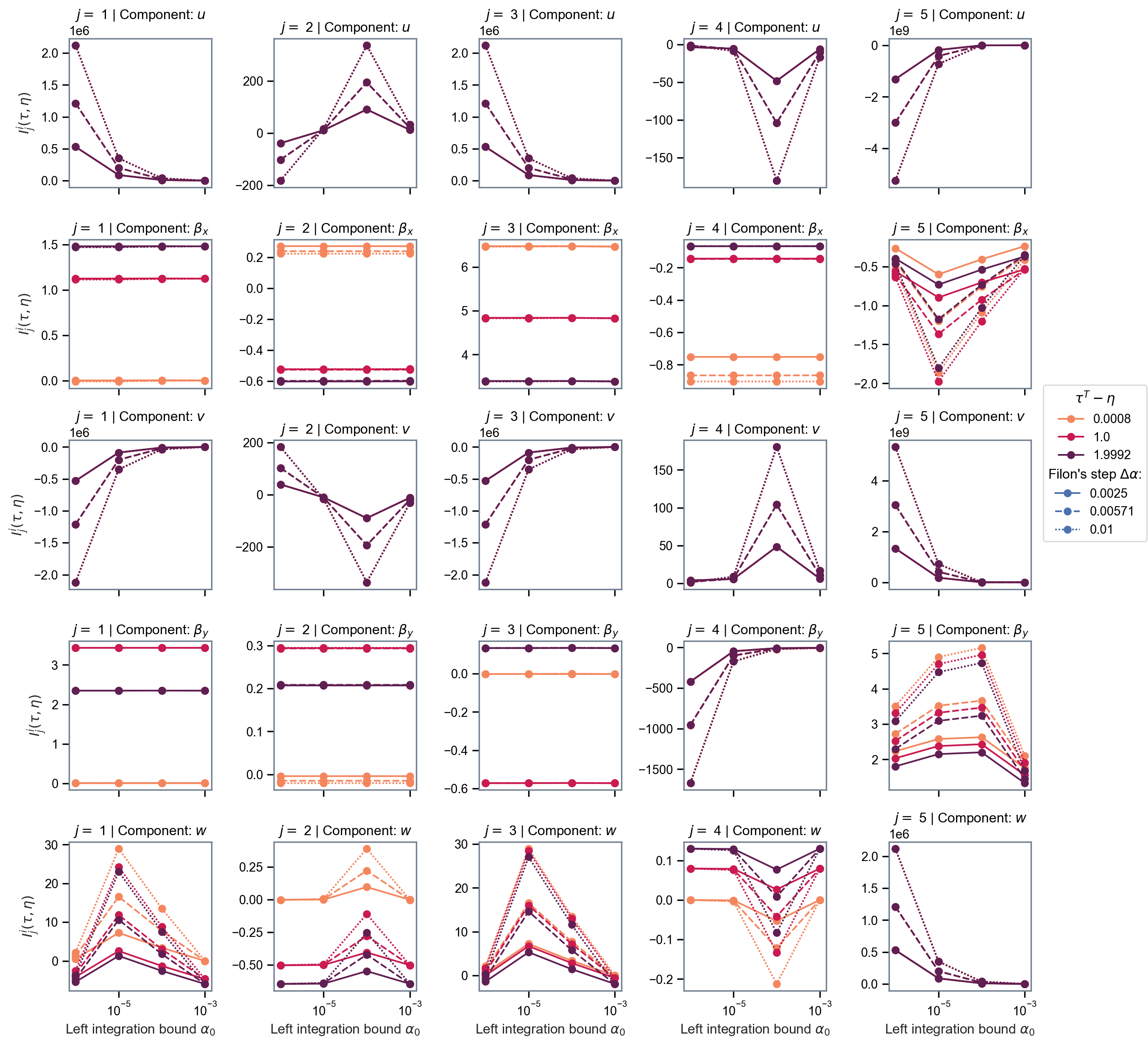}
    \caption{
    Relative error $\Delta I^k_{i}$ of the displacement kernel for a cylindrical shell as
    functions of the lower integration bound $\alpha_0$ and the Filon step $\Delta \alpha$. 
    The kernels exhibit a strong sensitivity to both parameters, preventing a clear
    identification of numerical convergence.  
    The dependence on the numerical parameters is even more pronounced than for the ellipsoid.
    }
    \label{fig:dispTransfCyl}
\end{figure}

\begin{figure}
    \centering
    \includegraphics[width =\linewidth]{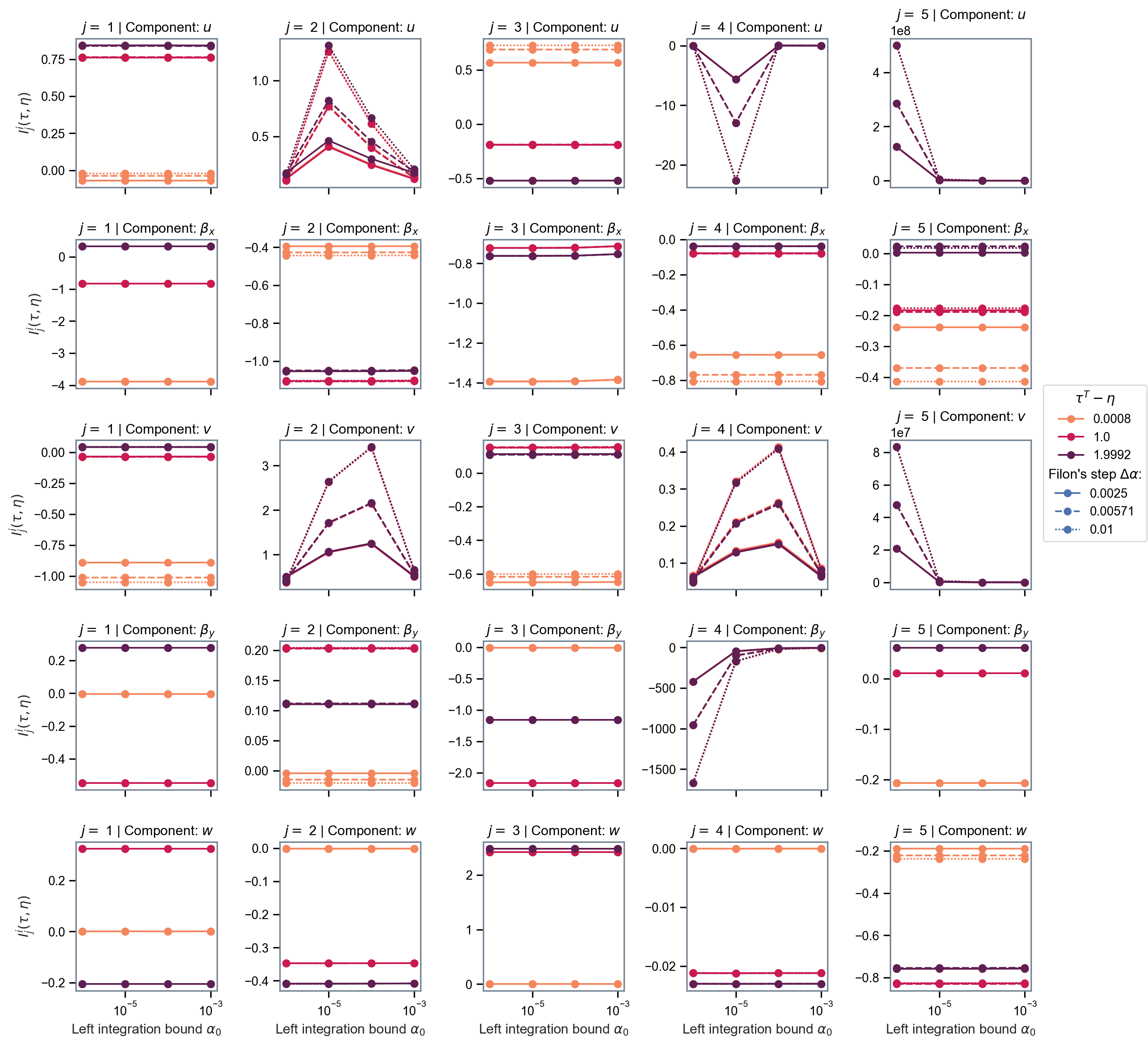}
    \caption{
    Relative error $\Delta I^k_{i}$ of the displacement kernel for a hyperboloidal shell as
    functions of the lower integration bound $\alpha_0$ and the Filon step $\Delta \alpha$.
    }
    \label{fig:dispTransfHyp}
\end{figure}

Once again, we perform an extensive numerical study of the effect of the lower integration bound
$\alpha_0$, of the discretization step $\Delta \alpha$ and of $\Theta_{ij}$ on the displacement 
transforms $I_l^k$. We define the relative error as: 

$$\Delta I^k_l = \frac{I^k_l - I^{k,ref}_l}{I^{k,ref}_l}$$

with respect to a reference solution obtained using for the smallest number of Filon nodes 
$n_F = \min(n_F)$ and the largest lower integration bound $\alpha_0 = \max(\alpha_0)$.
We also include the values for three representative values of $\Theta_{ij}$, being its minimum,
maximum and intermediate value. 
We show representative results for three shell geometries: an ellipsoid in 
Figure~\ref{fig:dispTransfEllip}, a cylinder in Figure~\ref{fig:dispTransfCyl} and a hyperboloid
in Figure~\ref{fig:dispTransfHyp}.

\clearpage

The strong singularities as $\alpha \to 0^+$ prevent the establishment of full numerical 
convergence for the displacement transform. 
This behavior changes with shell geometry, with convergence being the most difficult for the cylinder.
Consequently, each displacement solution was inspected individually to verify that it remained
physically meaningful and consistent with the prescribed loading. 
Such verification is presently required until the behavior of the displacement kernels near
$\alpha = 0$ is understood more rigorously. 
We found this complicated behavior of the displacement kernels as $\alpha\to 0^{+}$ to be
responsible for rigid body in-plane translations observed for the cylindrical shell in 
Figures~\ref{fig:displacements-tension}.h,~\ref{fig:DisplacementsShear}.h,
~\ref{fig:DisplacementsBending}.h,~\ref{fig:DisplacementsTwisting}.h. 
We interpret these translations as numerical artefacts for four reasons. 
First, in formulating the perturbation problem, we applied only self-equilibrating crack-surface tractions.
Second, the translation has the same magnitude at both crack tips within numerical precision. 
Thirdly, once the translation is removed, the crack deformation is fully consistent with the
stress responses shown in Figure~\ref{fig:KandERRTension}, Figure~\ref{fig:KandERRShear} and
Figure~\ref{fig:KandERRbending}, whose convergence has been independently established, as well
as with the in-plane displacements for the remaining shell geometries.
Finally, $\alpha = 0$ corresponds the zero frequency Fourier mode, which represents a rgid body 
translation in physical space. 
Consequently, the lack of convergence near this frequency is consistent with the appearance of a
spurious rigid body translation in the computed displacement field.

\section{Verification} \label{appendixC}

We verified our implementation against a the benchmark solutions for inclined cracks in cylindrical shells under general loading from
\citet{Yahsi:1983}. 
Some noticeable discrepancies are observed for the geometries with smallest thickness. 
As argued in Figure~\ref{fig:convSlantedIso}, we attribute these differences primarily to insufficient numerical
convergence in the original computations.

We present the normalized stress intensity factors $\tilde{K}_{ij}$ for all fracture modes,
evaluated at their respective through-thickness locations under general loading conditions.
The results are given as a function of the crack orientation $\beta$ and the normalized 
thickness $h/a$. 
We first show the entire set of normalized stress intensity factors:

\begin{figure}
    \centering
    \includegraphics[width =\linewidth]{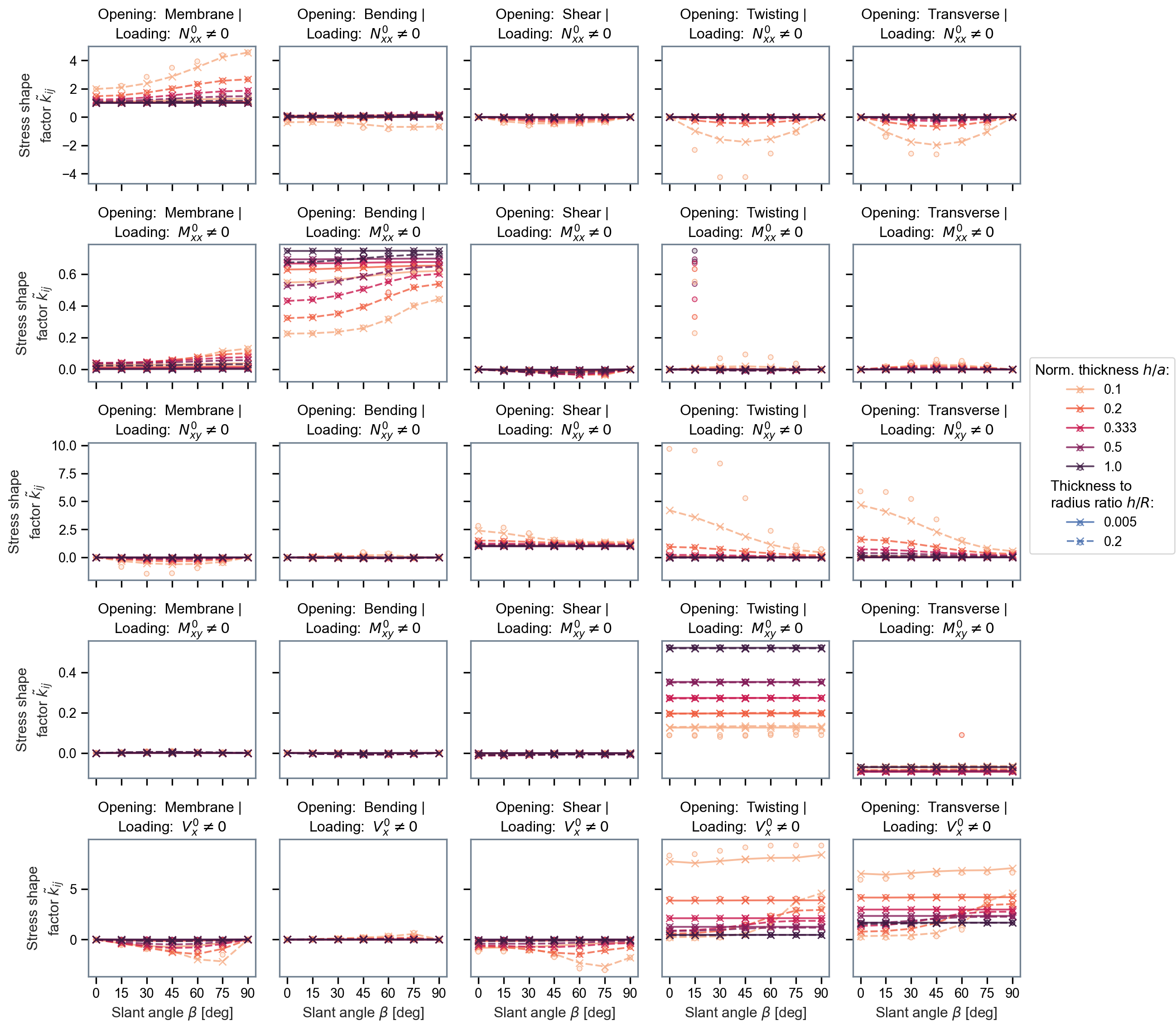}
    \caption{
    Normalized stress intensity factors $\tilde{K}_{ij}$ for an inclined crack in a cylindrical
    shell as functions of the crack orientation $\beta$, of the thickness to radius ratio $h/R$
    and of the normalized thickness $h/a$.
    Good agreement is obtained except for the thinnest and most highly curved shells.
    We attribute these discrepancies primarily to insufficient numerical convergence in the
    original computations. 
    A few isolated discrepancies in the twisting factor under bending loading and the transverse
    factor under twisting loading are likely due to typographical errors in \cite{Yahsi:1981}.
    }
    \label{fig:ovSlantedIso}
\end{figure} 

We further show the results for a larger set of $h/R$ values in the range $[0.005,0.2]$, while restricting the analysis to
$\beta = 45^{\circ}$ and $\nu = 0.3$.

\begin{figure}
    \centering
    \includegraphics[width =\linewidth]{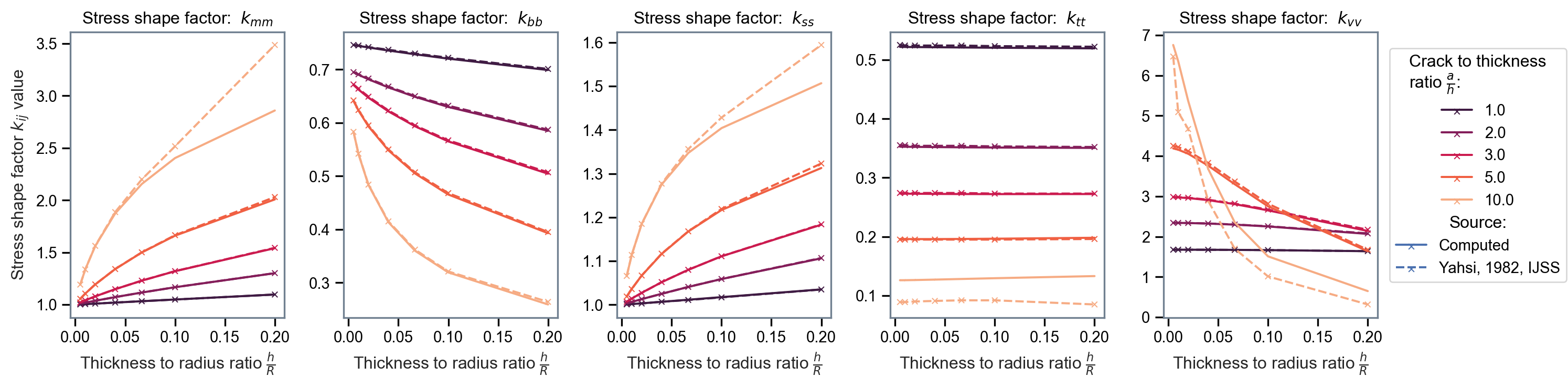}
    \caption{
    Normalized stress intensity factors $\tilde{K}_{ij}$ for inclined cracks in a 
    cylindrical shell ($\beta = 45^{\circ}$) as functions of the thickness to radius ratio $h/R$
    and the normalized thickness $h/a$. 
    Good agreement is obtained over most of the parameter range. 
    The largest discrepancies again occur for the thinnest and most highly curved shells.
    }
    \label{fig:hrSlantedIso}
\end{figure}

\section{Appendix D: Twisting and transverse loading} \label{app:incompatible-displacements}

\begin{figure}
    \centering
    \includegraphics[width=\textwidth]{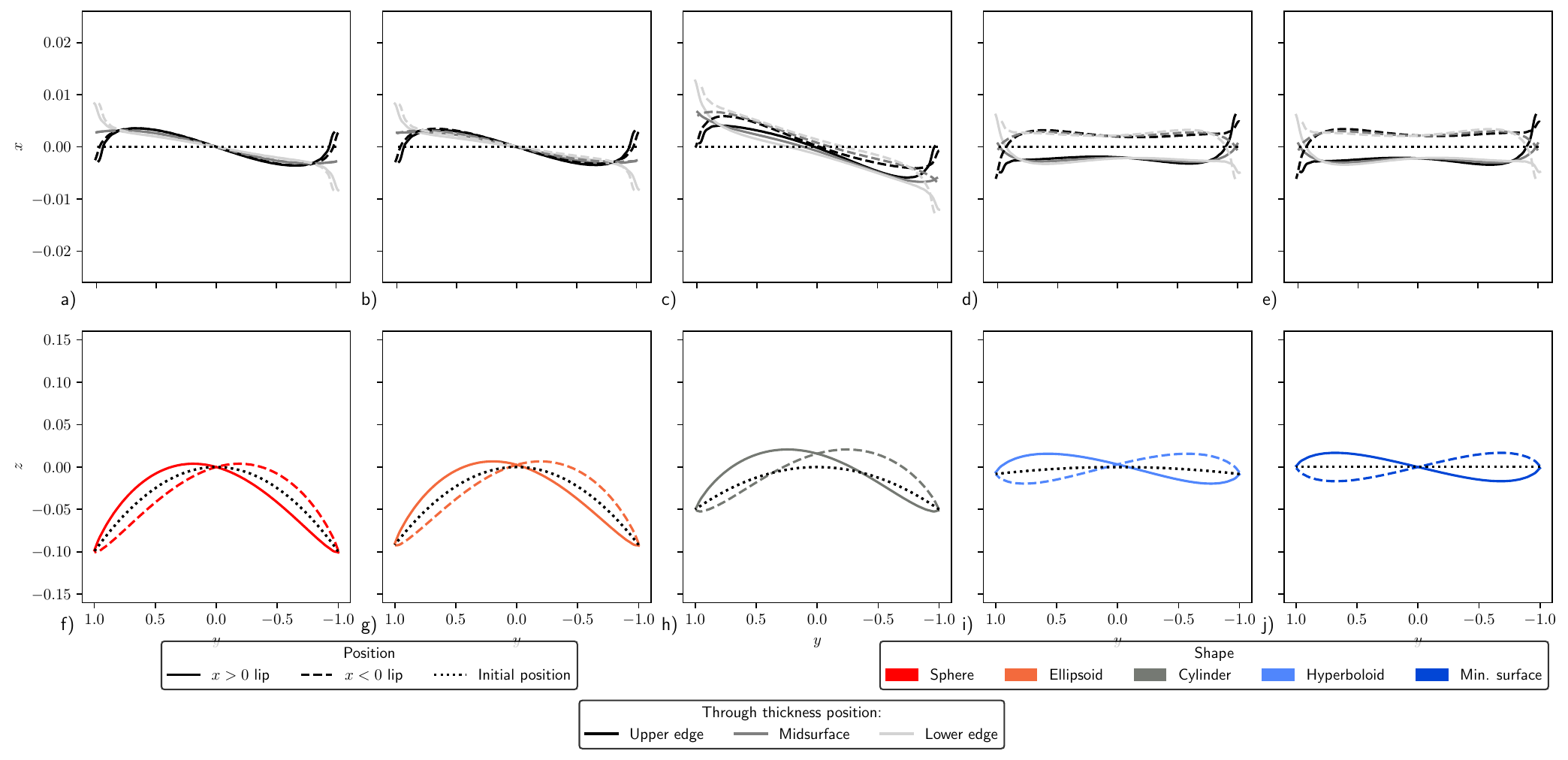}
    \caption{
    \emph{Crack-lip displacement fields under twisting loading for five representative 
    shells with crack orientation $\beta=45^\circ$.
    From left to right, the shell curvature profiles are: sphere ($\kappa_X/\kappa_Y=1$), ellipsoid ($\kappa_X/\kappa_Y=0.833$), cylinder ($\kappa_X/\kappa_Y=0$), hyperboloid 
    ($\kappa_X/\kappa_Y=-0.833$), and minimal surface ($\kappa_X/\kappa_Y=-1$).
    (a--e) The projections of the computed crack-lip displacement fields onto the $yx$ plane. 
    (f--j) The corresponding projections onto the $yz$ plane. 
    All geometries show a similar anti-symmetric out-of-plane profile.
    The sphere and ellipsoid show no in-plane opening, together with a twisting of the crack tips. 
    The hyperboloid and the minimal surface, instead, shows in-plane crack closing. 
    The cylinder shows a combination of the two regimes, as for the other loading cases.
    All geometries show incompatibility at the crack tips, with the two crack lips not meeting exactly.}}
    \label{fig:DisplacementsTwisting}
\end{figure}

The two remaining loading components to be investigated are twisting moment 
$\tilde{M}^0_{xy}>0$ and transverse shear loading $\tilde{V}_x^0>0$. 
As shown by \citet{Hui:1993}, the responses to these loadings are inherently coupled, even in
flat Reissner plates, and thus we consider them together.
We omit the stress intensity factors and the energy release rate for these loading cases, as the
solution obtained using the shallow shell model produces incompatible displacements.
In Figure~\ref{fig:DisplacementsTwisting}, we show the same reference geometries considered for
the membrane tension, in-plane shear, and bending cases, with a crack embedded at $\beta = 45^\circ$.

The response is largely as expected from a qualitative standpoint. 
All geometries exhibit a similar out-of-plane deformation profile, which is distinctly
antisymmetric. 
The in-plane response, however, differs between geometries.
Ellipsoids show little in-plane opening, analogous to the in-plane shear case, with a distinct
twisting of the crack tips and a slight rotation of the crack around the $z$-axis.\
Hyperboloids, instead, exhibit an opening of the crack away from the crack tips, which is nearly
constant through the shell thickness. 
Their cracks remain aligned with the initial crack orientation. 
At the crack tips, the same antisymmetric through-the-thickness twist observed in the ellipsoids
is recovered.\
The cylinder again shows a combination of the two regimes.
The membrane opening of the hyperboloid is recovered, but it is not constant through the
thickness. 
At the same time, the crack rotates around the $z$-axis, similar to what we observed in the
ellipsoids. 
Unlike the doubly curved geometries, the center of the crack also exhibits a small, symmetric
displacement.

\begin{figure}
    \centering
    \includegraphics[width=\textwidth]{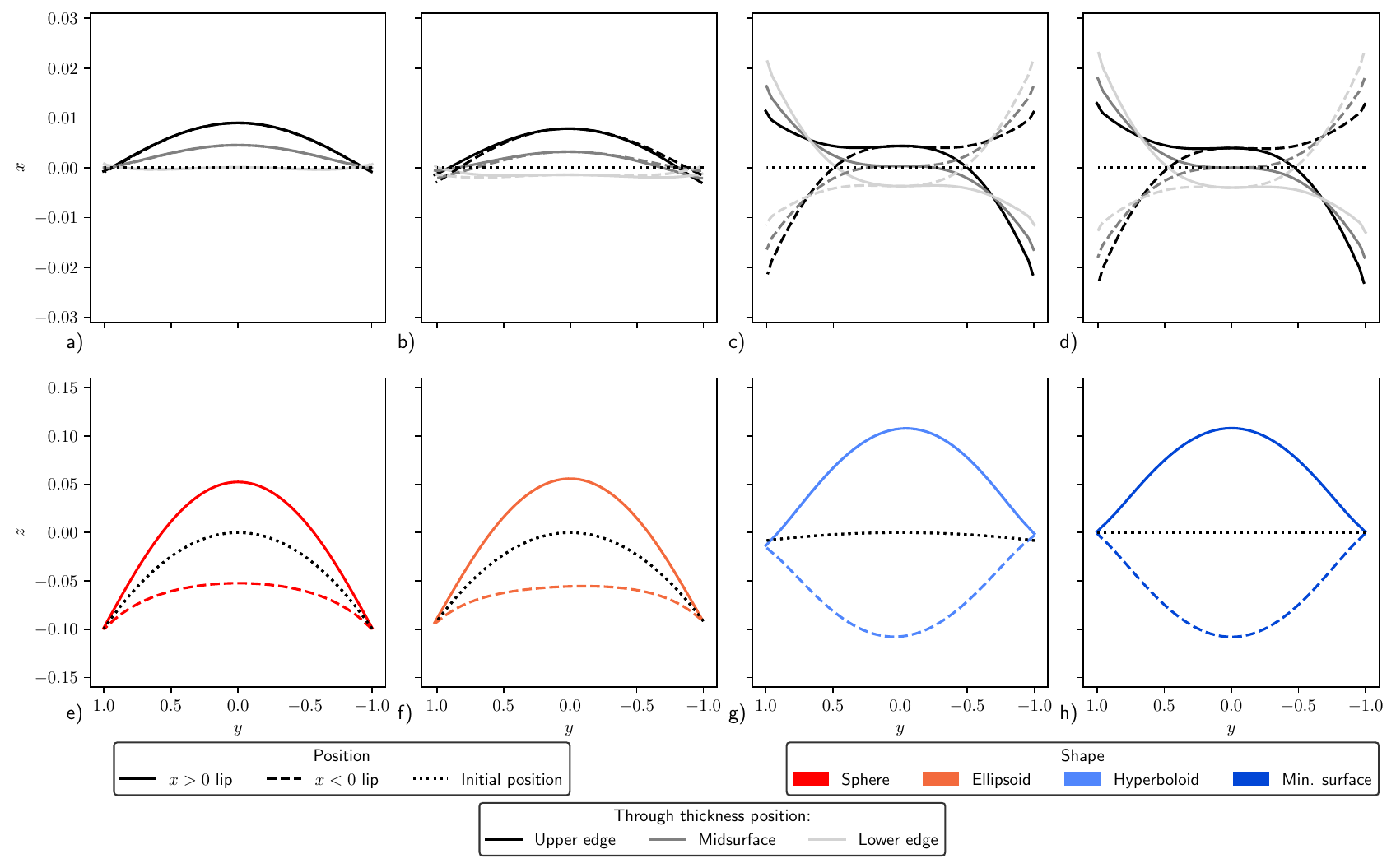}
    \caption{\emph{Crack-lip displacement fields under transverse shear loading for four representative 
    shells with crack orientation $\beta=45^\circ$.
    From left to right, the shell curvature profiles are: sphere ($\kappa_X/\kappa_Y=1$), ellipsoid ($\kappa_X/\kappa_Y=0.833$), hyperboloid 
    ($\kappa_X/\kappa_Y=-0.833$), and minimal surface ($\kappa_X/\kappa_Y=-1$).
    (a--d) The projections of the computed crack-lip displacement fields onto the $yx$ plane. 
    (e--h) The corresponding projections onto the $yz$ plane. 
    The ellipsoid and sphere show no discernible opening, and the expected curvature effect on the out-of-plane profile.
    The hyperboloid and minimal surface show incompatible in-plane displacements, with the crack tips not meeting.
    The results for the cylinder are omitted, since convergence was not reached during the calculations.}}
    \label{fig:DisplacementsTransverse}
\end{figure}

We then show in Figure~\ref{fig:DisplacementsTransverse} the results for the transverse shear
loading $\tilde{V}^0_{x}$.
The load is applied as a parabolic stress distribution through the shell thickness, consistent
with the generalized plane stress assumptions.
This is the only loading case where the out-of-plane displacement is directly induced by the
applied loading, rather than arising as a consequence of the membrane-bending coupling generated
by shell curvature.
Analogously to the twisting loading, we obtained incompatible displacement fields.
Therefore, we omit presenting stress intensity factors and energy release rates.
Additionally, we also omit the cylinder results since convergence was not reached.

In a center-cracked infinite flat plate, one expects an antisymmetric displacement profile around
the crack.
Each side assumes a half-elliptic shape with the opposite sign.
For a positive $K_{III}$, the crack lip with $x>0$ deforms upwards, whereas that with $x<0$ deforms
downwards \citep{Knott:1973}.

The out-of-plane profiles we obtain here are consistent with this behavior in all geometries once
superposed onto the curved midsurface of the shell.
The ellipsoids, where the crack is curved along its length, show one crack lip flattening while the
other becomes more curved.
The hyperboloids, instead, where the crack is nearly flat along its length, exhibit a deformation
profile that closely resembles that of a flat plate.

The in-plane profiles are markedly different in magnitude. 
The hyperboloids show a much larger displacement compared to the ellipsoids.
All are incompatible, with the crack lips failing to meet at the crack tips.
This is particularly evident for the hyperboloid and minimal surface, where the crack lips remain far apart rather than being only slightly misaligned.

For all geometries, both twisting and transverse loading produced incompatible in-plane
displacements.
Figures~\ref{fig:DisplacementsTwisting},\ref{fig:DisplacementsTransverse} show crack lips that do
not meet at the crack tips.
As argued by \citet{Hui:1993}, and corroborated by our previous results, the presence of the 
stresses induced by one of these loads must produce the other one.
The transverse load remains a peculiar case since it is the only load for which the out-of-plane
displacement is enforced, rather than arising as a consequence of the kinematic coupling induced 
by curvature.
From Figures~\ref{fig:DisplacementsTwisting},\ref{fig:DisplacementsTransverse}, we observe that the
incompatibility is more pronounced under transverse loading and in hyperboloids.
These two observations allowed us to trace the origin of this incompatibility to the shallow shell
theory itself. \
The shallowness assumption neglects the in-plane contributions of the transverse shear stress
resultants, thereby simplifying the rotational equilibrium equations.
Specifically, the equilibrium equation around the $z$-axis becomes trivial and is thus omitted in
Equations~\eqref{eqn:Equilibrium}.
Moreover, the twisting moment stress resultants become symmetric, meaning that $M_{xy}\sim M_{yx}$
\citep{elias:1967}. \
Especially when we apply transverse loads, the geometric coupling present in cracks embedded at an
angle with the principal directions of the shell clashes with these assumptions, leading to a
solution that is not physically admissible.
Indeed, observing Figures~\ref{fig:DisplacementsTransverse}.c,d, we see that the applied transverse
load produces an out-of-plane profile similar to that of a center-cracked infinite flat plate
subjected to a transverse load.
However, the geometric coupling induces a rotation around the $z$-axis of the crack lips.
The rotations have equal magnitude and opposite signs, and have no counteracting action because of the shallowness assumptions, leading to an unphysical deformation of the shell.

This rotation is exactly analogous to that observed for the twisting loading, although much larger in magnitude.

Given that the two sides of the crack in hyperboloids give rise to large interpenetrations, it is
reasonable to infer that accounting for crack closure in inclined cracks in curved shells could help
in obtaining a physical solution.
However, most of the semi-analytical solutions accounting for crack closure are developed for 
bending loading.
Therefore, they assume line contact between either the upper or the lower edges of the crack, but
not between the upper edge on one side and the lower edge on the other side of the crack.
Our displacement results qualitatively hint at this phenomenon being likely to occur when transverse
loading is dominant.
Therefore, the currently available closure models would not be applicable either to scenarios where
transverse loading $\tilde{V}^0_{x}$ is dominant.

These deficiencies in the model were not noticed in the source material \citep{Yahsi:1981,Yahsi:1983}. Therefore, the tabulated results there available for twisting and transverse loading should not be used as a reference for convergence of finite element models and verification of experimental set-ups.

\printcredits

\bibliography{Bib}

\end{document}